\documentclass[acmsmall,screen,nonacm]{acmart}
\setcopyright{none}

\AtBeginDocument{%
  }

\usepackage{amsmath,amsfonts,amssymb,amsthm}
\usepackage{enumitem}
\usepackage{xcolor}
\usepackage{graphicx}
\usepackage{pgfplots}
\pgfplotsset{compat=1.18}
\usepackage{wrapfig}
\usepackage{mathpartir}

\usepackage{mathabx}
\usepackage{cancel}
\usepackage{stmaryrd}
\usepackage{verbatim}
\usepackage{fancyvrb}
\usepackage{tcolorbox}

\renewcommand{\sec}[1]{Section~\ref{sec:#1}}

\newtheorem{theorem}{Theorem}[section]
\newtheorem{proposition}[theorem]{Proposition}
\newtheorem{corollary}[theorem]{Corollary}
\newtheorem{lemma}[theorem]{Lemma}

\newtheorem{definition}[theorem]{Definition}

\definecolor{ACMBlue}{RGB}{0, 150, 255}
\definecolor{ACMRed}{RGB}{220, 20, 60}

\newcommand{\share}[0]{\curlyveedownarrow{}}
\newcommand{\SATS}[0]{\hookrightarrow}
\newcommand{\SATSP}[0]{\hookrightarrow _ \perp}
\newcommand{\SATSL}[0]{\hookrightarrow \bigwedge}
\newcommand{\SATSR}[0]{\hookrightarrow _ \perp \bigvee}

\newcommand{\TP}[2]{\langle {#1}, {#2}\rangle}
\newcommand{\VALS}[1]{\llbracket{#1}\rrbracket}
\newcommand{\TERMS}[1]{\mathcal{T}\llbracket{#1}\rrbracket}
\newcommand{\TERMSP}[1]{\mathcal{T}_\perp\llbracket{#1}\rrbracket}

\newcommand{\tunit}[0]{\texttt{unit}}
\newcommand{\ttrue}[0]{\texttt{tt}}
\newcommand{\tfalse}[0]{\texttt{ff}}
\newcommand{\tfix}[3]{\texttt{fix } #1(#2).\ #3}
\newcommand{\tfixn}[4]{\texttt{fix}^{#1} \texttt{ } #2(#3).#4}
\newcommand{\tif}[3]{\texttt{if } #1 \texttt{ then } #2 \texttt{ else } #3}
\newcommand{\tmatchlist}[5]{\texttt{match } #1 \ \{\texttt{nil} \hookrightarrow #2\ |\ \texttt{cons} (#3, #4) \hookrightarrow #5 \}}
\newcommand{\tmatchlistabbr}[1]{\texttt{match } #1 \ \{\dots \}}
\newcommand{\tmatchpair}[4]{\texttt{match } #1 \ \{ (#2, #3) \hookrightarrow #4 \}}
\newcommand{\tlet}[3]{\texttt{let } #1 = #2 \texttt{ in } #3}
\newcommand{\ttick}[1]{\texttt{tick } #1}

\newcommand{\Tunit}[0]{\text{Unit}}
\newcommand{\Tok}[0]{\text{Ok}}
\newcommand{\Tbool}[0]{\text{Bool}}
\newcommand{\Tlist}[1]{\text{List}[{#1}]}

\font\stixarr=stix-mathsfit at 10pt
\newcommand{\onlyto}[0]{\mathrel{\text{\stixarr\char"B3}}}

\newcommand{\stack}[2]{\genfrac{}{}{0pt}{}{#1}{#2}}

\newcommand{\ltletunit}[2]{\texttt{let } #1 = () \texttt{ in } #2}
\newcommand{\ltlambda}[2]{\lambda #1.#2}
\newcommand{\ltletpair}[4]{\texttt{let } #1 = (#2, #3) \texttt{ in } #4}
\tcbset{
linearbox/.style={width=\linewidth, sharp corners, boxrule=0.4pt, colback=white, colframe=black, left=\fboxsep, right=\fboxsep, top=\fboxsep, bottom=\fboxsep, before skip=0pt, after skip=0pt, fontupper=\small, before upper={\setlength{\abovedisplayskip}{0pt}\setlength{\belowdisplayskip}{0pt}}},
linearnontyping/.style={linearbox, colbacktitle=white, coltitle=black},
lineartyping/.style={linearbox, colbacktitle=black, coltitle=white}
}

\AtBeginDocument{%
  \setlength{\textfloatsep}{8pt plus 2pt minus 2pt}%
  \setlength{\floatsep}{8pt plus 2pt minus 2pt}%
  \setlength{\intextsep}{8pt plus 2pt minus 2pt}%
  \setlength{\belowcaptionskip}{0pt}%
}

\title{Two-Sided Amortized Resource Analysis: Verifying Resource Lower Bounds via Ill-Typing}

\author{Zimu Chen}
\affiliation{%
  \department{Key Lab of High Confidence Software Technologies, Ministry of Education Department of Computer Science and Technology, School of Computer Science}
  \institution{Peking University}
  \city{Beijing}
  \country{China}}
\author{Di Wang}
\authornote{Corresponding author.}
\affiliation{%
  \department{Key Lab of High Confidence Software Technologies, Ministry of Education Department of Computer Science and Technology, School of Computer Science}
  \institution{Peking University}
  \city{Beijing}
  \country{China}}

\begin{abstract}
This paper presents a \emph{two-sided type system with potential}, a type-based technique for verifying resource lower bounds of functional programs. Automatic amortized resource analysis (AARA) verifies upper bounds: types carry potential, and a typing derivation certifies that the available resources are sufficient for evaluation. A lower bound, however, is a necessity statement---the program evaluates only if enough resources are available---and prior type-based approaches obtain it by switching the substructural discipline of potential, so that upper and lower bounds live in separate type systems. Yet the two kinds of bounds complement each other: a matching pair characterizes a program's cost exactly, and the gap between them limits how much the cost can vary across inputs of the same size.

Our type system combines AARA-style potential with the two-sided type systems of Ramsay and Walpole, whose central guarantee is that ill-typed programs do not evaluate. In a two-sided judgment, typing formulas may constrain arbitrary terms and may appear on both sides of the turnstile; the judgment $t : A \vdash q$ states that if $t$ evaluates to a result of type $A$, then at least $q$ units of resource must be available---that is, $q$ is a lower bound. Because potential makes typing formulas non-replicable, our system follows an affine discipline instead of the replicable contexts of the original framework. We define the meaning of types by semantic typing over \emph{value states}, pairs of a value and a remaining resource amount, and we prove soundness on a safety fragment; corollaries recover the classical two-sided guarantees and the soundness of both kinds of bounds in a single proof system. A prototype type checker in OCaml verifies, on a suite of 19 benchmarks, matching pairs of bounds that characterize costs exactly, as well as lower bounds for programs whose upper bounds are unavailable. Finally, as a parallel exploration, we study a two-sided linear type system in which every value is a non-replicable resource; there, closures that capture linear resources call for dual typing formulas, and the exact accounting explains the affine design of the main system.
\end{abstract}



\begin{document}

\maketitle

\section{Introduction}
\label{sec:intro}

\begin{wrapfigure}{r}{0.45\linewidth}
\vspace{-1em}
\centering
\begin{tikzpicture}
\begin{axis}[
    width=0.98\linewidth, height=4.7cm,
    xlabel={input size}, ylabel={cost},
    xlabel style={at={(axis description cs:0.97,0.07)}, anchor=south east},
    xmin=8, xmax=51, ymin=0, ymax=165,
    axis lines=left,
    tick label style={font=\scriptsize}, label style={font=\scriptsize},
]
\addplot[color=red, dashed, thick, domain=10:49, samples=50] {0.05*x^2 + 32};
\addplot[color=green!45!black, semithick, domain=10:49, samples=2] {2*x + 10};
\addplot[color=green!45!black, semithick, densely dashdotted, domain=10:49, samples=2] {x + 10};
\addplot[only marks, mark=*, mark size=1pt, color=green!55!black, opacity=0.4] coordinates {
(10,29.7) (10,20.2) (10,23.8) (10,22.9) (11,30.0) (11,27.5) (11,31.1) (11,29.2)
(12,24.1) (12,29.3) (12,31.7) (12,29.6) (13,35.9) (13,28.2) (13,25.6) (13,24.7)
(14,35.4) (14,35.9) (14,34.5) (14,24.9) (15,31.6) (15,30.1) (15,35.0) (15,28.1)
(16,40.8) (16,32.8) (16,41.3) (16,33.8) (17,34.4) (17,31.0) (17,34.7) (17,34.5)
(18,39.6) (18,45.8) (18,44.7) (18,30.9) (19,35.1) (19,42.5) (19,38.5) (19,29.5)
(20,45.2) (20,47.3) (20,40.5) (20,32.7) (21,42.7) (21,48.9) (21,41.5) (21,44.1)
(22,46.5) (22,33.0) (22,34.2) (22,49.4) (23,42.6) (23,47.4) (23,34.9) (23,37.4)
(24,37.3) (24,56.4) (24,43.9) (24,38.2) (25,43.3) (25,40.7) (25,56.9) (25,37.4)
(26,48.9) (26,55.9) (26,51.0) (26,61.8) (27,37.0) (27,41.2) (27,41.0) (27,38.9)
(28,42.5) (28,63.5) (28,53.3) (28,60.5) (29,51.2) (29,44.2) (29,61.3) (29,50.2)
(30,69.6) (30,50.3) (30,40.4) (30,62.6) (31,63.4) (31,58.5) (31,60.5) (31,67.0)
(32,51.1) (32,53.8) (32,43.2) (32,57.9) (33,56.8) (33,56.3) (33,56.0) (33,48.1)
(34,50.8) (34,63.4) (34,50.6) (34,57.2) (35,65.2) (35,49.0) (35,70.5) (35,69.3)
(36,81.3) (36,69.0) (36,72.0) (36,79.5) (37,81.1) (37,82.5) (37,77.0) (37,52.1)
(38,65.9) (38,57.5) (38,65.6) (38,73.6) (39,79.0) (39,59.2) (39,66.0) (39,63.0)
(40,73.9) (40,81.3) (40,67.1) (40,64.7) (41,89.3) (41,86.0) (41,68.2) (41,78.4)
(42,71.0) (42,78.9) (42,56.1) (42,55.4) (43,62.3) (43,56.4) (43,74.5) (43,55.3)
(44,84.6) (44,75.9) (44,75.5) (44,64.4) (45,90.8) (45,91.4) (45,97.6) (45,73.5)
(46,69.0) (46,90.0) (46,67.0) (46,69.9) (47,81.2) (47,61.2) (47,102.9) (47,74.3)
(48,64.8) (48,78.8) (48,89.2) (48,77.1) (49,64.6) (49,107.3) (49,95.9) (49,88.9)
};
\addplot[color=blue, dashed, thick, domain=10:49, samples=2] {0.6*x + 12};
\end{axis}
\end{tikzpicture}
\setlength{\abovecaptionskip}{4pt}
\caption{Per-input costs (points), worst-case and best-case costs (solid, dash-dotted), and sound upper and lower bounds (dashed).}
\label{fig:bounds}
\vspace{-1em}
\end{wrapfigure}
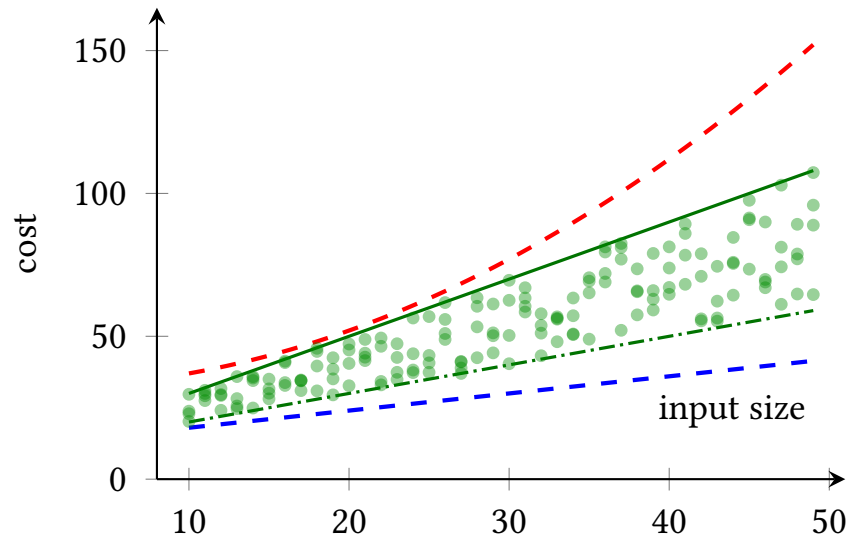
Resource analysis reasons about how much resource---time, memory, energy---a program consumes. An \emph{upper bound} is a (possibly symbolic) value that the actual cost never exceeds; a \emph{lower bound} is a value that the actual cost never falls below. Figure~\ref{fig:bounds} illustrates the two kinds of bounds. Upper-bound analyses are well studied; in particular, automatic amortized resource analysis (AARA)~\cite{POPL:HJ03, JMSCS:HJ22} is a well-established type-based technique: types carry \emph{potential} in the sense of amortized analysis.
AARA has grown from linear potential functions to polynomial~\cite{hoffmann2010amortized}, multivariate~\cite{POPL:HAH11}, and exponential~\cite{FOSSACS:KH20} ones, and has been adapted to higher-order, imperative, and probabilistic programs~\cite{JMSCS:HJ22}.

Lower bounds have received much less attention in type-based resource analysis. At first glance, a lower bound looks less useful, because it promises only that a program is at least so expensive. However, lower bounds complement upper bounds. A matching pair of bounds characterizes a program's cost exactly and certifies that neither bound can be improved. Even a non-matching pair is informative: every cost lies between the two bounds, so the gap between them limits how much the cost can vary across inputs of the same size. This variation is what constant-time programming seeks to eliminate, because a cost that depends on secret data leaks information.

\paragraph{Sufficiency versus necessity.}
A classical typing judgment is a \emph{sufficiency} statement. The slogan ``well-typed programs do not go wrong,'' combined with potential, yields ``well-typed programs are given \emph{sufficient} resources'': the assumptions on the left of $\vdash$ are enough---possibly more than enough---for the conclusion on the right. For example, let $\underline{\texttt{t}} := \tif{x}{\ttick{2}}{\ttick{3}}$, where the primitive $\ttick{q}$ consumes $q \ge 0$ units of resource and returns a unit value, and is the only source of cost in our language. Consider the AARA-style judgment
$$
x : \TP{\Tbool}{3} \vdash \underline{\texttt{t}} : \Tunit
$$
Here, $\TP{\Tbool}{3}$ annotates the Boolean type with $3$ units of \emph{potential}: resource credit that AARA associates with values through their types and uses to pay for evaluation. Thus $x : \TP{\Tbool}{3}$ requires a Boolean value for $x$ and at least $3$ available resource units. In our system, potential contributes to a global budget; it is not tied to individual variables. For example, $x : \TP{\Tbool}{2}, y : \TP{\Tunit}{3}$ and $x : \TP{\Tbool}{3}, y : \TP{\Tunit}{2}$ impose the same requirements: a Boolean $x$, a unit value $y$, and at least $5$ resource units in total.

Intuitively, the former typing judgment reads: \emph{if} $x$ is a Boolean and $3$ units of resource are available, \emph{then} $\underline{\texttt{t}}$ evaluates to a unit value.\footnote{More precisely, the term evaluates to a unit value \emph{unless it diverges}: upper-bound judgments are partial-correctness statements (\sec{semantics}). The example terms in this section all terminate, so we elide this qualification.} This is exactly an upper bound of $3$. A lower bound needs the converse, a \emph{necessity} statement: the program evaluates \emph{only if} enough resource is available. One-sided typing can express necessity, as \citet{SP:NDF17} demonstrate: varying the substructural discipline of AARA's potential yields all three kinds of bounds---\emph{affine} potential, which may be discarded, gives upper bounds; \emph{relevant} potential, which must all be consumed even though it need not cover the whole cost, gives lower bounds; and \emph{linear} potential gives constant resource usage. The judgment form never changes; each discipline fixes one reading of the judgment---sufficiency or necessity---once and for all, so each kind of bound lives in its own type system, and the two readings cannot interact within one derivation. Moreover, necessity stays confined to resources: such a system cannot refute a typing, and it cannot state a necessary condition on a program's input given its result. This raises a natural question: \emph{can sufficiency and necessity coexist in a single type system, so that one proof system derives upper bounds, lower bounds, and refutations alike?}

\paragraph{Our approach: ill-typing with potential.}
We draw the necessity abstraction from the \emph{two-sided type systems} recently introduced by \citet{ramsay2024ill}, which make the reading of a formula a matter of \emph{position} in the judgment rather than a global choice. A two-sided type system treats typing judgments as sequents $\Gamma \vdash \Delta$: typing formulas $t : A$ may constrain arbitrary terms rather than only variables, formulas may appear on both sides of $\vdash$, the left-hand side is a conjunction, and the right-hand side is a disjunction. An empty right-hand side denotes absurdity, so proving $t : \Tok \vdash$ refutes the typability of $t$, where $\Tok$ is the type containing all values. The resulting guarantee is the logical dual of the classical one: \emph{ill-typed programs do not evaluate}. Our key observation is that this duality is precisely what lower bounds need. Combining ill-typing with potential turns ``ill-typed programs do not evaluate'' into ``under-resourced programs do not evaluate.''

Concretely, this paper presents a \emph{two-sided type system with potential}. To state a lower bound, we move the term of interest to the left of $\vdash$ and place the necessary condition on the right, reversing the direction of the implication:
$$
\underline{\texttt{t}} : \Tunit \vdash x : \TP{\Tbool}{2} .
$$
This judgment reads: \emph{if} $\underline{\texttt{t}}$ evaluates to a unit value, \emph{then} $x$ is a Boolean and at least $2$ units of resource are available. That is, $2$ is a lower bound on the cost of $\underline{\texttt{t}}$. The quantity matters: with $\TP{\Tbool}{3}$ on the right, the judgment would be false---and is not provable in our system---because when $x = \ttrue$, the term evaluates with only $2$ units.

The same reversal extends to functions through the \emph{necessity arrow} $A \onlyto B$ of \citet{ramsay2024ill}: a function has type $A \onlyto B$ if, whenever an application of it returns a result of type $B$, the argument must have had type $A$. Consider the recursive function
\begin{align*}
    \underline{\texttt{s}} = \tfix{f}{\ell}{\tmatchlist{\ell}{\ttick{5}}{x}{xs}{\underline{\texttt{t}}; f\ xs}} .
\end{align*}
Classical AARA proves $\vdash \underline{\texttt{s}} : \TP{\Tlist{\TP{\Tbool}{3}}}{5} \to \Tunit$, an upper bound of $3n + 5$ where $n$ is the length of the input list. The inner annotation accounts for $3$ units of potential per Boolean list element, while the outer $5$ contributes constant potential independent of the list's length. Together, they specify a total potential of $3n+5$. Our system proves
$$
\vdash \underline{\texttt{s}} : \TP{\Tlist{\TP{\Tbool}{2}}}{5} \onlyto \Tunit ,
$$
a lower bound of $2n + 5$: for the function to return a unit value, the argument must be a list of Booleans, and at least $2n + 5$ units of resource must be available. Upper and lower bounds are thus derived in a single proof system, differing only in which side of $\vdash$ the term occupies and in the choice between $\to$ and $\onlyto$.

\paragraph{Challenges.}
Instantiating two-sided typing with potential is not a matter of re-annotating the rules of \citet{ramsay2024ill}. We highlight two challenges.

First, potential breaks replicability. In the original two-sided system, typing formulas carry no resources, so the context $\Gamma$ may be shared freely among the premises of a rule. In our setting, a formula such as $x : \TP{\Tbool}{2}$ carries potential, and duplicating it would duplicate resources and invalidate the bounds. Our typing rules therefore follow an \emph{affine} discipline: contexts are split among premises rather than shared, except that premises describing mutually exclusive branches of the same conditional or match may receive the same portion, because at most one branch evaluates. This discipline governs both the right rules and the left rules---the rules that analyze a term on the left of $\vdash$ by case analysis on its evaluation.
Second, the soundness proof requires a safety restriction. We define the meaning of types by semantic typing over \emph{value states}, i.e., pairs of a value and a remaining resource amount, and we interpret a judgment $\Gamma \vdash \Delta$ as a resource-aware implication between the two sides. Soundness for recursive programs relies on types denoting safety properties, so that a violation is already visible in some finite unfolding of a fixpoint. Necessity arrows whose codomains are higher-order types violate this requirement. Following \citet{ramsay2024ill}, we prove soundness for a \emph{safety fragment} that restricts the codomain of a necessity arrow to first-order observable types; the fragment retains all the types used in our examples and benchmarks.

\paragraph{Implementation.}
We implement the idea as a prototype type checker, which is guided by user-provided type annotations at the points where type inference would have to guess intermediate types. It verifies both upper-bound judgments $q \vdash t : A$ and lower-bound judgments $t : A \vdash q$. We evaluate the checker on a suite of 19 benchmark programs, ranging from simple tests that isolate individual typing rules to list-processing functions such as string comparison, zipping, and stack operations. The checker verifies lower bounds for all 19 programs and upper bounds for 14 of them; for three programs, the two bounds match and thus characterize the cost exactly. The experiments also demonstrate that lower bounds remain informative where upper bounds fail: the checker verifies a constant lower bound for a higher-order program that has no meaningful upper bound, a linear lower bound for a nested loop whose quadratic cost lies outside the linear potential functions implemented in our prototype, and matching bounds that distinguish a constant-time string comparison from a variant that leaks timing information.

\paragraph{Contributions.}
This paper makes the following contributions:
\begin{itemize}
    \item We design a \emph{two-sided type system with potential} for a call-by-value functional language with recursion (\sec{language}). A single proof system derives upper bounds, lower bounds, and refutations, following an affine context discipline that departs from the replicable contexts of prior two-sided type systems.
    \item We define a semantic-typing model over value states and prove the soundness of the type system on a safety fragment (\sec{semantics}). Corollaries recover the classical two-sided guarantees as well as the soundness of both kinds of resource bounds.
    \item We implement a prototype type checker in OCaml that verifies both kinds of bounds through annotation-guided proof search, and we evaluate it on a suite of 19 benchmarks (\sec{checking}).
\end{itemize}

In addition, we develop a two-sided \emph{linear} type system where all values are non-replicable resources (summarized in \sec{linear}, with the full study in our technical report). The linear setting serves as a parallel exploration: exact resource accounting exhibits the two-sided context discipline in its purest form, and supporting closures that capture linear resources calls for \emph{dual} typing formulas. The comparison explains why the main system pairs its affine discipline with ordered numeric potential rather than exact linear accounting.



\section{Language and Typing}
\label{sec:language}

\begin{figure}[!t]
\centering
\tcbset{before skip=0pt, after skip=0pt, fontupper=\small}
\begin{tcolorbox}[title=Terms and Values,colbacktitle=white, coltitle=black, width=\linewidth, sharp corners, boxrule=0.4pt, colback=white, colframe=black, left=\fboxsep, right=\fboxsep, top=\fboxsep, bottom=\fboxsep, before upper={\setlength{\abovedisplayskip}{0pt}\setlength{\belowdisplayskip}{0pt}}]

\begin{center}
\begin{minipage}[t]{0.55\linewidth}
\setlength{\abovedisplayskip}{0pt}
\setlength{\belowdisplayskip}{0pt}
\begin{equation*}\begin{aligned}[t]
    t := \
    & \ \ x\\
    & |\ \tunit \ |\ \tfalse \ |\ \ttrue \\
    & |\ \tfix{f}{x}{t}\\
    & |\ t \ t \\
    & |\ \tif{t}{t}{t}\\
    & |\ [] \ |\ t :: t \\
    & |\ \tmatchlist{t}{t}{hd}{tl}{t}\\
    & |\ (t,t) \\
    & |\ \tmatchpair{t}{x}{y}{t}\\
    & |\ \tlet{x}{t}{t}\\
    & |\ \ttick{q}\\
    & \ 
\end{aligned}\end{equation*} 
\end{minipage}
\hfill
\begin{minipage}[t]{0.35\linewidth}
\setlength{\abovedisplayskip}{0pt}
\setlength{\belowdisplayskip}{0pt}
\begin{equation*}\begin{aligned}[t]
    v := \
    & \ \ \tunit \ |\ \tfalse \ |\ \ttrue \\
    & |\ \tfix{f}{x}{t}\\
    & |\ (v_1, v_2)\\
    & |\  [] \ | \ v_1 :: v_2
\end{aligned}\end{equation*}
\end{minipage}
\end{center}

\end{tcolorbox}
\caption{Terms and values.}
\label{fig:terms}
\end{figure}

We first introduce a call-by-value functional language; we then present its two-sided type system with potential.

\paragraph{Terms and values.} The \emph{terms} and \emph{values} of the language (typically $t$ and $v$, respectively) are defined by the grammar in Figure~\ref{fig:terms}. Here $x$, $hd$, and $tl$ range over term variables. The cost constant $q$ in $\ttick{q}$ is always non-negative.
We use the following syntactic sugar: $t_1 ;\ t_2$ for $\tlet{x}{t_1}{t_2}$ when $x$ does not appear in $t_2$; $\lambda x.\ t$ for $\tfix{f}{x}{t}$ when $f$ does not appear in $t$; and $[t_1, t_2, \dots, t_n]$ for $t_1 :: t_2 :: \dots ::t_n ::[]$. An $n$-tuple abbreviates nested pairs: $(t_1, t_2, \dots, t_n)$ stands for $(t_1, (t_2, (\dots, t_n)))$, $(t_1)$ for $t_1$, and $()$ for $\tunit$.

\begin{figure}[!t]
\centering
\tcbset{before skip=0pt, after skip=0pt, fontupper=\small}
\begin{tcolorbox}[title=Evaluation Rules,colbacktitle=white, coltitle=black, width=\linewidth, sharp corners, boxrule=0.4pt, colback=white, colframe=black, left=\fboxsep, right=\fboxsep, top=\fboxsep, bottom=\fboxsep, before upper={\setlength{\abovedisplayskip}{0pt}\setlength{\belowdisplayskip}{0pt}}]

\begin{mathpar}
\text{(E-App1)}\inferrule{t_1, q \rightarrow t_1', q'}{t_1\ t_2, q \rightarrow t_1'\ t_2, q'}
\and
\text{(E-App2)}\inferrule{t_2, q \rightarrow t_2', q'}{v_1\ t_2, q \rightarrow v_1\ t_2', q'}
\and
\text{(E-AppAbs)}\inferrule{\ }{(\tfix{f}{x}{t})\ v, q \rightarrow t[f\mapsto \tfix{f}{x}{t}][x\mapsto v], q}
\and
\text{(E-If)}\inferrule{t_1, q \rightarrow t_1', q'}{\tif{t_1}{t_2}{t_3}, q \rightarrow \tif{t_1'}{t_2}{t_3}, q'}
\and
\text{(E-IfTrue)}\inferrule{\ }{\tif{\ttrue}{t_2}{t_3}, q \rightarrow t_2, q}
\and
\text{(E-IfFalse)}\inferrule{\ }{\tif{\tfalse}{t_2}{t_3}, q \rightarrow t_3, q}
\and
\text{(E-List)}\inferrule{t_i, q \rightarrow t_i', q'}{[v_1 ,\dots ,v_{i-1},t_i, \dots, t_{n}], q \rightarrow [v_1 ,\dots ,v_{i-1},t_i', \dots, t_{n}], q'}
\and
\text{(E-MList)}\inferrule{t_1, q \rightarrow t_1', q'}{\tmatchlist{t_1}{t_2}{hd}{tl}{t_3}, q \rightarrow \tmatchlistabbr{t_1'}, q'}
\and
\text{(E-MListNil)}\inferrule{\ }{\tmatchlist{[]}{t_2}{hd}{tl}{t_3}, q \rightarrow t_2, q}
\and
\text{(E-MListCons)}\inferrule{\ }{\tmatchlist{v_1 :: v_2}{t_2}{hd}{tl}{t_3}, q \rightarrow t_3[hd \mapsto v_1][tl \mapsto v_2], q}
\and
\text{(E-PairL)}\inferrule{t_1, q \rightarrow t_1', q'}{(t_1,t_2), q \rightarrow (t_1',t_2), q'}
\and
\text{(E-PairR)}\inferrule{t_2, q \rightarrow t_2', q'}{(v_1,t_2), q \rightarrow (v_1,t_2'), q'}
\and
\text{(E-MPair1)}\inferrule{t_1, q \rightarrow t_1', q'}{\tmatchpair{t_1}{x}{y}{t_2}, q \rightarrow \tmatchpair{t_1'}{x}{y}{t_2}, q'}
\and
\text{(E-MPair2)}\inferrule{\ }{\tmatchpair{(v_1,v_2)}{x}{y}{t}, q \rightarrow t[x \mapsto v_1][y \mapsto v_2], q}
\and
\text{(E-Tick)}\inferrule{q \ge p}{\ttick{p}, q \rightarrow \tunit, q - p}
\and
\text{(E-Let1)}\inferrule{t_1, q \rightarrow t_1', q'}{\tlet{x}{t_1}{t_2}, q \rightarrow \tlet{x}{t'_1}{t_2}, q'}
\and
\text{(E-Let2)}\inferrule{\ }{\tlet{x}{v_1}{t_2}, q \rightarrow t_2[x \mapsto v_1], q}
\end{mathpar}

\end{tcolorbox}
\caption{Evaluation rules.}
\label{fig:evaluation}
\end{figure}

Many AARA systems need an explicit sharing construct to split the potential of a variable that is used twice. Our typing rules support such sharing without a dedicated construct; we discuss this in \sec{typesystem}.

\paragraph{Evaluation.}

We use small-step evaluation rules in the form $t,q \rightarrow t',q'$, listed in Figure~\ref{fig:evaluation}. This form means that if the \emph{closed} term $t$ is given $q$ units of resource, then it steps to $t'$ with \emph{exactly} $q'$ units of resource: resources cannot be discarded during evaluation. Evaluation is deterministic: for every pair $t, q$, there exists at most one pair $t', q'$ such that $t,q \to t', q'$.
Whenever we write $t,q \rightarrow t',q'$, we implicitly require that $q, q' \ge 0$.

The \emph{multistep} evaluation relation $\rightarrow^*$ is defined as the reflexive transitive closure of $\rightarrow$.
We also use $t,q \rightarrow^\omega$ to indicate that the term $t$ diverges when given $q$ units of resource. Formally: $\forall t', q'$, if $t, q \rightarrow^* t', q'$, then $\exists t'', q''$ such that $t',q' \rightarrow t'' , q''$.

Evaluation is defined for terms that need not be well typed. A configuration is \emph{normal} if it cannot step further; a normal configuration is \emph{stuck} if its term is not a value. For example, $(\ttick{2},1)$ is stuck for lack of resources, while $(\tif{\tunit}{\ttrue}{\tfalse},p)$ is stuck because its condition is not Boolean. Neither configuration represents successful evaluation.

The relation $t_1\lesssim t_2$, defined formally in \sec{safety}, compares how defined two terms are. It requires $t_2$ to preserve every normal outcome of $t_1$ at the same initial resource, including the final term and remaining resource. If $t_1$ diverges at a given initial resource, it imposes no condition on $t_2$ at that resource.

The evaluation relation satisfies a standard weakening property:

\begin{theorem}[Weakening in evaluation]
\label{thm:evaluation-weakening}
    For all $t, t', q, q', p$, if $t, q \rightarrow^* t', q'$ and $p \ge 0$, then $t, q + p \rightarrow^* t', q' + p$.
\end{theorem}

\subsection{Type System}
\label{sec:typesystem}

We introduce a \emph{two-sided type system with potential}, based on the two-sided framework of \citet{ramsay2024ill}. Types come in two layers: $T$ ranges over types without potential, and $A$ ranges over types with potential, which pair a type with a non-negative amount of potential. The types without potential comprise the base types $\Tunit$, $\Tbool$, and $\Tok$, as well as sufficiency arrow, necessity arrow, pair, and list types, whose components are types with potential.
\begin{align*}
    T := \ & \Tunit\ |\ \Tbool\ |\ \Tok \ |\ A\rightarrow B\ |\ A\onlyto B\ |\ (A,B) \ |\ \Tlist{A} \\
    A := \
    & \langle T, q\rangle \quad (q \ge 0)
\end{align*}

The syntactic sugar for pairs of values also applies to pairs of types, giving the following notation for $n$-tuple types: $(A_1,A_2, \dots, A_n)$ for $(A_1,(A_2, (\dots, A_n)))$, $(A)$ for $A$, and $()$ for $\Tunit$.
Unless stated otherwise, a type means a type with potential, and we write $T$ as shorthand for $\TP{T}{0}$.

\paragraph{Typing judgment.} In the two-sided type system of \citet{ramsay2024ill}, typing judgments have the form $\Gamma \vdash \Delta$, where $\Gamma$ and $\Delta$ are not limited to variable formulas $x : A$ as in classical type systems: each is a list of typing formulas of the same shape. We reuse this design, except that our typing formulas use types with potential, such as $\tunit : \TP{\Tunit}{3}$. Unlike classical AARA, we need no dedicated syntax for constant potential in a judgment, because the formula $\tunit : \TP{\Tunit}{q}$ already represents $q$ units of potential (\sec{semantics} makes this precise). For brevity, in typing judgments we write a single number $q$ as shorthand for $\tunit : \TP{\Tunit}{q}$, and we write $t:T$ for $t : \TP{T}{0}$.

Thus, a typing judgment has the form $\Gamma \vdash \Delta$, where $\Gamma, \Delta$ are called \emph{typing environments}, and their forms are similar:
\begin{align*}
\Gamma & :=\  \emptyset \ |\ \Gamma, t : A &
\Delta & :=\  \emptyset \ |\ \Delta, t : A
\end{align*}

Before introducing the typing rules, we clarify how to read a typing judgment such as $[x, y] : \Tlist{\TP{\Tbool}{2}}, z : \Tunit, 3 \vdash (y, z) : \TP{(\Tbool, \Tunit)}{7}$. In a classical AARA type system, a typing judgment $x : A, y : B \vdash t : C$ can be read as: if variable $x$ has type $A$ and variable $y$ has type $B$, then $t$ has type $C$, and the total potential on the left is no less than the potential required on the right. In our work, the reading is slightly different. For a judgment $s : A, t : B \vdash u : C$, because $s$, $t$, and $u$ are arbitrary terms, we read $s : A$ as ``$s$ evaluates to a result of type $A$'' rather than ``$s$ has type $A$''. We say ``result'' rather than ``value'' because semantic typing interprets types over \emph{value states}, pairs of a value and a remaining resource amount (\sec{type-semantics}). For example, $(\ttick{3} , 5)$ evaluates to $(\tunit, 2)$, which satisfies type $\TP{\Tunit}{2}$ but not $\TP{\Tunit}{3}$. Thus, evaluating $\ttick{3}$ to a result of type $\TP{\Tunit}{2}$ requires at least $5$ initial units of resource.
Now suppose we have a substitution $\theta$ that maps all free variables in the judgment to values, and suppose we have total resource $p$. The judgment $s : A, t : B \vdash u : C$ means: if we can split $p = p_1 + p_2$ such that $s\theta$, given initial resource $p_1$, evaluates to a result of $A$, and $t\theta$, given initial resource $p_2$, evaluates to a result of $B$, then $u\theta$, given initial resource $p$, evaluates to a result of $C$ unless it diverges.

Additionally, unlike classical AARA systems, we allow multiple typing formulas on the right. The meaning is disjunction: $\Gamma \vdash s : A, t : B$ states that, under every substitution and resource split that satisfies $\Gamma$, at least one of $s : A$ and $t : B$ is satisfied. A special but useful case is an empty $\Delta$, which represents \emph{absurdity}, the unit of disjunction: proving $\Gamma \vdash$ shows that the formulas in $\Gamma$ cannot hold simultaneously \cite{ramsay2024ill}. For example, $x : \TP{\Tunit}{2}, x : \TP{\Tbool}{3} \vdash$ is derivable because no value is both a unit and a Boolean. This refutation ability is less central in our work, because we focus on resource bounds; the judgments we prove more often look like $x : \TP{\Tunit}{2}, x : \TP{\Tunit}{3} \vdash 5$.

\sec{semantics} gives the formal semantics of types and judgments.

\paragraph{Sharing relation.}
Some classical AARA systems use $\texttt{share }x\texttt{ as }y,z\texttt{ in }\cdots$ to give two uses of a value distinct variable names while splitting its potential. Our contexts admit arbitrary terms, so repetitions arise even for values: the left-hand side may contain two occurrences of $\tunit:\Tunit$. We allow these repetitions and account for their combined potential. For example, $x:\TP{\Tunit}{2}, x:\TP{\Tunit}{3}$ is equivalent to $x:\TP{\Tunit}{5}$. The sharing relation makes this equivalence precise.

The sharing relation is a ternary relation $\alpha \share (\alpha_1,\alpha_2)$ describing how a type, a typing formula, or an environment can be split into two parts while splitting its potential. On types, the relation is defined by the rules in Figure~\ref{fig:sharing}.

\begin{figure}[!t]
\centering
\tcbset{before skip=0pt, after skip=0pt, fontupper=\small}
\begin{tcolorbox}[title=Sharing Rules,colbacktitle=white, coltitle=black, width=\linewidth, sharp corners, boxrule=0.4pt, colback=white, colframe=black, left=\fboxsep, right=\fboxsep, top=\fboxsep, bottom=\fboxsep, before upper={\setlength{\abovedisplayskip}{0pt}\setlength{\belowdisplayskip}{0pt}}]

\begin{mathpar}
\text{(SH-Unit)}\inferrule{\ }{\Tunit\share (\Tunit, \Tunit)}
\and
\text{(SH-Bool)}\inferrule{\ }{\Tbool\share (\Tbool, \Tbool)}
\and
\text{(SH-Pair)}\inferrule{A\share (A_1,A_2) \quad B\share (B_1,B_2)}{(A,B)\share ((A_1,B_1), (A_2,B_2))}
\and
\text{(SH-Arrow)}\inferrule{\ }{A \to B\share (A \to B, A \to B)}
\and
\text{(SH-List)}\inferrule{A \share (A_1, A_2)}{\Tlist{A}\share (\Tlist{A_1},\Tlist{A_2})}
\and
\text{(SH-ArrowN)}\inferrule{\ }{A\onlyto B \share (A\onlyto B,A\onlyto B)}
\and
\text{(SH-Pot)}\inferrule{T \share (T_1, T_2) \quad q = q_1 + q_2}{\TP{T}{q}\share (\TP{T_1}{q_1}, \TP{T_2}{q_2})}
\end{mathpar}
\end{tcolorbox}
\caption{Sharing rules.}
\label{fig:sharing}
\end{figure}

The type relation specifies how potential is divided. To share a typing formula, we also require its term to be value-like, as defined below.

\begin{definition}
\label{def:value-like}
    A term $t$ is \emph{value-like}, written $VL(t)$, if it is either a value or a variable.
\end{definition}

Because the language is call-by-value, every variable is bound to a value at run time, so values and variables can be treated uniformly. The underlying requirement for sharing is resource-constancy:

\begin{definition}
\label{def:resource-constant}
    A term $t$ is \emph{resource-constant}, written $RC(t)$, if $t$ evaluates to a value given $0$ initial resource.
\end{definition}

Every value-like term is resource-constant, so $VL(t)$ is an easily decidable sufficient condition for $RC(t)$. Resource-constancy ensures that sharing does not count the cost of evaluating a term twice. For example, if we applied sharing to $\ttick{2} : \Tunit \vdash 4$, we would obtain $\ttick{2} : \Tunit, \ttick{2} : \Tunit \vdash 4$, converting a non-provable judgment into a provable one by duplicating resources. It would be sound to use $RC(t)$ in place of $VL(t)$, but $RC(t)$ is hard to check directly. We therefore use the syntactic condition $VL(t)$; in practice, we have not found this restriction limiting (\sec{checking}).

With these requirements on types and terms, we extend sharing to typing formulas and environments.

On formulas, $t : A \share (t:A_1, t:A_2)$ if and only if $A \share (A_1,A_2)$ and $t$ is value-like.
On environments, $\Gamma \share (\Gamma_1, \Gamma_2)$ if and only if the three environments have the same length and, at every position, the formulas are related by sharing.

In a left-hand context, a formula and its sharing-related pair can replace each other while preserving value constraints and the total resource requirement, and hence semantic validity. Likewise, if $\Gamma\share(\Gamma_1,\Gamma_2)$, replacing a left-hand subcontext $\Gamma$ by $\Gamma_1,\Gamma_2$, or conversely, preserves semantic validity.

\paragraph{Typing rules.} With the sharing relation in place, we present the derivation rules in four categories: structural rules and right rules (Figure~\ref{fig:structural-right-rules}), left rules, and OK rules.

\begin{figure}[!t]
\centering
\tcbset{before skip=0pt, after skip=0pt, fontupper=\small}
\begin{tcolorbox}[title=Structural Rules, width=\linewidth, sharp corners, boxrule=0.4pt, colback=white, colframe=black, left=\fboxsep, right=\fboxsep, top=\fboxsep, bottom=\fboxsep, before upper={\setlength{\abovedisplayskip}{0pt}\setlength{\belowdisplayskip}{0pt}}]
\begin{mathpar}
\text{(T-Id)}\inferrule{\ }{\Gamma, t:A \vdash t:A, \Delta}
\and
\text{(T-Share)}\inferrule{VL(t) \quad A \share (A_1, A_2) \quad \Gamma, t:A_1, t: A_2 \vdash \Delta}{\Gamma, t:A \vdash \Delta}
\and
\text{(T-Relax1)}\inferrule{\Gamma, t: \langle T, q\rangle \vdash \Delta \quad q' \ge q}{\Gamma, t: \langle T, q'\rangle \vdash \Delta}
\and
\text{(T-Relax2)}\inferrule{\Gamma\vdash t: \langle T, q\rangle, \Delta \quad q' \le q}{\Gamma \vdash t: \langle T, q'\rangle, \Delta}
\and
\text{(T-Resc)}\inferrule{\Gamma, t:\TP{T}{q + p} \vdash t_1:\TP{T_1}{q_1 + p}, \dots, t_n:\TP{T_n}{q_n + p} \\ q+p\ge 0 \quad \forall i.\ q_i+p\ge 0}{\Gamma, t:\TP{T}{q} \vdash t_1:\TP{T_1}{q_1}, \dots, t_n:\TP{T_n}{q_n}}
\end{mathpar}
\end{tcolorbox}

\begin{tcolorbox}[title=Right Rules, width=\linewidth, sharp corners, boxrule=0.4pt, colback=white, colframe=black, left=\fboxsep, right=\fboxsep, top=\fboxsep, bottom=\fboxsep, before upper={\setlength{\abovedisplayskip}{0pt}\setlength{\belowdisplayskip}{0pt}}]
\begin{mathpar}
\text{(T-Unit)}\inferrule{\ }{\Gamma \vdash \tunit: \Tunit, \Delta}
\and
\text{(T-Bool)}\inferrule{b \in \{\ttrue,\tfalse\}}{\Gamma \vdash b: \Tbool, \Delta}
\and
\text{(T-Abs)}\inferrule{\Gamma\share (\Gamma,\Gamma) \quad \Gamma, f:A\rightarrow B, x:A \vdash t : B, \Delta}{\Gamma \vdash \tfix{f}{x}{t} : A \rightarrow B, \Delta}
\and
\text{(T-App)}\inferrule{\Gamma_1 \vdash t_1:A\rightarrow B, \Delta \quad \Gamma_2 \vdash t_2:A, \Delta}{\Gamma_1, \Gamma_2 \vdash t_1\ t_2 : B, \Delta}
\and
\text{(T-AbsN)}\inferrule{\Gamma\share (\Gamma,\Gamma) \quad \Gamma, f:A\onlyto B, t:B \vdash x:A, \Delta}{\Gamma\vdash \tfix{f}{x}{t} : A \onlyto B, \Delta}
\and
\text{(T-If)}\inferrule{\Gamma_1\vdash t_1:\langle \Tbool, q\rangle, \Delta \quad \Gamma_2,q\vdash t_2:A, \Delta \quad \Gamma_2,q\vdash t_3:A, \Delta}{\Gamma_1, \Gamma_2\vdash \tif{t_1}{t_2}{t_3}: A, \Delta}
\and
\text{(T-ListNil)}\inferrule{\ }{\Gamma \vdash []: \Tlist{A}, \Delta}
\and
\text{(T-ListCons)}\inferrule{\Gamma_1 \vdash t_1 :A, \Delta \quad \Gamma_2 \vdash t_2:\Tlist{A}, \Delta}{\Gamma_1, \Gamma_2 \vdash t_1 :: t_2:\Tlist{A}, \Delta}
\and
\text{(T-ListMatch)}\inferrule{\Gamma_1 \vdash t_1:\langle \Tlist{A}, q \rangle, \Delta \quad \Gamma_2, q\vdash t_2:B, \Delta \\ \Gamma_2, hd:A, tl:\langle \Tlist{A}, q \rangle \vdash t_3:B, \Delta}{\Gamma_1, \Gamma_2 \vdash \tmatchlist{t_1}{t_2}{hd}{tl}{t_3}: B, \Delta}
\and
\text{(T-Pair)}\inferrule{\Gamma_1 \vdash t_1:A_1,\Delta \quad \Gamma_2 \vdash t_2:A_2,\Delta}{\Gamma_1,\Gamma_2 \vdash (t_1, t_2):(A_1,A_2), \Delta}
\and
\text{(T-PairMatch)}\inferrule{\Gamma_1 \vdash t_1:\TP{(A_1,A_2)}{q},\Delta \quad \Gamma_2, x:A_1, y:A_2, q \vdash t_2:A,\Delta}{\Gamma_1,\Gamma_2 \vdash \tmatchpair{t_1}{x}{y}{t_2} :A, \Delta}
\and
\text{(T-Let)}\inferrule{\Gamma_1\vdash t_1 : B, \Delta \quad \Gamma_2, x : B \vdash t_2 : A, \Delta}{\Gamma_1,\Gamma_2 \vdash \tlet{x}{t_1}{t_2} :A, \Delta}
\and
\text{(T-Tick)}\inferrule{\Gamma \vdash q, \Delta}{\Gamma \vdash \ttick{q}: \Tunit, \Delta}
\end{mathpar}
\end{tcolorbox}
\caption{Structural rules and right rules.}
\label{fig:structural-right-rules}
\end{figure}

\paragraph{Structural rules.}
$\text{T-Id}$ is the axiom of the proof system: a typing formula derives itself. %
$\text{T-Share}$ implements the sharing relation on the left: a proof under the split formulas $t:A_1, t:A_2$ yields a proof under $t:A$ when $VL(t)$ and $A\share(A_1,A_2)$. The term and type requirements were introduced above in the discussion of sharing.

$\text{T-Relax1}$ and $\text{T-Relax2}$ allow us to discard resources. $\text{T-Relax1}$ may look redundant because $\text{T-Share}$ achieves the same effect for value-like terms; however, formulas of non-value-like terms cannot be split, so the rule is necessary.

$\text{T-Resc}$ (the \emph{rescaling} rule) allows us to add or remove an equal amount of resource on both sides. $p$ may be either positive or negative as long as $q + p$ and all $q_i + p$ are non-negative.

\paragraph{Right rules.} The right rules are a conservative extension of the classical AARA rules to the two-sided setting. $\text{T-Unit}$, $\text{T-Bool}$, and $\text{T-ListNil}$ type values that have no components. The type system is affine in resources: discarding resources is allowed, so $\Gamma$ may be non-empty in these rules. $\text{T-Tick}$ is equally direct: with $q$ units of resource, $\ttick{q}$ evaluates to a unit value. These rules are stated only for formulas whose type has no potential (e.g., not for $\tunit : \TP{\Tunit}{2}$); to reason about a term with potential, one must apply $\text{T-Resc}$ first.

$\text{T-Abs}$ types abstractions with sufficiency arrows. The premise $\Gamma \share (\Gamma, \Gamma)$ requires $\Gamma$ to carry no resources: for every $t : A$ in $\Gamma$, the type $A$ must have zero potential, and $t$ must be value-like.

$\text{T-AbsN}$ types abstractions with necessity arrows. The only difference from $\text{T-Abs}$ is that the implication is reversed: the body having type $B$ implies that the argument has type $A$, matching the meaning of necessity arrows.

$\text{T-If}$ is similar to the corresponding rule in classical AARA. The important point is that evaluation is ordered: $t_1$ is evaluated before $t_2$ or $t_3$, so resources remaining after evaluation of $t_1$ can be reused in the selected branch. The $\text{T-ListMatch}$, $\text{T-PairMatch}$, and $\text{T-Let}$ rules follow the same pattern.

$\text{T-ListCons}$ and $\text{T-Pair}$ type terms constructed by the cons and pair constructors. For a constructed term to have a type, its components must have the corresponding types under split environments.

\newcommand{\clo}[1]{{\ [#1]}}

\begin{figure}[!t]
\centering
\tcbset{before skip=0pt, after skip=0pt, fontupper=\small}
\begin{tcolorbox}[title=Left Rules, width=\linewidth, sharp corners, boxrule=0.4pt, colback=white, colframe=black, left=\fboxsep, right=\fboxsep, top=\fboxsep, bottom=\fboxsep, before upper={\setlength{\abovedisplayskip}{0pt}\setlength{\belowdisplayskip}{0pt}}]
\begin{mathpar}
\text{(T-Dis)}\inferrule{\Gamma \vdash t:A \quad A \parallel B}{\Gamma, t: B \vdash \Delta}
\and
\text{(T-TickL)}\inferrule{\Gamma, q \vdash \Delta}{\Gamma, \ttick{q}: \Tunit \vdash \Delta}
\and
\text{(T-ListL)}\inferrule{\Gamma, t_1:A, t_2:\Tlist{A}\vdash \Delta}{\Gamma, t_1 :: t_2 : \Tlist{A}\vdash \Delta}
\and
\text{(T-PairL)}\inferrule{\Gamma, t_1:A, t_2:B \vdash \Delta}{\Gamma, (t_1,t_2) : (A, B)\vdash \Delta}
\and
\text{(T-LetL)}\inferrule{\Gamma_2, t_2: B \vdash x : A; \Phi
\quad \Gamma_1, t_1:A, \Phi \vdash \Delta}{\Gamma_1, \Gamma_2, \tlet{x}{t_1}{t_2} : B\vdash \Delta}
\and
\text{(T-AppNL)}\inferrule{\Gamma_1\vdash t_1:A\onlyto B; \Phi
\quad \Gamma_2,t_2:A,\Phi\vdash \Delta}{\Gamma_1, \Gamma_2,t_1\ t_2 : B \vdash \Delta}
\and
\text{(T-IfL)}\inferrule{\Gamma_2, t_2: A \vdash \Phi
\quad \Gamma_2, t_3: A \vdash \Phi
\quad \Gamma_1, t_1:\Tbool, \Phi\vdash \Delta}{\Gamma_1, \Gamma_2, \tif{t_1}{t_2}{t_3}: A\vdash \Delta}
\and
\text{(T-MatchL)}\inferrule{\Gamma_2, t_2: B\vdash \Phi
\quad \Gamma_2, t_3: B\vdash hd :: tl : \Tlist{A}; \Phi
\quad \Gamma_1, t_1 : \Tlist{A}, \Phi \vdash \Delta}{\Gamma_1, \Gamma_2, \tmatchlist{t_1}{t_2}{hd}{tl}{t_3}: B\vdash \Delta}
\and
\text{(T-MatchPairL)}\inferrule{\Gamma_2, t_2:A \vdash (x, y) : (A_1,A_2); \Phi \quad \Gamma_1, t_1 : (A_1,A_2), \Phi \vdash \Delta}{\Gamma_1,\Gamma_2, \tmatchpair{t_1}{x}{y}{t_2}: A\vdash \Delta}
\end{mathpar}
\end{tcolorbox}
\caption{Left rules.}
\label{fig:left-rules}
\end{figure}

\paragraph{Left rules.} The left rules (Figure~\ref{fig:left-rules}) derive typing judgments in which a complex (non-value-like) term appears on the left of $\vdash$. In most cases, the goal is either a \emph{lower bound}---the right-hand side states a resource amount that the term's evaluation necessitates---or a \emph{refutation}, where the right-hand side is empty.

$\text{T-Dis}$ refutes formulas that assign a term an impossible type, such as $\tunit : \Tbool$. The $\parallel$ relation is the \emph{disjoint} relation defined as follows:

\begin{definition}
    (Disjointness) For non-$\Tok$ types, two types $A$ and $B$ are \emph{disjoint}, written $A \parallel B$, if and only if one of the following holds: (1) $A$ is a unit type and $B$ is not; (2) $A$ is a Boolean type and $B$ is not; (3) $A$ is a list type and $B$ is not; (4) $A$ is an arrow type (sufficiency or necessity) and $B$ is neither; (5) $A = (A_1, A_2)$ is a pair type, and $B$ is either not a pair type, or is a pair type $(B_1,B_2)$ with $A_1 \parallel B_1$ or $A_2 \parallel B_2$. The type $\Tok$ is not disjoint from any type.
\end{definition}

Intuitively, $A \parallel B$ means that no value has both types, so a term that evaluates to a result of $A$ cannot evaluate to a result of $B$; it is therefore sound to refute $t:B$ once $t$ is known to have a disjoint type $A$. Disjointness is independent of potential: $A$ and $B$ may carry any amounts.

We retain $\text{T-Dis}$ primarily for compatibility with the system of \citet{ramsay2024ill}, since resource analysis usually starts from programs that pass ordinary type checking. The rule also enables additional resource proofs: a lower bound conditional on a result type may require ruling out branches that return an incompatible type. Multi-Type in \sec{evaluation} gives a concrete example where the proof needs $\text{T-Dis}$.

$\text{T-LetL}$ is the left-rule counterpart of $\text{T-Let}$. This rule and several rules after it share two features:

\begin{itemize}
    \item A connective ``;'', distinct from ``,'', appears on the right-hand side of some judgments.
    \item A symbol $\Phi$, representing a single typing formula, appears on the right-hand side of one premise and on the left-hand side of another.
\end{itemize}

We first ignore these features and consider a simplified form, $\text{T-LetL1}$, which is sound but incomplete:
$$
\text{(T-LetL1)}\inferrule{\Gamma_2, t_2: B \vdash x : A
\quad \Gamma_1,  t_1:A \vdash \Delta}{\Gamma_1, \Gamma_2 , \tlet{x}{t_1}{t_2} : B\vdash   \Delta}  
$$
To prove $\Gamma_1, \Gamma_2, \tlet{x}{t_1}{t_2} : B\vdash \Delta$, we need to reason from a condition containing a let term. Before looking at the rule, consider the difference between a $\texttt{let}$ term appearing as a conclusion and as a condition. On the right-hand side, to prove that a $\texttt{let}$ term has type $B$, we only need to (1) find \emph{some} type $A$ such that $t_1 : A$; and (2) prove that $t_2 : B$ under the assumption $x : A$. On the left-hand side, the situation is different. To simplify the intuition, consider the special case where $\Delta = \emptyset$, i.e., we want to refute that the $\texttt{let}$ term has type $B$. We cannot refute it just by finding some type $A$ such that $t_1 : A$ and $x : A \vdash t_2 : B$ do not hold simultaneously, because there may be another type for $t_1$ under which the $\texttt{let}$ term is well-typed. A complete refutation would require proving that, for all types $A$, $t_1 : A$ and $x : A \vdash t_2 : B$ do not hold simultaneously, which is impossible in this system. Inspired by the $\text{T-LetL2}$ rule of \citet{ramsay2024ill}, we adopt a useful special case: we refute that the $\texttt{let}$ term has type $B$ by proving that (1) $t_2 : B$ implies $x : A$ for some specific $A$; and (2) $t_1 : A$ leads to the desired conclusion. This loses some completeness but covers the cases that arise in our examples and benchmarks.
The same idea applies to other terms such as $\texttt{if}$.

Now let us construct the rule for $\texttt{let}$ terms. Once again we take part of the assumptions, namely $\Gamma_2$, and use it to derive an intermediate result by destructing the $\texttt{let}$ term. We choose $t_1 : A$ as the intermediate result, because the let binding identifies $x$ with $t_1$. The split is: (1) prove that together with $\Gamma_2$, $t_2:B$ implies $x : A$; and (2) prove that together with the remaining $\Gamma_1$, $t_1 : A$ implies the original conclusion $\Delta$. This rule extends the preceding intuition from the special case $\Delta = \emptyset$ to general $\Delta$.

Now we explain the omitted features. The ``;'' connective is not new syntax but syntactic sugar: $t_1 : A_1 ; t_2 : A_2$ abbreviates $(t_1, t_2) : (A_1, A_2)$, representing conjunction on the right-hand side of a judgment. By convention, $;$ binds more tightly than the comma, so a right-hand side containing both is parsed as a disjunction of conjunctions. Because the comma already means conjunction on the left, ``;'' is not needed there.
Meanwhile, $\Phi$ can be any \emph{single} typing formula $t : A$, possibly written with the sugar ``;'': for example, $t_1:A_1;t_2:A_2$ is by definition the single formula $(t_1,t_2):(A_1,A_2)$. Thus $\Phi$ may represent a conjunction of formulas after desugaring, but not a disjunction. The simplified form $\text{T-LetL1}$ cannot handle the following case:

\begin{itemize}
    \item $\Delta$ is to be split into two subgoals, for example, it has the form $(t_1,t_2) : (A_1,A_2)$ or $t_1 :: t_2 : \Tlist{A}$;
    \item One of its subgoals is to be satisfied by the body of the $\texttt{let}$ term;
    \item The other is to be satisfied by the variable of the $\texttt{let}$ term.
\end{itemize}

In the cons branch of Example 2 in \sec{examples}, the goal is $x::xs:\Tlist{\TP{\Tbool}{2}}$. The bound conditional supplies the head's Boolean type and $2$ resource units, while the recursive call supplies the tail's list type and potential. Without $\Phi$, decomposing the let first leaves the entire cons formula in one subgoal; decomposing the cons first assigns the entire let formula to one subgoal. Neither order combines the two contributions. Choosing $\Phi=xs:\Tlist{\TP{\Tbool}{2}}$ transfers the tail information from the body premise to the bound-expression premise, where it is combined with the head information. A constant resource formula cannot express this requirement, because the tail's potential depends on its length.

The different meanings of commas explain the asymmetry with the right rules. On the left, commas express conjunction, so context partitioning distributes jointly available resources. On the right, commas express disjunction, so partitioning the context cannot provide two pieces that are needed together. The conjunctive transfer through $\Phi$ supplies those pieces without duplicating resources.

$\text{T-IfL}$ has a similar resource reading. If the conditional returns a result of type $A$, then its selected branch does so. The first two premises establish the same necessary condition $\Phi$ for either branch; the last combines this condition with $t_1:\Tbool$ and the remaining context $\Gamma_1$ to derive $\Delta$. The branches share $\Gamma_2$ because they are mutually exclusive. In particular, with empty $\Gamma_2$ and $\Phi=q$, the branch premises state a common lower bound $q$ on the resources required for either branch to return a result of type $A$.

$\text{T-MatchL}$ and $\text{T-MatchPairL}$ are the analogous rules for match terms on the left.

$\text{T-AppNL}$ is the left-rule counterpart of $\text{T-App}$. Suppose we want to prove $\Gamma_1, \Gamma_2,t_1 \ t_2 : B \vdash  \Delta$, where the left-hand side contains an application term. In the simplified case without $\Phi$, we divide the goal into two subgoals: (1) part of the assumptions, namely $\Gamma_1$, implies the intermediate formula $t_1 : A\onlyto B$; and (2) under the remaining assumptions $\Gamma_2$, the assumption $t_2:A$ implies the original conclusion $\Delta$. The reason is that $t_1:A\onlyto B$ and $t_1 \ t_2:B$ together provide the necessary condition $t_2:A$. Thus, after extracting this necessary condition from the application, it is enough to prove the goal from $\Gamma_2,t_2:A$.

$\text{T-PairL}$ and $\text{T-ListL}$ destruct formulas whose terms are pair or list constructions: such a formula can be replaced directly by its components. Because multiple formulas on the left-hand side represent conjunction, this equivalence is straightforward.

Finally, $\text{T-TickL}$ reads a formula $\ttick{q}:\Tunit$ on the left as the availability of $q$ units of resource.
As with the right rules, some left rules are defined only on formulas whose type has zero potential. To apply them to types with potential, one must first apply $\text{T-Resc}$ or $\text{T-Relax1}$.


\begin{figure}[!t]
\centering
\tcbset{before skip=0pt, after skip=0pt, fontupper=\small}
\begin{tcolorbox}[title=OK Rules, width=\linewidth, sharp corners, boxrule=0.4pt, colback=white, colframe=black, left=\fboxsep, right=\fboxsep, top=\fboxsep, bottom=\fboxsep, before upper={\setlength{\abovedisplayskip}{0pt}\setlength{\belowdisplayskip}{0pt}}]
\begin{mathpar}
\text{(T-OkVar)}\inferrule{\ }{\Gamma \vdash x:\Tok, \Delta}
\and
\text{(T-Ok)}\inferrule{\Gamma \vdash t:A, \Delta \quad A \neq \Tok}{\Gamma \vdash t:\Tok, \Delta}
\and
\text{(T-TickOkL)}\inferrule{\Gamma, q \vdash \Delta}{\Gamma, \ttick{q}: \Tok \vdash \Delta}
\end{mathpar}
\end{tcolorbox}
\caption{OK rules.}
\label{fig:ok-rules}
\end{figure}

\paragraph{OK rules.}
The type $\Tok$ expresses that a term evaluates successfully to some value, without specifying a more precise value type.
The two core rules in Figure~\ref{fig:ok-rules} follow this interpretation.
$\text{T-OkVar}$ says that a variable may always be given type $\Tok$, because substitutions map variables to values in the call-by-value setting.
$\text{T-Ok}$ says that any successful typing by a non-$\Tok$ type also gives an $\Tok$ typing, because every ordinary type denotes values.

The auxiliary rule $\text{T-TickOkL}$ is the $\Tok$-typed counterpart of $\text{T-TickL}$.
When $\ttick{q}$ appears on the left with type $\Tok$, evaluating it still consumes exactly $q$ units of resource and produces a value, so the rule exposes the same resource formula $q$.

\subsection{Examples}
\label{sec:examples}

\paragraph{Example 1.} To show how the type system derives resource bounds, consider a simple term:
$$\underline{\texttt{t}} := \tif{x}{\ttick{2}}{\ttick{3}}$$

Under the environment $\Gamma = x:\Tbool$, this term has a tight upper bound of $3$ and a tight lower bound of $2$. We can prove both by deriving the following typing judgments:
\begin{align*}
	\Gamma,3 & \vdash \underline{\texttt{t}}: \Tunit &
	\Gamma, \underline{\texttt{t}}: \Tunit & \vdash 2
\end{align*}

The first judgment corresponds to a classical AARA judgment. Formally, under any substitution that replaces the only free variable $x$ with some value $v$, if we can evaluate $v$ to a $\Tbool$ given resource $q$ (i.e., $v$ is either $\ttrue$ or $\tfalse$), then we can evaluate $\underline{\texttt{t}}$ to a $\Tunit$ given resource $q+3$. Because $v$ is a value, it can be evaluated with resource $0$. Thus, we can evaluate $\underline{\texttt{t}}$ to a $\Tunit$ with resource $3$, giving the upper bound.

In the proof displays below, premises that follow immediately using $\text{T-Id}$, $\text{T-Relax1}$, $\text{T-Relax2}$, $\text{T-Resc}$, or $\text{T-Unit}$ are left unexpanded. To keep the displays compact, a subscripted $\Pi$ may name a subgoal whose full judgment and justification are given below the display.

In the following derivation, $\text{T-If}$ uses $q=0$, because $x:\Tbool$ is short for $x:\TP{\Tbool}{0}$. We omit the zero-resource assumption in its branch premises. Applying $\text{T-Tick}$ first in each branch leaves an immediate resource judgment:
\begin{mathpar}\small
\inferrule{
    \Gamma\vdash x:\Tbool
    \quad
    \inferrule{
        3 \vdash 2
    }{3 \vdash \ttick{2}: \Tunit}(\text{T-Tick})
    \quad
    \inferrule{
        3 \vdash 3
    }{3\vdash \ttick{3}: \Tunit}(\text{T-Tick})
}{\Gamma, 3 \vdash \underline{\texttt{t}}: \Tunit}(\text{T-If})
\end{mathpar}

The second judgment is less direct. It means that under any substitution, if we can evaluate $v$ to a $\Tbool$ given resource $q$ (which can be $0$, as above), and we can evaluate $\underline{\texttt{t}}$ to a $\Tunit$ given resource $q'$, then we can evaluate $\tunit$ to a $\TP{\Tunit}{2}$ given resource $q + q'$ (or in fact $q'$), i.e., we have at least $2$ units of resource.

To show this is true, we apply $\text{T-IfL}$ with $\Gamma_1 = \emptyset$, $\Gamma_2 = \Gamma$, $A=\Tunit$, $\Delta=2$, and $\Phi=2$. Then the two branch premises say that if either branch evaluates to $\Tunit$, then the available resource is at least $2$. Apply $\text{T-TickL}$ first in each branch; its resource premise and the final premise of $\text{T-IfL}$ are immediate:
\begin{mathpar}\small
\inferrule{
    \inferrule{
        \Gamma,2 \vdash 2
    }{\Gamma,\ttick{2}:\Tunit \vdash 2}(\text{T-TickL})
    \quad
    \inferrule{
        \Gamma,3 \vdash 2
    }{\Gamma,\ttick{3}:\Tunit \vdash 2}(\text{T-TickL})
    \quad
    x:\Tbool,2 \vdash 2
}{\Gamma, \underline{\texttt{t}}: \Tunit \vdash 2}(\text{T-IfL})
\end{mathpar}

Note that $\Gamma$ is unused in the second proof tree: once $\underline{\texttt{t}} : \Tunit$ appears on the left, it already implies that $x$ evaluates to a Boolean, so the assumption in $\Gamma$ is redundant. The derivation introduces copies of $x : \Tbool$ through $\Gamma$ and through the destruction of $\texttt{if}$; all can be discarded when closing the leaves.

\paragraph{Example 2.} The second example is a recursive list-processing function that reuses the term $\underline{\texttt{t}}$ from Example 1 inside the cons branch; \sec{checking} shows its type-checker version:
\[
\underline{\texttt{f2}} = \tfix{f}{\ell}{\tmatchlist{\ell}{\tunit}{x}{xs}{\tlet{u}{\underline{\texttt{t}}}{f\ xs}}}.
\]
Here the free variable $x$ of $\underline{\texttt{t}}$ is the head variable introduced by the cons branch of the list match. The function evaluates $\underline{\texttt{t}}$ once for each element before recursively processing the tail.

For the proofs, name the cons branch $\underline{\texttt{b}}$ and the match body $\underline{\texttt{m}}$, and recall the definition of $\underline{\texttt{t}}$:
\begin{align*}
\underline{\texttt{b}} & = \tlet{u}{\underline{\texttt{t}}}{f\ xs}, \\
\underline{\texttt{m}} & = \tmatchlist{\ell}{\tunit}{x}{xs}{\underline{\texttt{b}}}, \\
\underline{\texttt{t}} & = \tif{x}{\ttick{2}}{\ttick{3}}.
\end{align*}
We follow depth-first proof search, grouping consecutive rule applications and using the same convention for immediate premises.

\emph{Upper-bound proof.} The target judgment is
\[
\vdash \underline{\texttt{f2}}:\Tlist{\TP{\Tbool}{3}}\rightarrow\Tunit.
\]

Apply $\text{T-Abs}$ followed by $\text{T-ListMatch}$, with zero potential on the outer list type. We name the scrutinee premise $\Pi_1$ and the nil-branch premise $\Pi_2$. As before, we omit the zero-resource assumption of the match:
\begin{mathpar}\small
\inferrule{
    \inferrule{
        \Pi_1
        \quad
        \Pi_2
        \quad
        f:\Tlist{\TP{\Tbool}{3}}\rightarrow\Tunit,x:\TP{\Tbool}{3},xs:\Tlist{\TP{\Tbool}{3}} \vdash \underline{\texttt{b}}:\Tunit
    }{f:\Tlist{\TP{\Tbool}{3}}\rightarrow\Tunit,\ell:\Tlist{\TP{\Tbool}{3}} \vdash \underline{\texttt{m}}:\Tunit}(\text{T-ListMatch})
}{\vdash \underline{\texttt{f2}}:\Tlist{\TP{\Tbool}{3}}\rightarrow\Tunit}(\text{T-Abs})
\end{mathpar}

The named premises are
\begin{align*}
\Pi_1 &: \quad \ell:\Tlist{\TP{\Tbool}{3}} \vdash \ell:\Tlist{\TP{\Tbool}{3}}, \\
\Pi_2 &: \quad f:\Tlist{\TP{\Tbool}{3}}\rightarrow\Tunit \vdash \tunit:\Tunit.
\end{align*}
They follow immediately by $\text{T-Id}$ and $\text{T-Unit}$, respectively.

We continue with the cons branch. Applying $\text{T-Let}$ separates the conditional from the recursive call:
\begin{mathpar}\small
\inferrule{
    x:\TP{\Tbool}{3} \vdash \underline{\texttt{t}}:\Tok
    \quad
    f:\Tlist{\TP{\Tbool}{3}}\rightarrow\Tunit,u:\Tok,xs:\Tlist{\TP{\Tbool}{3}} \vdash f\ xs:\Tunit
}{f:\Tlist{\TP{\Tbool}{3}}\rightarrow\Tunit,x:\TP{\Tbool}{3},xs:\Tlist{\TP{\Tbool}{3}} \vdash \underline{\texttt{b}}:\Tunit}(\text{T-Let})
\end{mathpar}

For the first premise, apply $\text{T-Ok}$ and then $\text{T-If}$ with $q=3$. Applying $\text{T-Tick}$ in each branch leaves only immediate resource premises:
\begin{mathpar}\small
\inferrule{
    \inferrule{
        \Pi_3
        \quad
        \inferrule{
            3 \vdash 2
        }{3 \vdash \ttick{2}:\Tunit}(\text{T-Tick})
        \quad
        \inferrule{
            3 \vdash 3
        }{3 \vdash \ttick{3}:\Tunit}(\text{T-Tick})
    }{x:\TP{\Tbool}{3} \vdash \underline{\texttt{t}}:\Tunit}(\text{T-If})
    \quad
    \Tunit\neq\Tok
}{x:\TP{\Tbool}{3} \vdash \underline{\texttt{t}}:\Tok}(\text{T-Ok})
\end{mathpar}

Here
\[
\Pi_3:\quad x:\TP{\Tbool}{3} \vdash x:\TP{\Tbool}{3}
\]
is an instance of $\text{T-Id}$.

Returning to the second premise of $\text{T-Let}$, apply $\text{T-App}$ to the recursive call:
\begin{mathpar}\small
\inferrule{
    \Pi_4
    \quad
    xs:\Tlist{\TP{\Tbool}{3}} \vdash xs:\Tlist{\TP{\Tbool}{3}}
}{f:\Tlist{\TP{\Tbool}{3}}\rightarrow\Tunit,u:\Tok,xs:\Tlist{\TP{\Tbool}{3}} \vdash f\ xs:\Tunit}(\text{T-App})
\end{mathpar}

The function premise is
\[
\Pi_4:\quad f:\Tlist{\TP{\Tbool}{3}}\rightarrow\Tunit,u:\Tok \vdash f:\Tlist{\TP{\Tbool}{3}}\rightarrow\Tunit.
\]
It and the tail premise are instances of $\text{T-Id}$.

This completes the upper-bound judgment: each list element provides $3$ units of potential.

\emph{Lower-bound proof.} We now start from the lower-bound target:
\[
\vdash \underline{\texttt{f2}}:\Tlist{\TP{\Tbool}{2}}\onlyto\Tunit.
\]

Apply $\text{T-AbsN}$ followed by $\text{T-MatchL}$ with $\Phi=0$. The annotation selects the list type $\Tlist{\TP{\Tbool}{2}}$ for $\ell$. We name the nil-branch and scrutinee premises $\Pi_5$ and $\Pi_6$:
\begin{mathpar}\small
\inferrule{
    \inferrule{
        \Pi_5
        \quad
        f:\Tlist{\TP{\Tbool}{2}}\onlyto\Tunit,\underline{\texttt{b}}:\Tunit \vdash x::xs:\Tlist{\TP{\Tbool}{2}};0
        \quad
        \Pi_6
    }{f:\Tlist{\TP{\Tbool}{2}}\onlyto\Tunit,\underline{\texttt{m}}:\Tunit \vdash \ell:\Tlist{\TP{\Tbool}{2}}}(\text{T-MatchL})
}{\vdash \underline{\texttt{f2}}:\Tlist{\TP{\Tbool}{2}}\onlyto\Tunit}(\text{T-AbsN})
\end{mathpar}

The named premises are
\begin{align*}
\Pi_5 &: \quad f:\Tlist{\TP{\Tbool}{2}}\onlyto\Tunit,\tunit:\Tunit \vdash 0, \\
\Pi_6 &: \quad \ell:\Tlist{\TP{\Tbool}{2}},0 \vdash \ell:\Tlist{\TP{\Tbool}{2}}.
\end{align*}
Both follow by $\text{T-Id}$, using $0=\tunit:\Tunit$ for $\Pi_5$. We continue with the cons branch, where the let annotation selects $\Phi=xs:\Tlist{\TP{\Tbool}{2}}$. We name the recursive-call subgoal $\Pi_{\mathrm{rec}}$:
\begin{mathpar}\small
\inferrule{
    \Pi_{\mathrm{rec}}
    \quad
    \underline{\texttt{t}}:\Tok,xs:\Tlist{\TP{\Tbool}{2}} \vdash x::xs:\Tlist{\TP{\Tbool}{2}};0
}{f:\Tlist{\TP{\Tbool}{2}}\onlyto\Tunit,\underline{\texttt{b}}:\Tunit \vdash x::xs:\Tlist{\TP{\Tbool}{2}};0}(\text{T-LetL})
\end{mathpar}

Here the named subgoal is
\[
\Pi_{\mathrm{rec}}:\quad
f:\Tlist{\TP{\Tbool}{2}}\onlyto\Tunit,f\ xs:\Tunit \vdash u:\Tok;xs:\Tlist{\TP{\Tbool}{2}}.
\]
Apply $\text{T-AppNL}$ without an additional intermediate formula, followed by $\text{T-Pair}$. The function and tail premises follow by identity. The remaining component requires $u:\Tok$: because $u$ is unused in $f\ xs$, the left context contains no assumption about its type. Thus $u:\Tunit$ cannot be derived, whereas $\text{T-OkVar}$ gives $u:\Tok$. Since $\text{T-LetL}$ uses the same type for $u$ and $\underline{\texttt{t}}$ in its two premises, the conditional must be assigned $\underline{\texttt{t}}:\Tok$ in this derivation even though it returns unit. We use the same type in the upper-bound proof for consistency.

Returning to the second premise of $\text{T-LetL}$, destruct the conditional on the left. The annotation selects $x:\TP{\Tbool}{2}$; as in Example 1, choose $\Phi=2$. We name the two tick-branch subgoals $\Pi_{\mathrm{tick2}}$ and $\Pi_{\mathrm{tick3}}$:
\begin{mathpar}\small
\inferrule{
    \Pi_{\mathrm{tick2}}
    \quad
    \Pi_{\mathrm{tick3}}
    \quad
    x:\Tbool,2,xs:\Tlist{\TP{\Tbool}{2}} \vdash x::xs:\Tlist{\TP{\Tbool}{2}};0
}{\underline{\texttt{t}}:\Tok,xs:\Tlist{\TP{\Tbool}{2}} \vdash x::xs:\Tlist{\TP{\Tbool}{2}};0}(\text{T-IfL})
\end{mathpar}

The named subgoals are
\begin{align*}
\Pi_{\mathrm{tick2}} &: \quad \ttick{2}:\Tok \vdash 2, \\
\Pi_{\mathrm{tick3}} &: \quad \ttick{3}:\Tok \vdash 2.
\end{align*}
Applying $\text{T-TickOkL}$ reduces these subgoals to $2\vdash2$ and $3\vdash2$, respectively; both resource premises are immediate.

The remaining premise reconstructs the cons-cell formula and the zero-resource component. Apply $\text{T-Pair}$ and $\text{T-ListCons}$; the head, tail, and zero-resource premises are left unexpanded:
\begin{mathpar}\small
\inferrule{
    \inferrule{
        x:\Tbool,2 \vdash x:\TP{\Tbool}{2}
        \quad
        xs:\Tlist{\TP{\Tbool}{2}} \vdash xs:\Tlist{\TP{\Tbool}{2}}
    }{x:\Tbool,2,xs:\Tlist{\TP{\Tbool}{2}} \vdash x::xs:\Tlist{\TP{\Tbool}{2}}}(\text{T-ListCons})
    \quad
    \vdash 0
}{x:\Tbool,2,xs:\Tlist{\TP{\Tbool}{2}} \vdash x::xs:\Tlist{\TP{\Tbool}{2}};0}(\text{T-Pair})
\end{mathpar}

This completes the lower-bound proof: successful evaluation requires at least $2$ units of resource per list element.

\subsection{Summary}
\label{sec:summary}

This section introduced the target language, its resource-aware operational semantics, and the two-sided type system with potential. The main point is that the same judgment form supports both directions of resource reasoning: right rules recover standard upper-bound arguments, while left rules derive lower bounds and refutations by reading typing formulas as necessary conditions. The examples show how these two readings coexist in a single proof system.

\section{Semantics of Typing}
\label{sec:semantics}

This section develops the semantic counterpart of the type system.
\sec{type-semantics} interprets types over value states, and
\sec{judgment-semantics} defines when a judgment $\Gamma \vdash \Delta$ is
true. \sec{safety} identifies a \emph{safety fragment} of the type grammar,
which the soundness argument needs to handle recursion. Finally,
\sec{soundness} establishes that every derivable judgment in the safety
fragment is true; the classical two-sided guarantees and the soundness of both
kinds of resource bounds follow as corollaries.

\subsection{Semantics of Types}
\label{sec:type-semantics}

Let $\text{Vals}_0$ be the set of all closed values. A \emph{value state} is a pair $(v, p)$ of a closed value $v$ and a remaining resource amount $p$.

Figure~\ref{fig:type-semantics} defines the semantics of types: $\VALS{A}$ is the set of value states $(v, p)$ such that the type is correct for $v$ and the resource $p$ is sufficient.

\begin{figure}[!t]
\centering
\tcbset{before skip=0pt, after skip=0pt, fontupper=\small}
\begin{tcolorbox}[title=Semantics of Types, width=\linewidth, sharp corners, boxrule=0.4pt, colback=white, colframe=black, left=\fboxsep, right=\fboxsep, top=\fboxsep, bottom=\fboxsep, before upper={\setlength{\abovedisplayskip}{0pt}\setlength{\belowdisplayskip}{0pt}}]

\begin{align*}
    \VALS{\langle \Tunit, q\rangle} & = & \{(\tunit, p)\ | \ p \ge q\} \\
    \VALS{\langle \Tbool, q\rangle} & = & \{(\ttrue, p), (\tfalse, p)\ | \ p \ge q\} \\
    \VALS{\langle \Tok, q\rangle} & = & \{(v, p)\ | \ v \in \text{Vals}_0, p \ge q\} \\
    \VALS{\langle A \rightarrow B, q\rangle} & = & \{ (\tfix{f}{x}{t}, p) \ | \ p \ge q, \forall v, p_1, (v, p_1) \in \VALS{A} \rightarrow \\
    & & (t[f\mapsto \tfix{f}{x}{t}][x\mapsto v], p_1) \in \TERMSP{B}\} \\
    \VALS{\langle A \onlyto B, q\rangle} & = & \{ (\tfix{f}{x}{t}, p) \ | \ p \ge q, \forall v, p_1,\\
    \ & \ & (t[f\mapsto \tfix{f}{x}{t}][x\mapsto v], p_1) \in \TERMS{B} \rightarrow (v, p_1) \in \VALS{A} \} \\
    \VALS{\langle (A,B), q\rangle} & = & \{((v_1, v_2), p)\ | \ (v_1,p_1) \in \VALS{A}, (v_2,p_2) \in \VALS{B}, p \ge q + p_1+ p_2\}
    \\
    \VALS{\langle\Tlist{A}, q \rangle} & = & \{ ([a_1, a_2, \dots, a_n],p)\ |\ n \in \mathbb{N}, \exists p_1,\dots,p_n. \\
    & & \qquad \forall 1 \le i \le n.\ (a_i, p_i)\in \VALS{A},\ p \ge q + \sum_{i=1}^n p_i \}
\end{align*}
    
\par\nointerlineskip
\rule{\linewidth}{0.4pt}
\par\nointerlineskip

\begin{align*}
    \TERMS{A} &=& \{(t, p) \ | \ \exists v, p', \text{ s.t. } t,p \rightarrow^* v,p' \wedge (v, p') \in \VALS{A} \}\\
    \TERMSP{A} &=& \TERMS{A} \cup \{(t, p) \ | \ t,p \rightarrow^\omega\}
\end{align*}

\end{tcolorbox}
\caption{Semantics of types.}
\label{fig:type-semantics}
\end{figure}

$\TERMS{A}$ and $\TERMSP{A}$ lift the interpretation from values to terms: their elements are pairs $(t, p)$ of a term $t$ and an initial resource amount $p \in \mathbb{R}_{\ge 0}$. Membership $(t, p) \in \TERMS{A}$ means that, given $p$ units of resource, $t$ evaluates to a result of type $A$; $\TERMSP{A}$ additionally contains all diverging pairs. The two interpretations thus correspond to total and partial correctness, respectively.

For types without potential, we define $\TERMS{T} := \TERMS{\langle T, 0\rangle}$ and $\TERMSP{T} := \TERMSP{\langle T, 0\rangle}$.

The three interpretations are related as follows:

\begin{theorem}[Basic semantic inclusions]
\label{thm:basic-semantic-inclusions}
    For all types $A$, $\VALS{A} \subseteq \TERMS{A} \subseteq \TERMSP{A}$.
\end{theorem}

Both term interpretations are closed under backward evaluation:

\begin{lemma}[Reverse preservation]
\label{lem:reverse-preservation}
    Suppose $t, p \rightarrow^* t', p'$.
    \begin{itemize}
        \item If $(t', p') \in \TERMS{A}$, then $(t, p) \in \TERMS{A}$.
        \item If $(t', p') \in \TERMSP{A}$, then $(t, p) \in \TERMSP{A}$.
    \end{itemize}
\end{lemma}

\subsection{Semantics of Typing Relations}
\label{sec:judgment-semantics}

We first present two auxiliary definitions.

\begin{definition}[Valuation]
    A \emph{valuation} $\theta$ is a substitution that maps variables to closed values. We write $t\theta$ for the term obtained by applying $\theta$ to the term $t$; whenever we write $t\theta$, we assume that $\theta$ covers all free variables of $t$, so $t\theta$ is closed.
\end{definition}

\begin{definition}[Satisfaction]\
    \begin{itemize}
    \item A valuation $\theta$ \emph{satisfies} a formula on the left $t : A$ under initial resource $p$, written $\theta, p \SATS t :A$, if $(t\theta, p) \in \TERMS{A}$.
    \item A valuation $\theta$ \emph{satisfies} a formula on the right $t : A$ under initial resource $p$, written $\theta, p \SATSP t :A$, if $(t\theta, p) \in \TERMSP{A}$.
    \item A valuation $\theta$ satisfies a \emph{list} of formulas on the \emph{left} $\Gamma=\{s_i: A_i\}_{i=1}^n$ under an initial resource vector $\vec p=(p_1,\dots,p_n)$, written $\theta, \vec p \SATSL \Gamma$, if $\theta$ satisfies $s_i: A_i$ under $p_i$ for all $i$.
    \item A valuation $\theta$ satisfies a \emph{list} of formulas on the \emph{right} $\Delta=\{t_j: B_j\}_{j=1}^m$ under initial resource $p$, written $\theta, p \SATSR \Delta$, if $\theta$ satisfies $t_j: B_j$ under $p$ for some $j$.
\end{itemize}
\end{definition}

Intuitively, a formula specifies a condition on both the valuation and the initial resource. For $p\geq 0$, $\theta,p\SATS x:\Tbool$ holds exactly when $\theta(x)$ is a Boolean value; assignments to other variables are irrelevant. By contrast, $\theta,p\SATS\ttick{2}:\Tunit$ holds exactly when $p\geq 2$, independently of $\theta$. The left-hand relation requires successful evaluation, while its right-hand counterpart also admits divergence.

We say that a judgment $\Gamma \vdash \Delta$ is true, written $\Gamma \vDash \Delta$, if the left side semantically implies the right side with the same total resource. Let $\Gamma=\{s_i: A_i\}_{i=1}^n$ and $\Delta=\{t_j: B_j\}_{j=1}^m$. Then:
\[
\Gamma \vDash \Delta
\quad \text{iff} \quad
\forall \theta,\vec p.
\theta,\vec p \SATSL \Gamma
\Longrightarrow
\theta,\sum_{i=1}^n p_i \SATSR \Delta .
\]

In words: whenever every formula $s_i:A_i$ on the left is satisfied under its own resource $p_i$, at least one formula $t_j:B_j$ on the right must be satisfied under the total resource $\sum_{i=1}^n p_i$. If $\Delta$ is empty, then $\theta,p \SATSR \Delta$ is false for every $\theta$ and $p$, so $\Gamma \vDash \emptyset$ means that the formulas in $\Gamma$ cannot be satisfied simultaneously.

\subsection{Safety Fragment}
\label{sec:safety}

The semantic soundness proof for recursive programs needs the semantic
interpretation of types to behave like a safety property. Following
\citet{ramsay2024ill}, we identify two key requirements. First, if a program
satisfies a type, replacing one of its subterms by a less-defined term should
not invalidate the type. Second, if a recursive program violates a type, this
violation should already be visible after finitely many unfoldings. These
requirements are natural for well-behaved type systems: divergence cannot be
used as positive evidence for a bad computation, and recursive definitions can
be reasoned about through their finite approximants.

In the rest of the paper, $\underline{\texttt{div}}$ denotes a fixed closed term
that diverges for every initial resource. The finite approximants of a fixpoint
are defined as follows:
\begin{align*}
    \tfixn{0}{f}{x}{t} & = \lambda x.\underline{\texttt{div}} \\
    \tfixn{n + 1}{f}{x}{t} & =
    \lambda x.t[f \mapsto \tfixn{n}{f}{x}{t}] .
\end{align*}
We write $t_1 \lesssim t_2$ when $t_1$ is \emph{less defined} than $t_2$: every
normal configuration reached by $t_1$ is also reached by $t_2$, where a
configuration is \emph{normal} if it cannot step further. Formally,
\[
    \forall p,s,p'.\ t_1,p \to^* s,p'
    \wedge s,p' \text{ is normal}
    \implies t_2,p \to^* s,p' .
\]

\begin{definition}[Safety property]
    A type $A$ defines a safety property if, for every term context $C$ (a term
    with one hole), the following two clauses hold:
    \begin{itemize}
        \item \textbf{S1.} If $(C[t_2],p) \in \TERMSP{A}$ and
        $t_1 \lesssim t_2$, then
        $(C[t_1],p) \in \TERMSP{A}$.
        \item \textbf{S2.} If $(C[\tfix{f}{x}{t}],p) \notin \TERMSP{A}$, then
        there is some $k$ such that
        $(C[\tfixn{k}{f}{x}{t}],p) \notin \TERMSP{A}$.
    \end{itemize}
\end{definition}

Not every type of the full grammar defines a safety property. The
problem is caused by necessity arrows whose codomain is an arbitrary type,
especially a function type. Let $A_0 = (\Tok,\Tok)$ and $B_0 = \Tunit$.
Consider the type $A_0 \onlyto (B_0 \rightarrow B_0)$. The term
$\lambda x.\lambda y.\tmatchpair{x}{a}{b}{\tunit}$ satisfies this type: when
$x$ is not a pair, the pair destructor gets stuck, so the result is not a
$B_0 \rightarrow B_0$ function. However, replacing the destructor by
$\underline{\texttt{div}}$ gives the less-defined term
$\lambda x.\lambda y.\underline{\texttt{div}}$, and this term does not satisfy
$A_0 \onlyto (B_0 \rightarrow B_0)$. For a non-pair value of $x$, the returned
function belongs to $B_0 \rightarrow B_0$, because its body diverges and
function types are interpreted using $\TERMSP{\cdot}$. Thus the first safety
clause fails.

The second clause fails for the same structural reason, but with nested
necessity arrows. Consider a type of the shape
\[
    A_0 \onlyto (B_0 \rightarrow (A_0 \onlyto (B_0 \rightarrow B_0))) .
\]
A recursive term can use a list as fuel and, after consuming the whole list,
test a later argument with the pair destructor above. The full fixpoint reaches
that test for every finite fuel, so a non-pair argument can make the outer
necessity condition fail. In contrast, for any fixed finite approximant, one can
choose a fuel list longer than the number of unfoldings; the approximant then
diverges before reaching the destructor. This divergence changes the relevant
inner function membership and prevents that approximant from witnessing the
failure. Hence no single finite approximant necessarily exposes the violation,
so S2 fails for the full type grammar.

We therefore restrict attention to the following safety fragment:
\begin{align*}
    F := \ & \Tunit\ |\ \Tbool\ |\ \Tok\ |\ (A_F, A_F)\ |\ \Tlist{A_F} \\
    T := \ & \Tunit\ |\ \Tbool\ |\ \Tok\ |\ (A, B)\ |\ \Tlist{A}
    \ |\ A \rightarrow B\ |\ A \onlyto A_F \\
    A_F := \ & \langle F, q\rangle \quad (q \ge 0) \\
    A,B := \ & \langle T, q\rangle \quad (q \ge 0)
\end{align*}

The grammar is a subset of the type system from \sec{typesystem}. It follows the same
pattern as the safety fragment of \citet{ramsay2024ill}: a finitely verifiable
fragment $A_F$ contains only first-order observable structure, and the codomain
of a necessity arrow must belong to that fragment. In particular, necessity
arrows cannot demand a function type on the right. This removes exactly the
higher-order codomains used by the two counterexamples while retaining all the
types used in our examples and benchmarks.

\subsection{Soundness}
\label{sec:soundness}

The semantics of \sec{judgment-semantics} gives a direct interpretation of
typing judgments. The main theorem of this section states that derivable
judgments are true under that interpretation. Because the proof relies on the
safety-property argument for recursive terms, the theorem is stated for the
safety fragment from \sec{safety}.

\begin{theorem}[Soundness of type disjointness]
\label{thm:type-disjointness}
    For all types $A$ and $B$, if $A \parallel B$, then
    $\VALS{A} \cap \VALS{B} = \emptyset$.
\end{theorem}

\begin{corollary}
    For all types $A$ and $B$, if $A \parallel B$, then there is no term $t$ and
    no resources $p_1,p_2$ such that $(t,p_1) \in \TERMS{A}$ and
    $(t,p_2) \in \TERMS{B}$.
\end{corollary}

\begin{theorem}[Soundness of typing in the safety fragment]
    Let $\Gamma \vdash \Delta$ be a judgment whose types all belong to the
    safety fragment. If $\Gamma \vdash \Delta$ is derivable, then
    $\Gamma \vDash \Delta$.
\end{theorem}


We can instantiate the theorem with the type $\Tok$ to recover the standard
two-sided soundness consequences. Here, a closed term $t$ \emph{does not go
wrong} with resource $p$ if $(t,p) \in \TERMSP{\Tok}$, i.e., the term either
diverges or evaluates to a value; it \emph{does not evaluate} with resource $p$
if $(t,p) \notin \TERMS{\Tok}$.

\begin{corollary}[Well-typed and ill-typed programs]
    Let $t$ be a closed term, and suppose the following judgments have
    derivations whose types all belong to the safety fragment.
    \begin{itemize}
        \item If $\vdash t:\Tok$ is derivable, then $t$ does not go wrong with
        zero initial resource.
        \item If $t:\Tok \vdash$ is derivable, then for every $p \ge 0$, $t$
        does not evaluate with initial resource $p$.
    \end{itemize}
\end{corollary}

We next instantiate the theorem with resource formulas to obtain the soundness
of upper and lower bounds.

\begin{definition}[Upper-bounded and lower-bounded]
    Suppose $t$ is a closed term.
    \begin{itemize}
        \item We say that $t$ is \emph{upper-bounded} by $q$ if there exists a
        type $A$ such that $q \vdash t:A$ is derivable.
        \item We say that $t$ is \emph{lower-bounded} by $q$ if there exists a
        type $A$ such that $t:A \vdash q$ is derivable.
    \end{itemize}
\end{definition}

The next corollary spells out the resource meaning of these definitions. The
number $q$ is shorthand for the resource formula
$\tunit:\TP{\Tunit}{q}$, as in \sec{typesystem}.

\begin{corollary}[Soundness of resource bounds]
    Let $t$ be a closed term, and suppose the witnessing derivations use only
    types in the safety fragment.
    \begin{itemize}
        \item If $t$ is upper-bounded by $q$ with witness type $A$, then
        $(t,q) \in \TERMSP{A}$. Thus, with $q$ units of initial resource, $t$
        either diverges or evaluates to a result of type $A$.
        \item If $t$ is lower-bounded by $q$ with witness type $A$, then for
        every $p < q$, $(t,p) \notin \TERMS{A}$. Thus $t$ cannot evaluate to a
        result of the witness type without $q$ units of initial resource. In
        particular, when the witness type is $\Tok$, $t$ does not evaluate
        with any initial resource $p < q$.
    \end{itemize}
\end{corollary}

\section{A Prototype Two-Sided Checker}
\label{sec:checking}

This section describes an OCaml prototype for the language and type system. It implements annotation-guided proof search: a successful run gives a derivation, while failure does not show that the judgment is false. The prototype is a type checker, not a type inference algorithm. A small \texttt{typeinference} module recovers a few local types, but the user still supplies the intermediate types and auxiliary formulas that guide search.

\subsection{Type Checker Design}
\label{sec:checker-design}

\paragraph{User interface.}
The implementation provides a small command language. A command \texttt{def x := t} evaluates a closed term and stores the result as a global definition, and \texttt{norm q t} evaluates $t$ with $q$ units of initial resource. The checker is invoked by:
\[
\begin{array}{rcl}
\texttt{check right A t} &:& \vdash t:A,\\
\texttt{check left A t} &:& t:A \vdash,\\
\texttt{check upper q A t} &:& q \vdash t:A,\\
\texttt{check lower q A t} &:& t:A \vdash q.
\end{array}
\]
As in \sec{typesystem}, the number $q$ abbreviates $\tunit:\TP{\Tunit}{q}$, so the last two commands correspond to the upper- and lower-bound judgments in \sec{soundness}.

\paragraph{Implementation annotations.}
The implementation language adds annotations for choices that the formal rules leave implicit. The annotation \texttt{t as A} supplies the type used when $t$ is decomposed on the right; \texttt{t notas A} supplies the type used on the left. The latter name comes from ill-typability examples such as \texttt{unit notas Bool}, but in resource proofs it is better read as ``necessarily as'': for instance, \texttt{tick 3 notas <Unit, 7>} tells the checker which lower-bound type to place on the left.

Some constructs also accept an auxiliary term in braces, such as \texttt{f \{u\} x}, \texttt{if \{u\} b then t1 else t2}, and \texttt{let \{u\} x = t1 in t2}. This term supplies the formula $\Phi$ required by the corresponding left rule. It is not used by right rules; it only guides left-rule applications.

\begin{figure}[!t]
\begin{Verbatim}[fontsize=\small,frame=single,framesep=3pt]
def f2 := fix f(x). (
        (matchlist (x notas List <Bool, 1>) {
            nil -> unit notas Unit
          | cons(h, t) -> (
            let {t notas List <Bool, 1>} u =
              (if h notas <Bool, 1> then tick 1 else tick 2) notas Ok
            in
              f (t notas List <Bool, 1>)
            )}
        )  notas Unit
      ) notas Unit

check right (List <Bool, 1>) >- Unit f2
\end{Verbatim}
\caption{Type-checker version of the running list-processing example from
\sec{examples}, with costs $1$ and $2$ in place of $2$ and $3$.}
\label{fig:checker-example}
\end{figure}

Figure~\ref{fig:checker-example} shows the checker-level version of the running list-processing example from \sec{examples}. The final command asks for a necessity-arrow type for \texttt{f2}; the \texttt{notas} annotations provide the left-rule types for the match, conditional, let, and recursive call.

\paragraph{Simple judgments.}
The checker treats judgments with at most one non-value-like principal term as the common case.

\begin{definition}
    A typing judgment $\Gamma \vdash \Delta$ is \emph{simple} if there is at
    most one formula $t:A$ in $\Gamma,\Delta$ whose term $t$ is not value-like.
\end{definition}

The command forms above start with at most one formula on the right. If their derivations contained only simple judgments, proof search could be almost syntax-directed: expand the unique non-value-like formula when present, and otherwise try the value-like formulas. The full rules do not preserve this property. Both $\text{T-PairL}$ and $\text{T-ListL}$ can expose two non-value-like subterms on the left, so the checker must choose which formula to expand first.

Translating programs to let-normal (A-normal) form would make the exposed pair and list components variables and hence value-like, reducing this source of search. We omit this translation in the prototype because the benchmarks are small. The checker instead handles multiple non-value-like formulas with the search strategy below, without relying on the simple-judgment invariant.

\paragraph{Proof-search state.}
Internally, a formula is a pair $t:A$, and a judgment is represented by two formula lists. A search state also carries a priority queue, represented as a list of left- or right-side formulas to decompose next. Recursive calls place newly introduced formulas at the front in the order expected by the selected rule.

The priority queue determines the order of attempts without removing candidates. If it is empty or all suggested paths fail, the checker falls back to global backtracking search over formula-expansion paths under the supplied annotations. At each state, it tries every formula in $\Delta$ and then $\Gamma$, including value-like formulas, and retries with the next candidate after a failed expansion. This fallback preserves the available expansion choices; it does not establish completeness of the entire checker.

The checker memoizes exact annotated judgments independently of their priority order. If it encounters a judgment already active on the current search path, it rejects that path to avoid a circular proof. These measures reduce repeated work, but nested expansion choices may still require exponentially many attempts.

\paragraph{Resource accounting.}
Potential is stored inside types, as in the formal system. Right rules consume potential from the left; left rules expose resource requirements on the right. Each successful checker call returns a residual context containing the unused resources. Sequential premises pass this context to the next call, so potential is not spent twice. Mutually exclusive branches start from the same context; their compatible residuals are merged by taking the pointwise minimum potential, retaining only resources available after either branch. Sharing operations add or subtract compatible type potentials for value-like terms, including potential nested inside lists and pairs.

\paragraph{Structural rules.}
The checker applies structural rules during formula expansion and resource accounting. The right-formula routine handles $\text{T-Id}$, including variable lookup in the left context; $\text{T-Share}$ transfers the required potential and retains the unused portion. For example, before applying $\text{T-ListCons}$ to $t_1::t_2:\TP{\Tlist{\Tunit}}{2}$, the checker deducts $2$ units of outer potential from the left context and reduces the requested outer annotation to zero. This step implements $\text{T-Resc}$ and fails if the required potential is unavailable. Discarding surplus resources, when permitted, implements $\text{T-Relax1}$ and $\text{T-Relax2}$.

\paragraph{Right and left rules.}
On the right, the checker follows introduction structure: it checks values by matching the requested type, and decomposes applications, conditionals, matches, and let-bindings according to the corresponding right rules. For $\Tok$, variables are accepted directly, while non-variable terms are either lifted from a more precise inferred type or checked structurally with $\Tok$ as the result type.

On the left, the checker follows elimination structure. Tick terms expose resource consumption; pair and list constructors expose components; and conditional, match, application, and let-left rules use the auxiliary formula $\Phi$ and the types suggested by \texttt{notas}. Application on the left uses a necessity-arrow type for the function and then continues with the argument formula, the central mechanism for lower-bound proofs.

\paragraph{Implementation artifacts.}
The prototype contains auxiliary constructs not present in the core calculus, such as a direct \texttt{div} term that steps to itself, an internal \texttt{stuck} term, and surface syntax used by examples. These are implementation conveniences and do not change the typing rules studied in the paper.

\subsection{Experimental Evaluation}
\label{sec:evaluation}

We evaluate the checker on the benchmark programs in Table~\ref{tab:benchmark-programs}. The first group contains small tests for resource rules and higher-order functions; the second contains more realistic list-processing programs. The complete source terms, without implementation-only annotations, appear in our technical report. The artifact for reproducing the experiments is available on Zenodo~\cite{chen_2026_23662529}.

In the table, $n$ and $m$ are input sizes, and $n_i$ is the length of the $i$-th inner list. For 1-8, the bound is for rectangular inputs with $n+5$ rows of length $n+2$. All listed bounds are verified; a dash means that no bound is checked on that side. For 1-7, $\infty$ means any finite lower bound can be checked. For 2-6, the lower bound is $m$ for the Boolean case and $n$ for the Unit case, while the upper bound is checked through $\Tok$.

\begin{table}[!t]
\centering
{\small
\begin{tabular}{llp{0.38\linewidth}cc}
\toprule
No. & Name & Description & Lower & Upper \\
\midrule
1-1 & If & Branch costs 2 and 3. & $2$ & $3$ \\
1-2 & Matchlist & List-match branch costs. & $1$ & $4$ \\
1-3 & If-Fun & Chooses between functions. & $2$ & $4$ \\
1-4 & Higher-Order & Calls an argument function. & $2$ & -- \\
1-5 & Linear-Bool & Data-dependent list traversal. & $2n+5$ & $3n+5$ \\
1-6 & Early-Return & Stops at the first false. & $1$ & $n+1$ \\
1-7 & Infinite-Consumer & Consumes while diverging. & $\infty$ & -- \\
1-8 & Squared & Nested rectangular loop. & $7n+10$ & -- \\
1-9 & Stuck & Pays before a false guard. & $2n$ & -- \\
1-10 & Divergence & Diverges at zero extra cost. & $2n$ & -- \\
\midrule
2-1 & Strcmp & Compares two Boolean lists. & $0$ & $n$ \\
2-2 & Safe-Strcmp & Compares a fixed-length prefix. & $n$ & $n$ \\
2-3 & Zip & Zips lists with padding. & $\max(n,m)$ & $n+m$ \\
2-4 & Add & Concatenates two lists. & $n$ & $n$ \\
2-5 & Compare & Compares two unary naturals. & $0$ & $\min(n,m)$ \\
2-6 & Multi-Type & Returns different result types. & $n$ or $m$ & $\min(n,m)$ \\
2-7 & Max & Unit-cost maximum. & $n-1$ & $n-1$ \\
2-8 & Max-Potential & Potential on inner elements. & $0$ & $\sum_i n_i$ \\
2-9 & Stack & Processes push and pop operations. & $n$ & $2n$ \\
\bottomrule
\end{tabular}
}
\vspace{4pt}
\setlength{\abovecaptionskip}{8pt}
\setlength{\belowcaptionskip}{-4pt}
\caption{Benchmark programs checked by the prototype.}
\label{tab:benchmark-programs}
\end{table}

Several benchmarks isolate features that are difficult to express with an upper-bound-only AARA system.

\paragraph{Higher-order lower bounds.}
Benchmark 1-4 calls a function received as an argument. The supplied function may consume arbitrary resource, so the whole program has no useful upper bound, but the checker still verifies the fixed pre-call cost $2$.

\paragraph{Lower bounds beyond the upper-bound domain.}
Benchmark 1-8 is a nested loop with a quadratic cost component. A linear AARA instance cannot express this upper bound, but the checker verifies the linear lower bound $7n+10$ for the rectangular inputs.

\paragraph{Timing leakage in string comparison.}
Benchmarks 2-1 and 2-2 model the classic timing side channel in string comparison. The ordinary comparison stops at the first mismatch, whereas the safe comparison performs a fixed number of comparisons; only the safe version has matching bounds $n$.

\paragraph{Max-style lower bounds.}
Benchmark 2-3 returns a list whose length is $\max(n,m)$. The lower-bound proof shows each input length separately, giving $\max(n,m)$, while the current upper-bound proof uses the coarser bound $n+m$.

\paragraph{Result-type-sensitive lower bounds.}
Benchmark 2-6 (Multi-Type) uses necessity arrows to state lower bounds conditional on the result type. It returns unit when the first list is exhausted and $\ttrue$ when the second list runs out first. The upper-bound proof uses $\Tok$ to cover both results; necessity arrows with codomains $\Tunit$ and $\Tbool$ give the respective lower bounds $n$ and $m$. The Unit case requires $\text{T-Dis}$ to discharge the Boolean-returning branch, whose subgoal is $\ttrue:\Tunit\vdash\Phi$. Here $\text{T-Bool}$ proves $\vdash\ttrue:\Tbool$, and $\Tbool\parallel\Tunit$ allows $\text{T-Dis}$ to derive the required $\Phi$. The other rules cannot close this subgoal, making disjointness necessary for this conditional resource bound.

\paragraph{Potential for stack operations.}
Benchmark 2-9 processes stack operations. Push and successful pop each cost one unit, but unsuccessful pops cost nothing. The checker verifies an upper bound of $2n$ and, under the condition that every pop removes at least one element, a lower bound of $n$.

\subsection{Experimental Findings and Limitations}
\label{sec:implementation-discussion}

The prototype checks upper and lower bounds within one annotation-guided proof-search procedure. Right rules establish sufficient conditions for evaluation, while left rules extract necessary conditions from a hypothesized result. Sufficiency and necessity arrows use the same representation of formulas and potential. The benchmarks exercise these mechanisms on higher-order functions, branching, recursion, and data structures.

Table~\ref{tab:benchmark-programs} illustrates several uses of necessity reasoning. Higher-Order retains a fixed pre-call lower bound when the argument function's cost is unknown. Multi-Type restricts each lower bound to executions that return the specified type. Safe-Strcmp combines both directions: matching bounds establish a fixed tick cost for each input size, which a lower bound alone cannot guarantee. These properties are certified by typing judgments and apply to all inputs covered by their assumptions.

The experiments also distinguish the expressiveness of the judgments from that of the potential functions. Squared has a linear lower bound even though its quadratic cost is beyond the prototype's upper-bound domain; RaML recovers that cost with richer potentials (\sec{aara-comparison}). For Zip, two lower-bound proofs give $\max(n,m)$, while the checked upper bound is $n+m$. This gap reflects the precision of the available proofs and need not indicate variation in the program's cost.

\paragraph{Incompleteness of the type system.}
The rules do not derive every semantically valid resource bound. Consider
\[
\begin{aligned}
\underline{\texttt{c}} = {}&\texttt{let }u=(\tif{x}{\ttick{3}}{\ttick{2}})\\
                         &\texttt{in }\tif{x}{\ttick{2}}{\ttick{3}}.
\end{aligned}
\]
For either Boolean value of $x$, the cost is exactly $5$: both conditionals select corresponding branches, whose costs sum to $5$. Thus $x:\Tbool,\underline{\texttt{c}}:\Tunit\vDash 5$. The current rules derive lower bound $4$ and upper bound $6$, but cannot derive the exact bound $5$. In $\text{T-IfL}$, both branches must establish the same $\Phi$ without recording the value of $x$, so each conditional contributes only the common lower bound $2$. The upper-bound rule similarly budgets $3$ for each conditional. The rules lose the correlation between the two branch choices. Since the constant $5$ is expressible, this is a limitation of the rules, independent of the potential language and the priority-queue heuristic.

The evaluation covers 19 small benchmarks and explicit tick costs; it does not measure execution time or establish scalability on large programs. Users supply intermediate types and auxiliary formulas through \texttt{as}, \texttt{notas}, and brace annotations. The priority queue and fallback search provide no completeness guarantee, so a failed search leaves the judgment unresolved. Let-normalization could reduce the number of non-value-like formulas, while inference of auxiliary formulas and potential annotations could reduce user effort. These remain directions for extending the current prototype.

\section{Discussion}
\label{sec:discussion}

We first compare the benchmark bounds with an existing AARA implementation, then summarize a parallel exploration of two-sided linear typing that helps explain the design of the main system.

\subsection{Comparison with Existing AARA Systems}
\label{sec:aara-comparison}

We compare the bounds in Table~\ref{tab:benchmark-programs} with Resource Aware ML (RaML) version~1.5.0, which supports upper- and lower-bound AARA analyses~\cite{SP:NDF17}. The inputs are OCaml translations of the tick-instrumented benchmark implementations. The experiments use explicit ticks, the same input domains and successful-return conditions as the target bounds, and potential bases of degrees one and two. We distinguish default inference using the unmodified RaML analyzer from verification with prescribed input and output potentials. The latter uses an experimental adapter that adds constraints to the root annotations without changing RaML's analysis rules; it is not a bound-specification feature of the standard command-line interface.

\begin{table}[!htb]
  \centering
  \small
  \setlength{\tabcolsep}{4pt}
  \renewcommand{\arraystretch}{1.12}
  \caption{RaML comparison of benchmark bounds.
    $L$ and $U$ denote lower and upper bounds;
    $\leadsto$ shows an improvement from the paper's bound to RaML's bound.}
  \label{tab:raml-bound-classification}
  \begin{tabular}{@{}p{.24\linewidth}p{\dimexpr.76\linewidth-2\tabcolsep\relax}@{}}
    \toprule
    Classification & Benchmarks and bounds \\
    \midrule
    (1) Tighter bounds
      & 1-8~Squared: $L:7n+10\leadsto n^2+7n+10$;\newline
        2-1~Strcmp: $U:n\leadsto\min(n,m)$;\newline
        2-8~Max-Potential: $L:0\leadsto\sum_i n_i$. \\
    \midrule
    (2) Inferred bounds
      & All remaining bounds in Table~\ref{tab:benchmark-programs}:
        RaML infers the same bounds (25 targets). \\
    \midrule
    (3) Verification only
      & 2-7~Max: $L=U=n-1$. \\
    \midrule
    (4) After rewriting
      & 1-9~Stuck, 1-10~Divergence: $L=2n$;\newline
        2-6~Multi-Type: $L_{\mathsf{Unit}}=n$, $L_{\mathsf{Bool}}=m$. \\
    \bottomrule
  \end{tabular}
\end{table}

Table~\ref{tab:raml-bound-classification} classifies 34 bound targets, counting Multi-Type's two conditional lower bounds separately. The four categories range from stronger default-inference results to results requiring additional user input. \emph{Tighter bounds} records three targets for which default inference improves the paper's bound. \emph{Inferred bounds} records 25 targets for which RaML infers the same bound as Table~\ref{tab:benchmark-programs}; it reports the remaining targets, including their upper bounds, by reference to that table. \emph{Verification only} records two targets that RaML validates with prescribed input and output potentials but does not recover by default inference. \emph{After rewriting} records four targets recovered through source rewriting or result encoding, possibly with prescribed potentials.

The classification allows independent lower bounds to be combined by taking their maximum, and upper bounds by taking their minimum. Thus, for example, the reported $\min(n,m)$ for Strcmp combines two analyses; it is not the output of a single inference run. For Infinite-Consumer, included in the inferred category, the lower-bound optimization is unbounded rather than returning a numeric infinity. Checks of selected finite bounds support this classification but do not themselves enumerate or prove all finite instances.

\paragraph{Potential expressiveness and inference.}
RaML's richer potential basis recovers Squared's quadratic cost on the specified rectangular inputs and the sum of inner-list lengths for Max-Potential. Conversely, the exact bound $n-1$ for Max, on nonempty inputs, requires input potential $n$ and constant output potential $1$. Both bound directions can be verified with this contract, although default inference produces lower bound $0$ and upper bound $n$. The distinction is between a contract expressible by the analysis and the contract selected by its default optimization.

\paragraph{Nonreturning paths and result conditions.}
For Stuck and Divergence, RaML infers $2n$ when the nonreturning continuation is encoded as a zero-tick loop that carries the unprocessed list. The rewrite preserves the cost already paid on that branch. For Stuck, this comparison treats stuckness and divergence alike only in excluding successful returns; it does not establish observational equivalence.

For Multi-Type, a result tag paired with the unprocessed suffix makes conditional bounds expressible through output potential. Prescribing that potential permits verification or inference of the required input potential; decoding the tag recovers the original result. Alternatively, fused result projections allow automatic inference for each selected return case. These results depend on the additional encoding or annotations and do not come from default inference on the direct translation.

The comparison separates bound precision from the interface needed to obtain a bound. Necessity arrows express result-type-sensitive conditions directly. RaML recovers the same targets through suitable contracts or encodings and infers stronger bounds in several cases. Our prototype combines sufficiency, necessity, and type refutation in one proof system. These results concern the tested RaML version, potential degrees, and encodings; failure on one formulation does not rule out another formulation of the same property.

\subsection{Two-Sided Linear Typing}
\label{sec:linear}

The main system combines two-sided typing with affine, numeric potential. We also study a linear setting in which values themselves are resources that cannot be freely copied or discarded. This section summarizes the design; our technical report presents the \emph{Two-Sided Linear Type System}, its language and operational semantics, typing rules, semantic interpretation, and soundness discussion.

The linear calculus contains unit, Booleans, functions, and pairs. Evaluation uses an ordinary finite map from variables to values, and looking up a variable consumes its binding. A successful closed computation leaves no unused bindings. This exact accounting differs from the ordered numeric potential of the main system: a closure may retain concrete resources from its environment, which must be accounted for when the closure is used. Function values therefore cannot be treated as freely duplicable.

The key issue arises when such a closure receives a necessity-arrow type $A\onlyto B$. For an ordinary arrow $A\to B$, abstraction follows the resource equation ``captured environment plus argument equals result.'' Reading the arrow backward should recover the argument resources from the result, but the naive equation ``captured environment plus result equals argument'' counts the captured resources on the wrong side. To preserve the original equation, the necessity direction must instead read ``result minus captured environment equals argument.''

The linear system expresses this subtraction with \emph{dual formulas}: $\overline{t:T}$ is the dual of $t:T$, and $\overline{\Gamma}$ dualizes each formula in a context. Ordinary abstraction uses a premise $\Gamma,x:A\vdash t:B$; necessity abstraction instead uses $\overline{\Gamma},t:B\vdash x:A$. For example, the system derives
\[
  \vdash \lambda x.\lambda y.(x,y)
  : \Tunit\to(\Tbool\onlyto(\Tunit,\Tbool)).
\]
The inner closure captures $x$. Its necessity proof decomposes the result pair and cancels the recovered $x:\Tunit$ against $\overline{x:\Tunit}$, leaving the required argument $y:\Tbool$. This is why the dual context is needed even for a small higher-order example.

Runtime environments remain ordinary maps; signed resources appear only in the interpretation of proof formulas. A formula and its dual carry opposite signed contributions, with compatibility conditions that prevent duplication or cancellation against an unrelated witness. Captured contexts must still denote ordinary runtime maps. Our technical report spells out this distinction and the associated semantic conditions. Its soundness propositions describe the intended connection between rules and semantics; this parallel exploration does not provide a complete metatheoretic development comparable to the soundness proof for the main system's safety fragment.

Exact resource accounting does not make entailment reversible: $s:A\vdash t:B$ need not imply $t:B\vdash s:A$. The two systems support necessity reasoning under different structural disciplines. Reusable program values and ordered numeric potential support affine upper- and lower-bound reasoning in the main system; closures that capture linear resources require dual formulas and exact cancellation in the linear system.

\section{Related Work}
\label{sec:related}

\paragraph{Two-sided type systems.}

The two-sided type system of \citet{ramsay2024ill} is one of the main foundations of this paper. It expresses typability and ill-typability in a single sequent-style judgment, where formulas may occur on both sides of $\vdash$ and an empty right-hand side represents refutation. We keep this position-sensitive reading, including the necessity arrow, but add potential annotations so that refuting typability can express that an under-resourced program cannot evaluate. This adaptation requires an affine context discipline, a value-state semantics, and a safety fragment.
In the same line of work, \citet{li2026complementary} interpret types by sets of normal forms, making a complement operator sound and obtaining completeness. Our work stays closer to the value-based interpretation of \citet{ramsay2024ill}, because the main goal is to integrate two-sided typing with resource bounds rather than to obtain completeness for type refutation.

\paragraph{Automatic Amortized Resource Analysis.}
Automatic Amortized Resource Analysis (AARA), pioneered by \citet{POPL:HJ03}, infers symbolic resource bounds using amortized potential \cite{JMSCS:HJ22}: type annotations assign potential to data structures, and local typing constraints reduce bound inference to linear programming. AARA has been extended from linear to polynomial \cite{hoffmann2010amortized,POPL:HAH11}, exponential \cite{FOSSACS:KH20}, and logarithmic \cite{MSCS:HLO22} potentials, and adapted to higher-order \cite{POPL:JHL10}, imperative and object-oriented \cite{PLDI:CHS15}, lazy \cite{ESOP:VJF15}, session-typed \cite{LICS:DHP18}, probabilistic \cite{PLDI:NCH18}, and separation-logic \cite{LMCS:Atkey11} settings.

AARA is the other main foundation of this paper. Our system preserves its potential-based structure: types carry potential, resource consumption is local, and sufficiency-arrow judgments retain the usual upper-bound reading. The two-sided layer adds lower-bound and refutation reasoning while remaining compatible with standard AARA upper-bound analysis, so the system extends AARA along a two-sided dimension.

\paragraph{Lower-bound resource analysis.} Most AARA work focuses on upper bounds. The closest work to ours is that of \citet{SP:NDF17}, which verifies lower bounds and constant-resource behavior by varying the substructural discipline of potential: affine potential yields upper bounds, relevant potential yields lower bounds, and linear potential yields constant-resource guarantees. Our work addresses a similar problem and also relies on AARA potential, but obtains necessity from the two-sided position of formulas and the necessity arrow while keeping the main resource discipline affine.

Beyond AARA, lower bounds have been studied with different methods. \citet{vzikelic2022differential} use anti-potentials to bound the cost difference between two programs. \citet{frohn2020inferring} infer worst-case lower bounds for integer programs with metering functions and recurrence solving. \citet{PLDI:WFG19} use martingales for expected-cost bounds in probabilistic programs. These approaches rely on recurrence reasoning, martingales, or anti-potentials rather than a two-sided type system.

\paragraph{Incorrectness logic.}
Incorrectness logic, introduced by \citet{inclogic}, provides a program-logic foundation for under-approximate reasoning: instead of proving that all executions avoid an error, it proves the presence of concrete behaviors. This idea has been developed for local heap reasoning \cite{raad2020local}, large-scale bug finding \cite{OOPSLA:LRV22}, concurrency \cite{POPL:RBD22}, outcome logics \cite{zilberstein2023outcome,zilberstein2024outcomeseparationlogiclocal}, non-termination \cite{OOPSLA:RVO24}, and backward under-approximate reasoning \cite{ascari2024sufficientincorrectnesslogicsil}. The closest connection to resources is the quantitative incorrectness logic of \citet{jin2026programlogicunderapproximatingworstcase}, which under-approximates worst-case resource usage using resource functions. Our lower-bound judgments have a similar flavor, but derive resource facts compositionally from AARA-style potential and two-sided typing rather than from program-logic triples.

\paragraph{Linear logic and substructural type systems.}
Linear logic \cite{TCS:Girard87} is a substructural logic in which contraction and weakening are not freely available. Linear and affine type systems \cite{wadler1990linear,pierce2002types} use these restrictions to control resources and support resource analysis \cite{JIC:Hofmann03}. Linear logic also informs our rules, especially the exploratory two-sided linear type system (\sec{linear} and our technical report). Our systems are not presentations of linear logic with typing formulas as atoms: types carry ordered potential, the main system is affine, and the semantics is operational and resource-aware. Still, weakening, contraction, context splitting, and left--right symmetry follow the same broad structural intuition.

\section{Conclusion}

This paper presented a \emph{two-sided type system with potential}, in which a single proof system derives resource upper bounds, lower bounds, and refutations: sufficiency and necessity differ only in the position of formulas around $\vdash$. We defined a semantic-typing model over value states and proved the soundness of the type system on a safety fragment that retains all the types used in our examples and benchmarks. A prototype checker verifies matching pairs of bounds that characterize costs exactly, as well as lower bounds for programs whose upper bounds are unavailable. Finally, a parallel exploration of two-sided linear typing showed that closures capturing linear resources call for dual typing formulas, which explains the affine design of the main system. Automating the search for annotations---type inference for two-sided judgments---is a natural next step.

\begin{acks}
This work is sponsored by the National Natural Science Foundation of China under Grant No.~62672008.
\end{acks}

\bibliographystyle{ACM-Reference-Format}
\bibliography{db}

\newpage
\appendix

\section{Appendix}

\subsection{Full Type System Definition}
\label{subsec:full-type-system-definition}

\paragraph{OK Rules.}
This subsection records implementation-level rules that are not all part of
the core presentation in Section~\ref{sec:typesystem}.  The type checker treats $\Tok$ as a
coarse success type: a term of type $\Tok$ must evaluate successfully to some
value, but the value is not required to have a more precise shape.  The rules
below make only the genuinely $\Tok$-specific part of this full system
explicit.  Rules such as $\text{T-App}$, $\text{T-If}$,
$\text{T-ListMatch}$, $\text{T-PairMatch}$, and $\text{T-Let}$ already allow
their result type to be arbitrary, so their $\Tok$ cases are just ordinary
instances of the rules from Section~\ref{sec:typesystem} and are not repeated here.

\tcbset{before skip=0pt, after skip=0pt}
\begin{tcolorbox}[title=Full OK Rules, width=\linewidth, sharp corners, boxrule=0.4pt, colback=white, colframe=black, left=\fboxsep, right=\fboxsep, top=\fboxsep, bottom=\fboxsep, before upper={\setlength{\abovedisplayskip}{0pt}\setlength{\belowdisplayskip}{0pt}}]
\begin{align*}
\text{(T-OkVar)}\inferrule{\ }{\Gamma \vdash x:\Tok, \Delta}
\qquad
\text{(T-Ok)}\inferrule{\Gamma \vdash t:A, \Delta \quad A \neq \Tok}
{\Gamma \vdash t:\Tok, \Delta}
\end{align*}

\begin{align*}
\text{(T-OkCons)}\inferrule{\Gamma_1 \vdash t_1:\Tok, \Delta \quad
\Gamma_2 \vdash t_2:\Tok, \Delta}
{\Gamma_1,\Gamma_2 \vdash t_1 :: t_2:\Tok, \Delta}
\end{align*}

\begin{align*}
\text{(T-TickOkL)}\inferrule{\Gamma, q \vdash \Delta}
{\Gamma,\ttick{q}:\Tok \vdash \Delta}
\qquad
\text{(T-PairOkL)}\inferrule{\Gamma,t_1:\Tok,t_2:\Tok \vdash \Delta}
{\Gamma,(t_1,t_2):\Tok \vdash \Delta}
\end{align*}

\begin{align*}
\text{(T-ConsOkL)}\inferrule{\Gamma,t_1:\Tok,t_2:\Tok \vdash \Delta}
{\Gamma,t_1 :: t_2:\Tok \vdash \Delta}
\end{align*}

\begin{align*}
\text{(T-MatchListOkL)}
\inferrule{\stack{\Gamma_2,t_2:B \vdash \Phi \quad
\Gamma_2,hd:\Tok,tl:\Tok,t_3:B \vdash hd :: tl:\Tok;\Phi}
{\Gamma_1,t_1:\Tok,\Phi \vdash \Delta}}
{\Gamma_1,\Gamma_2,\tmatchlist{t_1}{t_2}{hd}{tl}{t_3}:B \vdash \Delta}
\end{align*}
\end{tcolorbox}

The rules are written at zero potential to keep the display compact.  Their
potential-carrying forms are obtained by the same resource conventions used by
the structural rules.  In particular, the implementation threads the potential
of a right-side $t_1 :: t_2:\TP{\Tok}{q}$ obligation to the tail
$t_2:\TP{\Tok}{q}$, rather than assigning the tail a list type such as
$\Tlist{\Tok}$.  This is not an instance of $\text{T-ListCons}$ followed by
$\text{T-Ok}$, because $\text{T-ListCons}$ would require the tail to have a
list type.  On the left, the implementation similarly decomposes
$(t_1,t_2):\Tok$ and $t_1::t_2:\Tok$ into $\Tok$ obligations for their
components, which is not covered by $\text{T-PairL}$ or $\text{T-ListL}$.
For match-list terms on the left, the scrutinee may be destructed at type
$\Tok$; in that case both bound variables are given type $\Tok$ and the
cons-branch evidence is $hd::tl:\Tok$.

The implementation also contains branches for applications, conditionals,
pair matches, list matches, and let expressions when the target type is
$\Tok$.  These branches are proof-search conveniences: they select the
intermediate types using annotations or simple local inference, but the
corresponding derivations use the general rules of Section~\ref{sec:typesystem} with the result
type instantiated to $\Tok$.  Hence they are intentionally not listed as new
rules.  The implementation's built-in diverging term is also omitted from this
paragraph, because it is a separate implementation convenience rather than an
$\Tok$-specific typing principle.

The disjointness relation is unchanged: $\Tok$ is not disjoint from any type.
Therefore $\text{T-Dis}$ never refutes a formula merely because one of the two
types is $\Tok$.


\subsection{Soundness}
\label{subsec:soundness-appendix}

\subsubsection{Preliminaries}
\label{subsubsec:soundness-preliminaries}

First, to make the proof concise, we extend the satisfaction relation to lists of formulas.

We then record a few semantic lemmas used by the soundness proof.

\paragraph{Proof of Proposition~\ref{thm:evaluation-weakening}.}
\begin{proof}
    We first show the one-step version: if $t,q\rightarrow t',q'$ and $p\ge 0$, then $t,q+p\rightarrow t',q'+p$. The proof is by cases on the rule deriving the one-step reduction. Congruence rules, such as $\text{E-App1}$, $\text{E-If}$, $\text{E-PairL}$, and $\text{E-Let1}$, follow immediately from the induction hypothesis on the reduced subterm. Rules that do not consume resource, such as $\text{E-AppAbs}$, $\text{E-IfTrue}$, $\text{E-MListNil}$, and $\text{E-Let2}$, are unchanged after adding $p$ to both sides. For $\text{E-Tick}$, if $\ttick{r},q\rightarrow \tunit,q-r$, then $q\ge r$; since $p\ge 0$, we also have $q+p\ge r$, and therefore
    $\ttick{r},q+p\rightarrow \tunit,(q+p)-r=(q-r)+p$.

    The stated multi-step property follows by induction on the derivation of $t,q\rightarrow^* t',q'$. The reflexive case is immediate. In the transitive case, write the derivation as $t,q\rightarrow s,r\rightarrow^* t',q'$. By the one-step result, $t,q+p\rightarrow s,r+p$. By the induction hypothesis applied to $s,r\rightarrow^* t',q'$, we have $s,r+p\rightarrow^* t',q'+p$. Combining the two reductions gives $t,q+p\rightarrow^* t',q'+p$.
\end{proof}

\paragraph{Proof of Proposition~\ref{thm:basic-semantic-inclusions}.}
\begin{proof}
    Let $A$ be any type. If $(v,p)\in\VALS{A}$, then $v,p\rightarrow^* v,p$ by zero-step evaluation. Hence $(v,p)\in\TERMS{A}$ by choosing the same value state as the final result. Therefore $\VALS{A}\subseteq\TERMS{A}$.

    The second inclusion is immediate from the definition
    $\TERMSP{A}=\TERMS{A}\cup\{(t,p)\mid t,p\rightarrow^\omega\}$.
    Thus $\TERMS{A}\subseteq\TERMSP{A}$, and the two inclusions together give the result.
\end{proof}

\paragraph{Proof of Proposition~\ref{lem:reverse-preservation}.}
\begin{proof}
    Suppose $t,p\rightarrow^* t',p'$.
    If $(t',p')\in\TERMS{A}$, then by definition there exist $v$ and $p''$ such that
    $t',p'\rightarrow^* v,p''$ and $(v,p'')\in\VALS{A}$.
    By transitivity of multi-step evaluation, $t,p\rightarrow^* v,p''$.
    Hence $(t,p)\in\TERMS{A}$.

    Now suppose $(t',p')\in\TERMSP{A}$.
    If $(t',p')\in\TERMS{A}$, the previous argument gives $(t,p)\in\TERMS{A}$ and therefore $(t,p)\in\TERMSP{A}$ by Proposition~\ref{thm:basic-semantic-inclusions}.
    Otherwise, by the definition of $\TERMSP{A}$, $t',p'\rightarrow^\omega$.
    Prefixing this infinite evaluation with $t,p\rightarrow^* t',p'$ gives $t,p\rightarrow^\omega$.
    Therefore $(t,p)\in\TERMSP{A}$.
\end{proof}

\paragraph{Proof of Proposition~\ref{thm:type-disjointness}.}
\begin{proof}
    We prove the stronger statement that if $A\parallel B$, then no closed value $v$ can satisfy both $A$ and $B$ under any two resource amounts. The theorem follows by taking the two resource amounts to be equal.

    The proof is by cases on the definition of $A\parallel B$, ignoring potential annotations because disjointness is independent of potential. If one type is $\Tunit$ and the other is not, then the first type contains only the unit value, while every non-unit, non-$\Tok$ type contains values with a different outer constructor. The Boolean, list, and function cases are analogous: Boolean values are exactly $\ttrue$ and $\tfalse$, list values are exactly list constructors, and function values are exactly fixpoint abstractions. The disjointness relation deliberately does not make $\Tok$ disjoint from any type, so there is no $\Tok$ case.

    It remains to consider pair types. If $A=(A_1,A_2)$ and $B$ is not a pair type, then the outer constructor again distinguishes their values. Otherwise $B=(B_1,B_2)$ and either $A_1\parallel B_1$ or $A_2\parallel B_2$. A value of both pair types must have the form $(v_1,v_2)$. By the semantics of pair types, the corresponding component values satisfy the corresponding component types. If $A_1\parallel B_1$, this contradicts the induction hypothesis for $A_1$ and $B_1$; the case $A_2\parallel B_2$ is identical. Hence no value can belong to both disjoint types.
\end{proof}

\begin{lemma}
(Semantic Weakening Lemma, SWL) For all valuations $\theta$, resources $p$, and terms $t$:

\begin{itemize}
    \item For every type $\langle T, q\rangle$, if $\theta, p \SATS t:\TP{T}{q}$ and $p' \ge p$, $p'-q'\ge p-q$, then $\theta, p' \SATS t:\TP{T}{q'}$.
    \item For every type $\langle T, q\rangle$, if $\theta, p \SATSP t:\TP{T}{q}$ and $p' \ge p$, $p'-q'\ge p-q$, then $\theta, p' \SATSP t:\TP{T}{q'}$.
    \item For every type $\langle T, q\rangle$, if $\theta, p \SATS t:\TP{T}{q}$, then $\theta, p \SATSP t:\TP{T}{q}$.
\end{itemize}
\end{lemma}

The first two statements also hold for $p' < p$.

\begin{lemma}
(Sharing Lemma)

\begin{itemize}
    \item If $\theta, p \SATS t : T$, $t$ is resource-constant, and $T\share(T_1,T_2)$, then there exist $p_1,p_2$ such that $p_1 + p_2 \le p$ and $\theta, (p_1, p_2) \SATSL t : T_1, t:T_2$.
\end{itemize}
\end{lemma}

\begin{lemma}
(destruction of ";")
If $\theta, p \SATSP \Delta; \Phi$, then there are $p_1$ and $p_2$ such that $p_1 + p_2 = p$, $\theta, p_1 \SATSP \Delta$, and $\theta, p_2 \SATSP \Phi$.
\end{lemma}

We also clarify the notation used below:

\begin{itemize}
    \item For a vector $\vec p$, if $\theta, \vec p \SATSL \Gamma$ is available in the context, then we may index $\vec p$ by a formula $t:T$ in $\Gamma$.

    $\vec p[t]$ refers to the $i$-th component of $\vec p$, if $t : T$ is the $i$-th formula in $\Gamma$.

    Similarly, if $\Gamma' \subseteq \Gamma$, then $\vec p[\Gamma']$ refers to the subvector of resources corresponding to the formulas of $\Gamma'$.

    We use $(\vec p_1, \vec p_2)$ to refer to the concatenation of two (or more) vectors. We may write a single resource value $p$ as a single-element vector.
\end{itemize}

\subsubsection{Proof of Soundness}
\label{subsubsec:proof-of-soundness}

Proof of the soundness theorem:

\begin{theorem}
    (Soundness) Let $\Gamma \vdash \Delta$ be a judgment whose types all belong to the safety fragment. If $\Gamma \vdash \Delta$ is derivable, then $\Gamma \vDash \Delta$.
\end{theorem}

\begin{proof}
    The proof is by induction on the typing rule deriving $\Gamma \vdash \Delta$ in the safety fragment.
    
    \begin{itemize}

        \item[] \textbf{Structural rules.}
    
        \item $\text{T-Id}$. $\Gamma \vdash \Delta$ can be written as $\Gamma', t:A  \vdash t:A,  \Delta'$.
        
        Suppose $\theta, \vec p \SATSL \Gamma$. Thus $\theta, \vec p[t : A] \SATS t : A$.  Since $\vec p[t : A] \le \sum\vec p$, SWL gives $\theta, \sum\vec p \SATS t : A$, and hence $\theta, \sum \vec p \SATSP t:A$. Thus $\theta, \sum \vec p \SATSR \Delta$.

        \item $\text{T-Share}$. $\Gamma\vdash\Delta$ can be written as $\Gamma', t : A\vdash\Delta$. IH: $\Gamma', t:A_1, t :A_2\vDash\Delta$. Extra conditions are $VL(t)$ and $A\share(A_1,A_2)$.

        Suppose $\theta, (\vec p, p) \SATSL \Gamma' ,t : A$. From the component for $t:A$, we have $\theta, p\SATS t : A$. From $VL(t)$, we know $RC(t)$. By the sharing lemma, there exist $p_1,p_2$ such that $p_1+p_2\le p$ and $\theta, (p_1, p_2)\SATSL t : A_1,t:A_2$. Combining this with the $\Gamma'$ part gives $\theta, (\vec p, p_1, p_2) \SATSL \Gamma', t:A_1,t:A_2$. By IH, $\theta, \sum\vec p + p_1 + p_2 \SATSR \Delta$. By SWL, the proof is concluded.

        \item $\text{T-Relax1}$. $\Gamma \vdash \Delta$ can be written as $\Gamma', t : \TP{T}{q'} \vdash \Delta$. IH: there exists $q \le q'$ such that $\Gamma', t : \TP{T}{q} \vdash \Delta$.

        Suppose $\theta, \vec p \SATSL \Gamma', t : \TP{T}{q'}$. By SWL (on the component of $t : \TP{T}{q}$), $\theta, \vec p \SATSL \Gamma', t : \TP{T}{q}$. By IH, $\theta, \sum \vec p\SATSR \Delta$, which is exactly the conclusion. 
        
        \item $\text{T-Relax2}$. $\Gamma \vdash \Delta$ can be written as $\Gamma \vdash t : \TP{T}{q'}, \Delta'$. IH: there exists $q \ge q'$ such that $\Gamma \vdash t : \TP{T}{q}, \Delta'$.

        Suppose  $\theta, \vec p \SATSL \Gamma$. By IH, $\theta, \sum \vec p\SATSR t : \TP{T}{q}, \Delta$. If $\theta, \sum \vec p\SATSR \Delta$, then the conclusion is immediate. Otherwise,  $\theta, \sum \vec p\SATSP t : \TP{T}{q}$. By SWL, $\theta, \sum \vec p\SATSP t : \TP{T}{q'}$, which immediately gives the conclusion.

        \item $\text{T-Resc}$. $\Gamma \vdash \Delta$ can be written as $\Gamma', t : \TP{T}{q} \vdash t_1 : \TP{T_1}{q_1}, t_2 : \TP{T_2}{q_2}, \dots, t_n: \TP{T_n}{q_n}$. IH: $\Gamma', t : \TP{T}{q + p} \vDash t_1 : \TP{T_1}{q_1 + p}, t_2 : \TP{T_2}{q_2 + p}, \dots,  t_n: \TP{T_n}{q_n + p}$.

        Suppose $\theta, (\vec p, p_1) \SATSL \Gamma', t : \TP{T}{q}$. By SWL, $\theta, (\vec p, p_1 + p) \SATSL \Gamma', t : \TP{T}{q + p}$. By IH, $\theta, \sum\vec p + p_1 + p \SATSR t_1 : \TP{T_1}{q_1 + p}, t_2 : \TP{T_2}{q_2 + p}, \dots,  t_n: \TP{T_n}{q_n + p}$. 
        
        By definition, there exists $i$ such that $\theta, \sum\vec p + p_1 + p \SATSP t_i : \TP{T_i}{q_i + p}$. By SWL, $\theta, \sum\vec p + p_1 \SATSP t_i : \TP{T_i}{q_i}$, which immediately gives the conclusion.

        \item[] \textbf{Right typing rules.}
        
        \item $\text{T-Unit}$. $\Gamma \vdash \Delta$ can be written as $\Gamma  \vdash \tunit : \Tunit ,\Delta$.

        Suppose  $\theta, \vec p \SATSL \Gamma$. By definition, $(\tunit, 0) \in  \VALS{\Tunit}$. Immediately, $(\tunit, 0) \in \TERMSP{\Tunit}$. Also, for every valuation $\theta$, $\tunit \theta = \tunit$.
        
        By definition, $\theta, 0 \SATSP \tunit : \Tunit$.
        
        By SWL, $\theta,  \sum\vec p \SATSP \tunit : \Tunit$.
        
        \item $\text{T-Bool}$. $\Gamma \vdash \Delta$ can be written as $\Gamma  \vdash \ttrue : \Tbool ,\Delta$. (The $\tfalse$ case is identical, so we omit it.)

        Suppose  $\theta, \vec p \SATSL \Gamma$. By definition, $(\ttrue, 0) \in  \VALS{\Tbool}$. Immediately, $(\ttrue, 0) \in \TERMSP{\Tbool}$. Also, for every valuation $\theta$, $\ttrue \theta = \ttrue$.
        
        By definition, $\theta, 0 \SATSP \ttrue : \Tbool$.
        
        By SWL, $\theta,  \sum\vec p \SATSP \ttrue : \Tbool$.

        \item $\text{T-Abs}$ and $\text{T-AbsN}$. See the following subsection.
        
        \item $\text{T-App}$. $\Gamma\vdash \Delta$ can be written as  
        $\Gamma_1, \Gamma_2 \vdash t_1 \ t_2 : B, \Delta'$. IH: $\Gamma_1 \vDash t_1:A\rightarrow B, \Delta'$ and $\Gamma_2 \vDash t_2:A, \Delta'$.

        Suppose $\theta, (\vec p_1, \vec p_2) \SATSL \Gamma_1, \Gamma_2$. Taking the two subvectors gives $\theta, \vec p_1 \SATSL \Gamma_1$ and $\theta, \vec p_2 \SATSL \Gamma_2$.

        By IH, $\theta, \sum \vec p_1 \SATSP t_1:  A \to B$ and $\theta, \sum \vec p_2 \SATSP t_2: A$. (Suppose neither premise satisfies $\Delta'$; otherwise the proof is complete.)

        By definition, $(t_1\theta, \sum\vec p_1) \in \TERMSP{A \to B}$. If it diverges, then the application diverges and the case is complete. Otherwise, there exist $v_1, p_1'$ such that $t_1\theta, \sum\vec p_1 \to^* v_1, p_1'$ and $(v_1, p_1') \in \VALS{A \to B}$.

        Similarly, if $(t_2\theta,  \sum\vec p_2)$ diverges, then the application diverges. Otherwise, there exist $v_2, p_2'$ such that $t_2\theta,  \sum\vec p_2 \to^* v_2, p_2'$ and $(v_2, p_2') \in \VALS{A}$.

        By definition of $\VALS{\cdot}$, $v_1 = \tfix{f}{x}{t}$ and $\forall v', p', (v', p') \in \VALS{A} \to (t[f \mapsto \tfix{f}{x}{t}][x \mapsto v'], p') \in \TERMSP{B}$. (The approximant and lambda-abstraction cases introduced in the $\text{T-Abs}$ proof are analogous.)

        By the evaluation rules, $t_1\theta \ t_2\theta, \sum\vec p_1 + \sum\vec p_2 \to^* v_1\ t_2\theta, p_1' + \sum\vec p_2 \to^* v_1 \ v_2, p_1' + p_2' \to^* t[f \mapsto \tfix{f}{x}{t}][x \mapsto v_2], p_1' + p_2'$.

        As $(v_2, p_2') \in \VALS{A}$, we have $(v_2, p_1' + p_2') \in \VALS{A}$. Thus, $(t[f \mapsto \tfix{f}{x}{t}][x \mapsto v_2], p_1' + p_2')\in \TERMSP{B}$. By reverse preservation, $(t_1\theta \ t_2\theta,\sum\vec p_1 + \sum\vec p_2) \in \TERMSP{B}$, which means $\theta, \sum\vec p_1 + \sum\vec p_2 \SATSP  t_1 \ t_2 : B$.
        
        \item $\text{T-If}$. $\Gamma \vdash \Delta$ can be written as $\Gamma_1,\Gamma_2 \vdash  \tif{t_1}{t_2}{t_3}: A, \Delta'$. IH: $\Gamma_1 \vDash t_1:\langle \Tbool, q\rangle, \Delta'$, $\Gamma_2, q \vDash t_2:A, \Delta'$, and $\Gamma_2, q \vDash t_3:A, \Delta'$.

        Suppose $\theta, (\vec p_1, \vec p_2) \SATSL \Gamma_1, \Gamma_2$. Taking the two subvectors gives $\theta, \vec p_1 \SATSL \Gamma_1$ and $\theta, \vec p_2 \SATSL \Gamma_2$ independently.

        The resource formula $q$ abbreviates $\tunit:\TP{\Tunit}{q}$, so $\theta,q \SATS q$. Thus $\theta,(\vec p_2,q)\SATSL \Gamma_2,q$.
        
        By IH, $\theta, \sum \vec p_1 \SATSP t_1 : \TP{\Tbool}{q}$, $\theta, q + \sum \vec p_2 \SATSP t_2: A$, and $\theta, q + \sum \vec p_2 \SATSP t_3: A$ (supposing none of these satisfies $\Delta'$; otherwise the case is complete).

        To complete the proof we need to prove $\theta, \sum \vec p_1 + \sum \vec p_2 \SATSP \tif{t_1}{t_2}{t_3}: A$, i.e. $((\tif{t_1}{t_2}{t_3})\theta,  \sum \vec p_1 + \sum \vec p_2) \in \TERMSP{A}$.

        $(\tif{t_1}{t_2}{t_3})\theta = \tif{t_1\theta}{t_2\theta}{t_3\theta}$.

        By IH(1), $(t_1\theta,  \sum \vec p_1) \in \TERMSP{\TP{\Tbool}{q}}$. If $t_1\theta,  \sum \vec p_1$ diverges, then the original term diverges and the proof is complete. Otherwise, there exist $v_1, q'$ such that $ t_1\theta,  \sum \vec p_1 \rightarrow^* v_1, q'$ and $(v_1, q') \in \VALS{\TP{\Tbool}{q}}$. By definition, $v_1$ is either $\ttrue$ or $\tfalse$ and $q' \ge q$.

        By $\text{E-If}$ and weakening in evaluation we have $ \tif{t_1\theta}{t_2\theta}{t_3\theta},  \sum \vec p_1 + \sum \vec p_2 \rightarrow^*  \tif{v_1}{t_2\theta}{t_3\theta} ,  \sum \vec p_2 + q'$.

        Case analysis on $v_1$.

        \begin{itemize}
            \item If $v_1 = \ttrue$, then the above term further evaluates to $t_2\theta$. By IH(2) and SWL with $q' \ge q$, $(t_2\theta, q' + \sum \vec p_2) \in \TERMSP{A}$. By reverse preservation, $((\tif{t_1}{t_2}{t_3})\theta,  \sum \vec p_1 + \sum \vec p_2) \in \TERMSP{A}$.
            
            \item If $v_1 = \tfalse$, the argument is identical, except that the term evaluates to $t_3\theta$ and we use IH(3).
        \end{itemize}

        \item $\text{T-ListNil}$. $\Gamma\vdash \Delta$ can be written as $\Gamma \vdash [] : \Tlist{A} ,\Delta'$.

        Suppose  $\theta, \vec p \SATSL \Gamma$. By definition, $([], 0) \in  \VALS{\Tlist{A}}$. Immediately, $([], 0) \in \TERMSP{\Tlist{A}}$. Also, for every valuation $\theta$, $[] \theta = []$.
        
        By definition, $\theta, 0 \SATSP [] : \Tlist{A}$.
        
        By SWL, $\theta,  \sum \vec p \SATSP [] : \Tlist{A}$.

        \item $\text{T-ListCons}$. $\Gamma \vdash \Delta$ can be written as $\Gamma_1, \Gamma_2 \vdash t_1::t_2 : \Tlist{A}, \Delta'$. IH: $\Gamma_1 \vDash t_1 :A, \Delta'$ and $\Gamma_2 \vDash t_2 :\Tlist{A}, \Delta'$.
        
        Suppose $\theta, (\vec p_1, \vec p_2) \SATSL \Gamma_1, \Gamma_2$. Taking the two subvectors gives $\theta, \vec p_1 \SATSL \Gamma_1$ and $\theta, \vec p_2 \SATSL \Gamma_2$ independently.

        By IH, $\theta,  \sum\vec p_1 \SATSP t_1 : A$ and $\theta,  \sum\vec p_2 \SATSP t_2 : \Tlist{A}$ (supposing neither satisfies $\Delta'$; otherwise the case is complete).

        To complete the proof we need to prove $\theta, \sum\vec p_1 +\sum\vec p_2 \SATSP t_1::t_2 : \Tlist{A}$, i.e. $((t_1 :: t_2)\theta,  \sum\vec p_1 + \sum\vec p_2) \in \TERMSP{\Tlist{A}}$.

        $(t_1 :: t_2)\theta = t_1\theta :: t_2\theta$.

        By IH we know $(t_1\theta,  \sum\vec p_1) \in \TERMSP{A}$ and $(t_2\theta, \sum\vec p_2) \in \TERMSP{\Tlist{A}}$.

        If either $t_1\theta$ or $t_2\theta$ diverges, then $t_1\theta :: t_2\theta$ diverges and the proof is complete. Otherwise, there exist $v_1,q'$ and $v_2,q''$ such that $t_1\theta, \sum\vec p_1 \rightarrow^* v_1, q'$, $t_2\theta, \sum\vec p_2 \rightarrow^* v_2, q''$, $(v_1,q') \in \VALS{A}$, and $(v_2,q'') \in \VALS{\Tlist{A}}$.

        By definition of semantics of types, we know $(v_1 :: v_2, q' + q'') \in \VALS{\Tlist{A}}$.

        Also, by $\text{E-List}$, we have $t_1\theta :: t_2\theta, \sum\vec p_1 + \sum\vec p_2 \rightarrow^* v_1 :: t_2\theta, q' + \sum\vec p_2 \rightarrow^*  v_1 :: v_2, q' + q''$.

        By reverse preservation, $(t_1\theta :: t_2\theta,  \sum\vec p_1 + \sum\vec p_2) \in \TERMS{\Tlist{A}}$.

        This completes the case.

        \item $\text{T-ListMatch}$. $\Gamma \vdash \Delta$ can be written as $\Gamma_1, \Gamma_2 \vdash  \tmatchlist{t_1}{t_2}{hd}{tl}{t_3}: B, \Delta'$. IH: $\Gamma_1 \vDash t_1 : \TP{\Tlist{A}}{q}, \Delta'$, 
        $\Gamma_2, q \vDash t_2: B, \Delta'$ and $\Gamma_2, hd:A,tl:\TP{\Tlist{A}}{q} \vdash t_3:B, \Delta$.

        Suppose $\theta, (\vec p_1, \vec p_2) \SATSL \Gamma_1,\Gamma_2.$ Taking the two subvectors gives $\theta, \vec p_1 \SATSL \Gamma_1$ and $\theta, \vec p_2 \SATSL \Gamma_2$.

        By IH(1), the first implies $\theta, \sum\vec p_1 \SATSP t_1: \TP{\Tlist{A}}{q}$.(Suppose it does not satisfy $\Delta'$; otherwise the case is complete.)

        By definition, this means $(t_1\theta, \sum\vec p_1) \in \TERMSP{\TP{\Tlist{A}}{q}}$. Thus either $(t_1\theta, \sum\vec p_1)$ diverges, or there exist $v,p'$ such that $t_1\theta, \sum\vec p_1\to^* v, p'$ and $(v, p') \in \VALS{\TP{\Tlist{A}}{q}}$.

        To complete this case, we need to prove $\theta, \sum\vec p_1 + \sum\vec p_2 \SATSP  \tmatchlist{t_1}{t_2}{hd}{tl}{t_3}: B$, i.e. $( \tmatchlist{t_1\theta}{t_2\theta}{hd}{tl}{t_3\theta}, \sum \vec p_1 + \sum\vec p_2) \in \TERMSP{B}$.

        If $(t_1\theta, \sum\vec p_1)$ diverges, then this term also diverges and the proof is complete.

        Otherwise, by $\text{E-MList}$, we know $\tmatchlist{t_1\theta}{t_2\theta}{hd}{tl}{t_3\theta}, \sum \vec p_1 + \sum\vec p_2 \to^* \tmatchlist{v}{t_2\theta}{hd}{tl}{t_3\theta},p' + \sum\vec p_2$.

        Case analysis on $v$. 

        \begin{itemize}
            \item $v = []$. The term above further evaluates to $t_2\theta , p' + \sum\vec p_2$. We already have $\theta, \vec p_2 \SATSL \Gamma_2$. By definition of $\VALS{\cdot}$, we also know $p' \ge q$, which means $\theta, p' \SATS \tunit:\TP{\Tunit}{q}$. Combining these facts and applying IH(2), we get $\theta, p' + \sum\vec p_2 \SATSP t_2:B$. If $t_2\theta, p' + \sum\vec p_2$ diverges, the original term diverges; otherwise, by reverse preservation, the original term evaluates to a $B$.

            \item $v = v_1 :: v_2$. The term above further evaluates to $t_3[hd \mapsto v_1][tl \mapsto v_2], p' + \sum\vec p_2$. We already have $\theta, \vec p_2 \SATSL \Gamma_2$. By definition of $\VALS{\cdot}$, we also know $p' \ge p_1 + p_2$, where $(v_1, p_1) \in \VALS{A}$ and $(v_2, p_2) \in \VALS{\TP{\Tlist{A}}{q}}$. Let $\theta_1 = \theta[hd \mapsto v_1][tl \mapsto v_2]$; then $\theta_1, p_1 \SATS hd:A$ and $\theta_1, p_2 \SATS tl:\TP{\Tlist{A}}{q}$. Combining these facts with the $\Gamma_2$ part and applying IH(3), we get $\theta_1, p_1 + p_2 + \sum\vec p_2 \SATSP t_3:B$. If $t_3\theta_1, p' + \sum\vec p_2$ diverges, the original term diverges; otherwise, by reverse preservation, the original term evaluates to a $B$.
        \end{itemize}

        \item $\text{T-Pair}$. $\Gamma \vdash \Delta$ can be written as $\Gamma_1,\Gamma_2\vdash (t_1,t_2): (A_1,A_2), \Delta'$. IH: $\Gamma_1\vdash t_1:A_1,\Delta'$ and $\Gamma_2\vdash t_2:A_2,\Delta'$.
        
        Suppose $\theta, (\vec p_1, \vec p_2) \SATSL \Gamma_1,\Gamma_2.$ Taking the two subvectors and applying IH gives $\theta, \sum \vec p_1\SATSP t_1:A_1$ and $\theta, \sum \vec p_2\SATSP t_2:A_2$ (supposing neither satisfies $\Delta'$; otherwise the case is complete).

        If either $t_1\theta, t_2\theta$ diverges, then the term diverges and the proof is complete.

        Otherwise, by definition, $t_1\theta, \sum\vec p_1 \to^* v_1, p_1'$, $(v_1,p_1') \in \VALS{A_1}$ and $t_2\theta, \sum\vec p_2 \to^* v_2, p_2'$, $(v_2,p_2') \in \VALS{A_2}$. By definition of $\VALS{\cdot}$ of pair types, $((v_1,v_2), p_1'+p_2')\in \VALS{(A_1,A_2)}$.

        By $\text{E-PairL}$ and $\text{E-PairR}$, we have $(t_1\theta, t_2\theta), \sum\vec p_1 + \sum\vec p_2 \to^* (v_1, t_2\theta), p_1' + \sum\vec p_2 \to^*(v_1,v_2), p'_1 + p'_2$. By reverse preservation, $((t_1\theta, t_2\theta), \sum\vec p_1 + \sum\vec p_2) \in \TERMS{(A_1,A_2)}$, and by definition, the conclusion is satisfied.

        \item $\text{T-PairMatch}$. $\Gamma \vdash \Delta$ can be written as $
        \Gamma_1,\Gamma_2\vdash \tmatchpair{t_1}{x}{y}{t_2} : A, \Delta'$. IH: $\Gamma_1 \vdash t_1 : \TP{(A_1,A_2)}{q}, \Delta'$ and $\Gamma_2,x:A_1,y:A_2,q\vdash t_2:A,\Delta'$.

        Suppose $\theta, (\vec p_1, \vec p_2) \SATSL \Gamma_1,\Gamma_2$. Taking the two subvectors gives $\theta, \vec p_1 \SATSL \Gamma_1$ and $\theta,\vec p_2 \SATSL \Gamma_2$. The first, by IH(1), implies $\theta, \sum\vec p_1 \SATSP t_1:\TP{(A_1,A_2)}{q}$ (supposing it does not satisfy $\Delta'$; otherwise the case is complete).

        This means $t_1\theta, \sum\vec p_1 \to^* v, p'$ and $(v,p') \in \VALS{\TP{(A_1,A_2)}{q}}$ (or else $t_1$ diverges, in which case the proof is immediate). By definition of $\VALS{\cdot}$ for pair types, $v$ has the form $(v_1,v_2)$, and $p' \ge p_1' + p_2' + q$, where $(v_1,p_1') \in \VALS{A_1}$ and $(v_2,p_2')\in\VALS{A_2}$.

        $ \tmatchpair{t_1\theta}{x}{y}{t_2\theta}, \sum\vec p_1 + \sum\vec p_2 \to^* \tmatchpair{(v_1, v_2)}{x}{y}{t_2\theta}, p' + \sum\vec p_2\to^* t_2\theta[x\mapsto v_1][y\mapsto v_2], p' + \sum\vec p_2$.

        Let $\theta_1 = \theta[x\mapsto v_1][y\mapsto v_2]$. It is easy to derive that $\theta_1, p_1' \SATS x: A_1$, $\theta_1, p_2' \SATS y: A_2$, and $\theta_1, q \SATS \tunit:\TP{\Tunit}{q}$. Combining these facts with the $\Gamma_2$ subset gives the premise of IH(2), so $\theta_1, p' +  \sum\vec p_2 \SATSP t_2:A$. By reverse preservation, the original term evaluates to an $A$, and the case is complete.

        \item $\text{T-Let}$. $\Gamma \vdash \Delta$ can be written as $\Gamma_1, \Gamma_2 \vdash \tlet{x}{t_1}{t_2} : A, \Delta'$. IH: $\Gamma_1 \vDash t_1 : B, \Delta'$ and $\Gamma_2, x:B\vdash t_2 : A, \Delta'$.

        Suppose $\theta, (\vec p_1, \vec p_2) \SATSL \Gamma_1, \Gamma_2$. Taking the two subvectors gives $\theta, \vec p_1 \SATSL \Gamma_1$ and $\theta,\vec p_2 \SATSL \Gamma_2$. The first, by IH(1), implies $\theta, \sum\vec p_1 \SATSP t_1:B, \Delta'$. Suppose it does not satisfy $\Delta'$; otherwise the case is complete.

        By definition, this means $(t_1\theta, \sum\vec p_1) \in \TERMSP{B}$. If $t_1$ diverges, then the let term diverges and the proof is complete. By definition of $\VALS{\cdot}$, there exist $v_1, p_1'$ such that
        $t_1\theta, \sum\vec p_1 \to^* v_1, p_1'$ and $(v_1, p_1') \in \VALS{B}$.

        Let $\theta_1 = \theta[x \mapsto v_1]$. It is easy to derive $\theta_1, p_1' \SATS x : B$. Combining this with the second subset gives $\theta_1, (\vec p_2, p_1') \SATSL \Gamma_2, x : B$. By IH(2), $\theta_1, \sum\vec p_2 + p_1' \SATSP t_2 : A$. If $t_2$ diverges, then the original term diverges and the proof is complete. Otherwise, there exist $v_2,p_2'$ such that $t_2\theta_1, \sum\vec p_2 + p_1' \to^* v_2, p_2'$ and $(v_2, p_2') \in \VALS{A}$.

        Check how the let term evaluates: $\tlet{x}{t_1\theta}{t_2\theta}, \sum\vec p_1 + \sum\vec p_2 \to^* \tlet{x}{v_1}{t_2\theta}, p_1' + \sum\vec p_2 \to^* t_2\theta[x\mapsto v_1], p_1' + \sum\vec p_2 \to^* v_2, p_2'$. By reverse preservation, the original term evaluates to an $A$ given resources $\sum\vec p_1 + \sum\vec p_2$, thus the case is complete.

        \item $\text{T-Tick}$. $\Gamma \vdash \Delta$ can be written as $\Gamma' \vdash \ttick{q} : \Tunit,\Delta'$. IH: $\Gamma' \vDash q,\Delta'$.

        Suppose $\theta,\vec p \SATSL \Gamma'$. By IH, either $\theta,\sum\vec p \SATSR \Delta'$, in which case the conclusion is immediate, or $\theta,\sum\vec p \SATSP q$.

        In the latter case, since $q$ abbreviates $\tunit:\TP{\Tunit}{q}$, we have $\sum\vec p \ge q$.

        By $\text{E-Tick}$, $\ttick{q}, \sum\vec p \rightarrow^* \tunit, \sum\vec p - q$, where $\sum\vec p - q \ge 0$. By definition, $(\ttick{q}, \sum\vec p) \in \TERMS{\Tunit}$.

        Also, for every valuation $\theta$, $(\ttick{q})\theta = \ttick{q}$.
        
        By definition, $\theta,\sum\vec p \SATSP \ttick{q} :\Tunit$.

        This completes the proof.

        \item $\text{T-OkVar}$. $\Gamma \vdash \Delta$ can be written as
        $\Gamma \vdash x : \Tok, \Delta'$.

        Suppose $\theta,\vec p \SATSL \Gamma$.
        Since substitutions map variables to closed values, $\theta(x)$ is a value.
        The total resource $\sum\vec p$ is non-negative, so by the semantics of $\Tok$,
        $(\theta(x),\sum\vec p)\in \VALS{\Tok}$.
        Thus $\theta,\sum\vec p \SATSP x:\Tok$, which satisfies the right-hand side.

        \item $\text{T-Ok}$. $\Gamma \vdash \Delta$ can be written as
        $\Gamma \vdash t:\Tok,\Delta'$.
        IH: $\Gamma \vDash t:A,\Delta'$ with $A\neq \Tok$.

        Suppose $\theta,\vec p \SATSL \Gamma$.
        By IH, either $\theta,\sum\vec p \SATSR \Delta'$, in which case the conclusion is immediate, or
        $\theta,\sum\vec p \SATSP t:A$.
        In the second case, $(t\theta,\sum\vec p)\in\TERMSP{A}$.
        If $t\theta$ diverges, then $(t\theta,\sum\vec p)\in\TERMSP{\Tok}$ by definition of $\TERMSP{\cdot}$.
        Otherwise $t\theta$ evaluates to some value $(v,p')$.
        Since resources remain non-negative during evaluation and $\Tok$ accepts every value, $(v,p')\in\VALS{\Tok}$.
        Hence $(t\theta,\sum\vec p)\in\TERMSP{\Tok}$ by reverse preservation.
        This proves $\theta,\sum\vec p \SATSP t:\Tok$.

\item[] \textbf{Left typing rules.}

        \item $\text{T-Dis}$. $\Gamma \vdash \Delta$ can be written as $\Gamma', t:B \vdash \Delta$. IH: $\Gamma' \vDash t:A, \Delta$ and $A \parallel B$.

        Suppose $\theta, (\vec p, p_1) \SATSL \Gamma', t:B$. Immediately, $\theta, \vec p \SATSL \Gamma'$. By IH, $\theta,  \sum \vec p \SATSR t : A, \Delta$.

        Analyze by cases:

        \begin{itemize}
            \item  $\theta,  \sum \vec p \SATSP t : A$. This implies $(t\theta,  \sum\vec p) \in \TERMSP{A}$. By assumption, $(t\theta, p_1) \in \TERMSP{B}$. Since $A \parallel B$, soundness of type disjointness rules out this case.
            \item  $\theta,  \sum \vec p \SATSR t :  \Delta$. By SWL, $\theta, \sum \vec p + p_1 \SATSR t :  \Delta$, which completes the proof.
        \end{itemize}

        \item $\text{T-AppNL}$. $\Gamma \vdash \Delta$ can be written as $\Gamma_1, \Gamma_2, t_1\ t_2 : B \vdash \Delta$. IH: $\Gamma_1 \vDash t_1 : A\onlyto B; \Phi$ and $\Gamma_2 , t_2 : A, \Phi \vDash \Delta$.

        Suppose $\theta, (\vec p_1, \vec p_2, p) \SATSL \Gamma_1, \Gamma_2, t_1\ t_2 : B$. Taking subvectors gives $\theta, \vec p_1 \SATSL \Gamma_1$, $\theta, \vec p_2 \SATSL \Gamma_2$, and $\theta, p \SATSL t_1\ t_2 : B$.
        
        By IH(1), $\theta, \sum\vec p_1 \SATSP t_1 : A\onlyto B; \Phi$. By destruction of ";", there exist $p', p''$ such that $\theta, p' \SATSP t_1 : A\onlyto B$, $\theta, p'' \SATSP \Phi$ (1), and $p' + p'' = \sum\vec p_1$.

        If $t_1$ diverges, then $t_1\ t_2$ also diverges, contradicting the assumption $\theta, p \SATS t_1\ t_2 : B$. Also, $\Phi$ does not diverge. Hence $\theta, p' \SATS t_1 : A\onlyto B$ and $\theta, p'' \SATS \Phi$.

        Expand $\theta, p' \SATS t_1 : A\onlyto B$ by definition: $(t_1\theta, p') \to (\tfix{f}{x}{t'}, p^*) \in \VALS{A \onlyto B}$, which implies that for every $(v, p^\Delta)$, if $(t'[f \mapsto \tfix{f}{x}{t'}][x \mapsto v], p^\Delta) \in \TERMS{B}$, then $(v, p^\Delta) \in \VALS{A}$ (2).

        Expanding the third subset $\theta, p \SATSL t_1\ t_2 : B$ gives $(t_1\theta\ t_2\theta, p) \to^* (\tfix{f}{x}{t'}\ t_2\theta, p^{**}_1) \to^* (\tfix{f}{x}{t'}\ v_2, p^{**}_2) \to (t'[f \mapsto \tfix{f}{x}{t'}][x \mapsto v_2], p^{**}_2) \in \TERMS{B}$.

        By (2), $(v_2, p^{**}_2) \in \VALS{A}$. By reverse preservation, $(t_2\theta, p^{**}_2) \in \TERMS{A}$, which means $\theta, p^{**}_2 \SATS t_2 : A$. Combining this with (1) and the first subset gives $\theta, (\vec p_2, p^{**}_2, p'') \SATSL \Gamma_2 , t_2 : A, \Phi$. By IH(3), $\theta, \sum \vec p_2 + p^{**}_2 + p'' \SATSR \Delta$. Since the sum of resources is bounded by $\sum\vec p_1 + \sum\vec p_2 + p$, the conclusion holds.


        \item  $\text{T-IfL}$. $\Gamma \vdash \Delta$ can be written as $\Gamma_1, \Gamma_2, \tif{t_1}{t_2}{t_3}: A\vdash \Delta$. IH: $\Gamma_2, t_2:A \vDash \Phi$; $\Gamma_2, t_3:A \vDash \Phi$ and  $\Gamma_1, t_1:\Tbool, \Phi \vDash \Delta$.

        Suppose $\theta, (\vec p_1, \vec p_2, p) \SATSL \Gamma_1,\Gamma_2,\tif{t_1}{t_2}{t_3}: A$. Split them into subsets, we get $\theta,\vec p_1\SATSL \Gamma_1$ (1), $\theta,\vec p_2\SATSL \Gamma_2$ (2) and $\theta ,p\SATS \tif{t_1}{t_2}{t_3}: A$ (3). 

        Expand (3) by definition: $(\tif{t_1\theta}{t_2\theta}{t_3\theta}, p) \to^* (\tif{v_1}{t_2\theta}{t_3\theta}, p') \to (t_x\theta, p') \in \TERMS{A}$. By the evaluation rules, $v_1$ is either $\ttrue$ or $\tfalse$, so $(t_1\theta, p - p') \in \TERMS{\Tbool}$, i.e. $\theta, p - p' \SATS t_1:\Tbool$ (4).
        
        The choice of $x$ depends on the value of $v_1$:
        \begin{itemize}
            \item $x = \ttrue$. In this case we get $(t_2\theta, p') \in \TERMS{A}$, and therefore $\theta, p' \SATSL t_2 : A$. Combining this with (2) gives $\theta, (\vec p_2, p') \SATSL \Gamma_2, t_2 : A$. By IH(1), $\theta, \sum \vec p_2 + p' \SATSR \Phi$. Since $\Phi$ does not diverge, we have $\theta, \sum \vec p_2 + p' \SATS \Phi$. Combining this with (1) and (4) gives $\theta, (\vec p_1, p - p', \vec p_2, p') \SATSL \Gamma_1, t_1:\Tbool, \Phi$. By IH(3), we get the conclusion.
            \item $x = \tfalse$. This is similar to the previous case, except that we get $t_3$ instead of $t_2$ and use IH(2) instead of IH(1).
        \end{itemize}

        \item $\text{T-LetL}$.  $\Gamma \vdash \Delta$ can be written as $\Gamma_1, \Gamma_2, \tlet{x}{t_1}{t_2}: B\vdash \Delta$. IH: $\Gamma_2, t_2:B\vDash x : A; \Phi$ and $\Gamma_1, t_1:A,\Phi \vDash\Delta$.
        
        Suppose $\theta, (\vec p_1, \vec p_2, p) \SATSL \Gamma_1,\Gamma_2,\tlet{x}{t_1}{t_2}: B$. Split them into subsets, we get $\theta,\vec p_1\SATSL \Gamma_1$ (1), $\theta,\vec p_2\SATSL \Gamma_2$ (2) and $\theta ,p\SATS \tlet{x}{t_1}{t_2}: B$ (3). 

        Expand (3) by definition: $(\tlet{x}{t_1\theta}{t_2\theta}, p) \to^* (\tlet{x}{v_1}{t_2\theta}, p') \to (t_2\theta[x \mapsto v_1], p') \in \TERMS{B}$. Let $\theta_1 = \theta[x \mapsto v_1]$; then $\theta_1, p' \SATS t_2 : B$. Combining this with (2) gives $\theta_1, (\vec p_2, p') \SATSL \Gamma_2, t_2:B$. By IH(1), we get $\theta_1, \sum\vec p_2 + p' \SATSR x : A; \Phi$. Since $x$ is a variable, $x$ cannot diverge, and $\Phi$ does not diverge. Thus, by destruction of ";", there exist $p^*_1,p^*_2$ such that $p^*_1 + p^*_2 = \sum\vec p_2 + p'$ and $\theta_1, p^*_1, p^*_2 \SATSL x : A, \Phi$, equivalently $\theta, p^*_1, p^*_2 \SATSL t_1 : A, \Phi$. Combining this with (1), and then applying IH(2), gives the conclusion.

        \item $\text{T-MatchL}$ $\Gamma \vdash \Delta$ can be written as $\Gamma_1, \Gamma_2, \tmatchlist{t_1}{t_2}{hd}{tl}{t_3} : B\vdash \Delta$. IH: $\Gamma_2, t_2:B \vdash \Phi$, $\Gamma_2, t_3:B \vdash hd::tl : \Tlist{A}; \Phi$ and $\Gamma_1, t_1:\Tlist{A},\Phi\vdash\Delta$.
        
        Suppose $\theta, (\vec p_1, \vec p_2, p) \SATSL \Gamma_1,\Gamma_2,\tmatchlist{t_1}{t_2}{hd}{tl}{t_3} : B$. Split them into subsets, we get $\theta,\vec p_1\SATSL \Gamma_1$ (1), $\theta,\vec p_2\SATSL \Gamma_2$ (2) and $\theta ,p\SATS \tmatchlist{t_1}{t_2}{hd}{tl}{t_3} : B$ (3). 

        Expand (3) by definition: $(\tmatchlist{t_1\theta}{t_2\theta}{hd}{tl}{t_3\theta}, p) \to^* (\tmatchlist{v_1}{t_2\theta}{hd}{tl}{t_3\theta}, p') \in \TERMS{B}$. Since this then evaluates, $v_1$ must be a list value. The further evaluation depends on whether $v_1$ is nil:

        \begin{itemize}
            \item $v_1 = []$. This case is similar to the $\text{T-IfL}$ case, except that destructing $[]$ introduces no variables in the nil branch.
            \item $v_1 = v_2 :: v_3$. The term further evaluates to $(t_3\theta[hd \mapsto v_2][tl \mapsto v_3], p') \in \TERMS{B}$. Let $\theta_1 =\theta[hd \mapsto v_2][tl \mapsto v_3]$; then $\theta_1, p'\SATS t_3:B$. Combining this with (2), we get $\theta_1, (\vec p_2, p')\SATSL \Gamma_2, t_3:B$. By IH(2), $\theta_1, \sum\vec p_2 + p' \SATSR hd :: tl :  \Tlist{A}; \Phi$. Since $hd$ and $tl$ are variables, they cannot diverge, and $\Phi$ does not diverge. Thus, by destruction of ";", there exist $p^*_1,p^*_2$ such that $p^*_1 + p^*_2 = \sum\vec p_2 + p'$ and $\theta_1, p^*_1, p^*_2 \SATSL hd :: tl :  \Tlist{A}, \Phi$, equivalently $\theta, p^*_1, p^*_2 \SATSL t_1 :  \Tlist{A}, \Phi$. Combining this with (1), and then applying IH(3), gives the conclusion.
        \end{itemize}

        \item $\text{T-MatchPairL}$ $\Gamma \vdash \Delta$ can be written as $\Gamma_1, \Gamma_2, \tmatchpair{t_1}{x}{y}{t_2} : A\vdash \Delta$. IH: $\Gamma_2, t_2:A \vdash (x,y):(A_1,A_2);\Phi$ and $\Gamma_1, t_1:(A_1,A_2),\Phi\vdash\Delta$.
        
        Suppose $\theta, (\vec p_1, \vec p_2, p) \SATSL \Gamma_1,\Gamma_2,\tmatchpair{t_1}{x}{y}{t_2} : A$. Split them into subsets, we get $\theta,\vec p_1\SATSL \Gamma_1$ (1), $\theta,\vec p_2\SATSL \Gamma_2$ (2) and $\theta ,p\SATS \tmatchpair{t_1}{x}{y}{t_2} : A$ (3). 

        Expand (3) by definition: $(\tmatchpair{t_1\theta}{x}{y}{t_2\theta}, p) \to^* (\tmatchpair{(v_1, v_2)}{x}{y}{t_2\theta}, p') \to (t_2\theta[x\mapsto v_1][y \mapsto v_2],p') \in \TERMS{A}$. Let $\theta_1 = \theta[x\mapsto v_1][y\mapsto v_2]$; by definition, $\theta_1, p' \SATS t_2 : A$. Combining this with (2) and applying IH(1), we get $\theta_1, \sum\vec p_2 + p' \SATSR (x, y) : (A_1,A_2); \Phi$. By destruction of ";" (and again, since $x,y$ and $\Phi$ do not diverge), there exist $p^*_1,p^*_2$ such that $p^*_1 + p^*_2 = \sum\vec p_2 + p'$ and $\theta_1, p^*_1, p^*_2 \SATSL (x, y) : (A_1,A_2), \Phi$, equivalently $\theta, p^*_1, p^*_2 \SATSL t_1 :  (A_1,A_2), \Phi$. Combining this with (1), and then applying IH(2), gives the conclusion.

        \item $\text{T-ListL}$. $\Gamma \vdash \Delta$ can be written as $\Gamma_1, t_1 :: t_2 : \Tlist{A} \vdash \Delta$. IH: $\Gamma_1, t_1 : A, t_2 : \Tlist{A} \vdash \Delta$.
        
        Suppose $\theta, (\vec p_1, p) \SATSL \Gamma_1, t_1 :: t_2 : \Tlist{A}$. Split them into subsets, we get $\theta,\vec p_1\SATSL \Gamma_1$ (1) and $\theta, p\SATS t_1 :: t_2 : \Tlist{A}$ (2). 

        Expand (2) by definition: $(t_1\theta :: t_2\theta, p) \to^* (v_1 :: t_2\theta, p') \to^* (v_1 :: v_2, p'') \in \VALS{\Tlist{A}}$. By definition of $\VALS{\Tlist{A}}$, there exist $p^*, p^{**}$ such that $(v_1,p^*) \in \VALS{A}$, $(v_2,p^{**}) \in \VALS{\Tlist{A}}$, and $p^* + p^{**} = p''$. By reverse preservation, $(t_1\theta, p^* + p - p') \in \TERMS{A}$ and $(t_2\theta, p^{**} + p' - p'') \in \TERMS{\Tlist{A}}$. Thus $\theta, p^* + p - p', p^{**} + p' - p'' \SATSL t_1:A, t_2:\Tlist{A}$. Combining this with (1) and applying IH gives the conclusion.

        \item $\text{T-PairL}$. Similar to the $\text{T-ListL}$ case, except that we analyze evaluation of $(t_1\theta, t_2\theta)$ to $(v_1, v_2)$ and get that they have types $A$ and $B$ respectively.
        
        \item $\text{T-TickL}$. $\Gamma \vdash \Delta$ can be written as $\Gamma_1,\ttick{q} : \Tunit \vdash \Delta$. IH: $\Gamma_1, \tunit : \TP{\Tunit}{q} \vdash \Delta$.

        Suppose $\theta, (\vec p_1, p) \SATSL \Gamma_1, \ttick{q} : \Tunit$. Split them into subsets, we get $\theta,\vec p_1\SATSL \Gamma_1$ (1) and $\theta, p\SATS \ttick{q} : \Tunit$ (2). 

        Expand (2) by definition: $(\ttick{q}, p) \to (\tunit, p - q)$. From this we know $p \ge q$, and thus by definition $(\tunit, p) \in \TP{\Tunit}{q}$. Combining this with (1) and applying IH(1) completes the proof.

        \item $\text{T-TickOkL}$. $\Gamma \vdash \Delta$ can be written as
        $\Gamma_1,\ttick{q}:\Tok \vdash \Delta$.
        IH: $\Gamma_1,q \vDash \Delta$.

        Suppose $\theta,(\vec p_1,p)\SATSL \Gamma_1,\ttick{q}:\Tok$.
        From $\theta,p\SATS \ttick{q}:\Tok$, the term $\ttick{q}$ must evaluate to a value under resource $p$.
        By the operational rule for $\ttick{q}$, this implies $p\ge q$.
        Together with $\theta,\vec p_1\SATSL\Gamma_1$, the premise of the induction hypothesis is satisfied with the additional resource component $q$.
        Hence $\theta,\sum\vec p_1+q\SATSR\Delta$.
        Since $p\ge q$, semantic weakening gives $\theta,\sum\vec p_1+p\SATSR\Delta$, which is the desired conclusion.

    \end{itemize}
\end{proof}

\subsubsection{The $\text{T-Abs}$ and $\text{T-AbsN}$ Case}

\paragraph{Lambda abstractions.} We write $\lambda x. t$ for $\tfix{f}{x}{t}$ where $f$ does not appear in $t$. $\lambda x. t$ is not a new piece of grammar, but instead syntactic sugar. We present the semantic definition for lambda abstractions here. This is not an extension of the original semantic definition, but only a more convenient way to reason about the $\lambda$-subset of the fixpoint terms.

\begin{align*}
    \VALS{\langle A \rightarrow B, q\rangle} & \supset & \{ (\lambda x.t, p) \ | \ p \ge q, \forall v, p_1, (v, p_1) \in \VALS{A} \rightarrow 
    (t[x\mapsto v], p_1) \in \TERMSP{B}\} \\
    \VALS{\langle A \onlyto B, q\rangle} & \supset & \{ (\lambda x.t, p) \ | \ p \ge q, \forall v, p_1,
    (t[x\mapsto v], p_1) \in \TERMS{B} \rightarrow (v, p_1) \in \VALS{A} \} 
\end{align*}

\begin{definition}
    (Fixpoint Approximants)
\begin{align*}
    \tfixn{0}{f}{x}{t} & = \lambda x.\texttt{div} \\
    \tfixn{n + 1}{f}{x}{t} & = \lambda x.t[f \mapsto \tfixn{n}{f}{x}{t}]
\end{align*}
\end{definition}

\begin{definition}
    We write $t_1 \lesssim t_2$ if and only if:
    \[
        \forall p,n,p'.\ (t_1,p)\to^*(n,p') \wedge (n,p')\text{ is a normal configuration}
        \implies (t_2,p)\to^*(n,p').
    \]
\end{definition}

\begin{definition}
    (Safety Property) A type $A$ defines a safety property if:
    \begin{itemize}
        \item If $(C[t_2], p) \in \TERMSP{A}$ and $t_1 \lesssim t_2$, then $(C[t_1], p) \in \TERMSP{A}$.
        \item If $(C[\tfix{f}{x}{t}], p) \notin \TERMSP{A}$, then there is some $k$ for which $(C[\tfixn{k}{f}{x}{t}], p) \notin \TERMSP{A}$.
    \end{itemize}
\end{definition}

The strengthened S1 statement and proof are given as Theorem~\ref{thm:s1-enhanced}
in our technical report. As there, $F$ denotes an annotated finitely verifiable type, namely
an instance of the category $A_F$ in the safety-fragment grammar.

\begin{theorem}
    (Safety Property, part 2, enhanced) For all types $A$ and finitely verifiable types $F$: 
    \begin{itemize}
        \item (i) If $(C[\tfix{f}{x}{t}], p) \notin \TERMSP{A}$, then there is some $n_0$ such that for every $n \ge n_0$, $(C[\tfixn{n}{f}{x}{t}], p) \notin \TERMSP{A}$.
        \item (ii) If $(C[\tfix{f}{x}{t}], p) \in \TERMSP{F}$, then there is some $n_0$ such that for every $n \ge n_0$, $(C[\tfixn{n}{f}{x}{t}], p) \in \TERMSP{F}$.
    \end{itemize}
\end{theorem}

\begin{proof}
    The proof is by induction on $A$.

    \begin{itemize}
        \item $A = \TP{\Tunit}{q}$. \begin{itemize}
            \item $(C[\tfix{f}{x}{t}], p) \notin \TERMSP{A}$. Then $(C[\tfix{f}{x}{t}], p)$ either gets stuck, evaluates to a non-unit value, or evaluates to $(\tunit, p')$ with $p' < q$.
            Let $m$ be the number of times the E-AppAbs rule is applied during this reduction.
            Then for any $n \ge m$, the term $(C[\tfixn{n}{f}{x}{t}], p)$ can simulate the same reduction steps (because each E-AppAbs step requires a fixpoint abstraction with at least one unfolding, and $\tfixn{n}{f}{x}{t}$ provides exactly $n$ unfoldings). Thus $(C[\tfixn{n}{f}{x}{t}], p)$ gets stuck or evaluates to a non-unit value in the same way. Pick $n_0 = m$ and the case is complete.
            \item $(C[\tfix{f}{x}{t}], p) \in \TERMS{A}$. Then $C[\tfix{f}{x}{t}]$ evaluates to $\tunit$ in finite steps. 
            Let $m$ be the number of E-AppAbs reductions in this evaluation.
            For all $n \ge m$, $(C[\tfixn{n}{f}{x}{t}], p)$ also evaluates to $\tunit$ by the same simulation argument. Once again we pick $n_0 = m$.
        \end{itemize}
        \item $A = \TP{\Tbool}{q}$. Similar to the unit case.
        \item $A = \TP{\Tok}{q}$. \begin{itemize}
            \item $(C[\tfix{f}{x}{t}], p) \notin \TERMSP{A}$.
            Since $\Tok$ accepts every value, failure can only mean that the computation gets stuck, or that it evaluates to some value $(v,p')$ with $p' < q$.
            Let $m$ be the number of E-AppAbs reductions in this finite computation.
            For every $n \ge m$, the approximant can simulate the same finite reduction prefix and reaches the same stuck configuration or a value with the same final resource.
            Thus $(C[\tfixn{n}{f}{x}{t}], p) \notin \TERMSP{A}$.

            \item $(C[\tfix{f}{x}{t}], p) \in \TERMSP{A}$.
            If this holds by divergence, the usual approximant simulation of the divergent run gives an $n_0$ such that all $n \ge n_0$ also diverge, hence belong to $\TERMSP{A}$.
            Otherwise the full fixpoint evaluates in finite steps to some value $(v,p')$ with $p'\ge q$.
            Let $m$ be the number of E-AppAbs reductions in this evaluation.
            For all $n \ge m$, the approximant simulates the same evaluation and reaches a value with the same final resource, so it is in $\TERMSP{A}$.
        \end{itemize}
        \item $A = \TP{(A_1,A_2)}{q}$.  \begin{itemize}
            \item $(C[\tfix{f}{x}{t}], p) \notin \TERMSP{A}$. Then there is either: \begin{itemize}
                \item $(C[\tfix{f}{x}{t}], p)$ gets stuck;
                \item $(C[\tfix{f}{x}{t}], p)$ evaluates to a non-pair value;
                \item $(C[\tfix{f}{x}{t}], p)$ evaluates to $((v_1, v_2), p')$ but $p' < q$;
                    
                    These cases are similar to the unit case. The same simulation argument using $m$ shows that, for all $n \ge m$, the approximant behaves identically, so we can take $n_0 = m$.

                \item $(C[\tfix{f}{x}{t}], p)$ evaluates to a pair $((V_1[\tfix{f}{x}{t}], V_2[\tfix{f}{x}{t}]), p')$ where $p' - q \ge 0$ and, for every $p_1 + p_2 = p' - q$, at least one of the following holds: \begin{itemize}
                    \item $(V_1[\tfix{f}{x}{t}], p_1)$ does not satisfy $A_1$;
                    \item $(V_2[\tfix{f}{x}{t}], p_2)$ does not satisfy $A_2$.
                \end{itemize} 

                Let there be $r$ choices of $(p_1, p_2)$. For the $i$-th choice, by IH, at least one of the following holds: \begin{itemize}
                    \item there exists $n_{i,1}$ such that for every $n \ge n_{i,1}$, $(V_1[\tfixn{n}{f}{x}{t}], p_1)$ does not satisfy $A_1$;
                    \item there exists $n_{i,2}$ such that for every $n \ge n_{i,2}$, $(V_2[\tfixn{n}{f}{x}{t}], p_2)$ does not satisfy $A_2$.
                \end{itemize}

                Let $n_i$ be $n_{i,1}$ if the first holds, $n_{i,2}$ if the second holds, or $\max\{n_{i,1},n_{i,2}\}$ if both hold.
                
                Let $m$ be the number of E-AppAbs reductions in this evaluation. We pick $n_0 = \max_{i}\{n_i\} + m$ and the original conclusion holds, because for any $n \ge n_0$, the term $C[\tfixn{n}{f}{x}{t}]$ has at least $m$ unfoldings to simulate the original E-AppAbs steps, and the remaining unfoldings are enough to guarantee the component values do not satisfy their types by the induction hypothesis.
            \end{itemize} 

            \item $(C[\tfix{f}{x}{t}], p) \in \TERMS{A}$. Then $C[\tfix{f}{x}{t}]$ evaluates to the value $((V_1[\tfix{f}{x}{t}], V_2[\tfix{f}{x}{t}]), p')$ in finite steps.
            Let $m$ be the number of E-AppAbs reductions in this evaluation.
            By definition of $\VALS{\TP{(A_1,A_2)}{q}}$, there are $p_1,p_2$ such that $(V_1[\tfix{f}{x}{t}], p_1) \in \VALS{A_1}$, $(V_2[\tfix{f}{x}{t}], p_2) \in \VALS{A_2}$, and $p_1 + p_2 + q = p'$.
            Since $A$ is finitely verifiable, the subtypes $A_1$ and $A_2$ are finitely verifiable as well.
            By IH, there is some $n_1, n_2$ such that for all $n\ge \max\{n_1,n_2\}$, $(V_1[\tfixn{n}{f}{x}{t}], p_1) \in \TERMS{A_1}$ and $(V_2[\tfixn{n}{f}{x}{t}], p_2) \in \TERMS{A_2}$. 
            For similar reasons we pick $n_0 = \max\{n_1, n_2\} + m$.
        \end{itemize}

        \item $A = \TP{\Tlist{A_1}}{q}$.  \begin{itemize}
            \item $(C[\tfix{f}{x}{t}], p) \notin \TERMSP{A}$. Then there is either: \begin{itemize}
                \item $(C[\tfix{f}{x}{t}], p)$ gets stuck;
                \item $(C[\tfix{f}{x}{t}], p)$ evaluates to a non-pair value;
                \item $(C[\tfix{f}{x}{t}], p)$ evaluates to $(v_1 :: v_2 :: \dots :: v_\ell :: [], p')$ but $p' < q$;
                    
                    These cases are similar to the previous cases.

                \item $(C[\tfix{f}{x}{t}], p)$ evaluates to a list $(V_1[\tfix{f}{x}{t}] :: V_2[\tfix{f}{x}{t}] :: \dots :: V_\ell[\tfix{f}{x}{t}] ::  [], p')$ ($\ell \ge 0$) where $p' - q \ge 0$ and, for every choice of $\{p_i\}_{i=1}^\ell$ with $\sum_{i=1}^\ell p_i = p' - q$, there exists $i$ such that $(V_i[\tfix{f}{x}{t}], p_i)$ does not satisfy $A_1$.
                
                Note that the existential condition requires $\ell \ge 1$; otherwise no such $i$ exists.
                
                Let there be $r$ choices of $\{p_i\}_{i=1}^\ell$. For the $j$-th choice, by IH, there exists $i$ and an index $n_{j,i}$ such that for every $n \ge n_{j,i}$, $(V_i[\tfixn{n}{f}{x}{t}], p_i)$ does not satisfy $A_1$.

                Let $n_j$ be the maximum of all $n_{j,i}$ such that the above statement holds for such $i$.
                
                Let $m$ be the number of E-AppAbs reductions in this evaluation. We pick $n_0 = \max_{j}\{n_j\} + m$ and the original conclusion holds.
            \end{itemize} 

            \item $(C[\tfix{f}{x}{t}], p) \in \TERMS{A}$. Then $C[\tfix{f}{x}{t}]$ evaluates to the value $(V_1[\tfix{f}{x}{t}] :: V_2[\tfix{f}{x}{t}] :: \dots :: V_\ell[\tfix{f}{x}{t}] ::  [], p')$ ($\ell \ge 0$) in finite steps.
            Let $m$ be the number of E-AppAbs reductions in this evaluation.
            For $\ell = 0$, $C[\tfix{f}{x}{t}]$ evaluates to $[]$, which is invariant under replacing fixpoint with its approximant. Similar to the unit case, we pick $n_0 = m$.

            For $\ell \ge 1$, by definition of $\VALS{\TP{\Tlist{A_1}}{q}}$, for every $1 \leq i \leq \ell$ there is a resource $p_i$ such that $(V_i[\tfix{f}{x}{t}], p_i) \in \VALS{A_1}$ and $\sum_{i=1}^\ell p_i + q = p'$.
            Since $A$ is finitely verifiable, $A_1$ is finitely verifiable as well.
            By IH, for every $i$ there is $n_i$ such that for all $n\ge n_i$, $(V_i[\tfixn{n}{f}{x}{t}], p_i) \in \TERMS{A_1}$.

            For reasons similar to the previous cases, we pick $n_0 = \max_{i=1}^\ell\{n_i\} + m$.
        \end{itemize}

        \item $A = \TP{A_1 \to A_2}{q}$.
            Since arrow is not finitely verifiable type, we need only consider case (i): $(C[\tfix{f}{x}{t}], p) \notin \TERMSP{A}$. Then there is either: \begin{itemize}
            \item $(C[\tfix{f}{x}{t}], p)$ gets stuck;
            \item $(C[\tfix{f}{x}{t}], p)$ evaluates to a non-abstraction value;
            \item $(C[\tfix{f}{x}{t}], p)$ evaluates to $(\tfix{f'}{x'}{t'}, p')$ but $p' < q$;
                
                These cases are similar to the previous cases.

            \item $(C[\tfix{f}{x}{t}], p)$ evaluates to $(\tfix{f'}{x'}{t'}, p')$ where $p' \ge q$, but there exists $(v_1, p_1) \in \VALS{A_1}$ such that 
            $(t'[f' \mapsto \tfix{f'}{x'}{t'}][x' \mapsto v_1], p_1) \notin \TERMSP{A_2}$.
            
            Let $m$ be the number of E-AppAbs reductions in this evaluation to the abstraction.
            For any $n \ge m$, the term $C[\tfixn{n}{f}{x}{t}]$ evaluates to the abstraction 
            $(\tfix{f'}{x'}{t'_n}, p')$ where $t'_n$ is $t'$ with every occurrence of $\tfix{f}{x}{t}$ replaced by $\tfixn{n}{f}{x}{t}$.

            Now consider the same input $(v_1, p_1) \in \VALS{A_1}$.
            The resulting computation, which we denote as $D[\tfix{f}{x}{t}]$, satisfies
            $D[\tfix{f}{x}{t}] \notin \TERMSP{A_2}$ by assumption.

            Since $A_2$ is a smaller type (strict subterm of $A$), by the induction hypothesis (i) applied to $A_2$,
            there exists $n_1$ such that for all $n \ge n_1$, $D[\tfixn{n}{f}{x}{t}] \notin \TERMSP{A_2}$.

            Therefore, for all $n \ge m + n_1$, the abstraction $(\tfix{f'}{x'}{t'_n}, p')$ fails to satisfy $A$ (because the same input $v_1$ leads to a value not in $\TERMSP{A_2}$).

            Hence $C[\tfixn{n}{f}{x}{t}] \notin \TERMSP{A}$.
            Take $n_0 = m + n_1$.
        \end{itemize} 

        \item $A = \TP{A_1 \onlyto A_2}{q}$.
        Since necessity arrow is not a finitely verifiable type, we need only consider case (i): $(C[\tfix{f}{x}{t}], p) \notin \TERMSP{A}$. Then there is either:
        \begin{itemize}
            \item $(C[\tfix{f}{x}{t}], p)$ gets stuck;
            \item $(C[\tfix{f}{x}{t}], p)$ evaluates to a non-abstraction value;
            \item $(C[\tfix{f}{x}{t}], p)$ evaluates to $(\tfix{f'}{x'}{t'}, p')$ but $p' < q$;
                
            These cases are similar to the previous cases.
                
            \item $(C[\tfix{f}{x}{t}], p)$ evaluates to $(\tfix{f'}{x'}{t'}, p')$ where $p' \ge q$, but there exists $(v_1, p_1) \notin \VALS{A_1}$ such that 
            $(t'[f' \mapsto \tfix{f'}{x'}{t'}][x' \mapsto v_1], p_1) \in \TERMSP{A_2}$.
                
            Let $m$ be the number of E-AppAbs reductions in this evaluation to the abstraction.
            For any $n \ge m$, the term $C[\tfixn{n}{f}{x}{t}]$ evaluates to the abstraction 
            $(\tfix{f'}{x'}{t'_n}, p')$ where $t'_n$ is $t'$ with every occurrence of $\tfix{f}{x}{t}$ replaced by $\tfixn{n}{f}{x}{t}$.

            Now consider the same input $(v_1, p_1) \notin \VALS{A_1}$.
            The resulting computation, which we denote as $D[\tfix{f}{x}{t}]$, satisfies
            $D[\tfix{f}{x}{t}] \in \TERMSP{A_2}$ by assumption.

            Since $A_2$ is a smaller type and, in the safety fragment, $A_2$ is finitely verifiable,
            by the induction hypothesis (ii) applied to $A_2$,
            there exists $n_1$ such that for all $n \ge n_1$, $D[\tfixn{n}{f}{x}{t}] \in \TERMSP{A_2}$.

            Therefore, for all $n \ge m + n_1$, the abstraction $(\tfix{f'}{x'}{t'_n}, p')$ fails to satisfy $A$ (because the same input $v_1$, which is not in $A_1$, yields a result in $\TERMSP{A_2}$).

            Hence $C[\tfixn{n}{f}{x}{t}] \notin \TERMSP{A}$.
            Take $n_0 = m + n_1$.
        \end{itemize}

    \end{itemize}
\end{proof}

\begin{lemma}
    (Fixpoint Closure Lemma) In the safety fragment:
    \begin{itemize}
        \item if $(\lambda f. \lambda x. t, p) \in \VALS{(A \to B) \to A \to B}$, then $(\tfix{f}{x}{t}, p) \in \TERMSP{A \to B}$.
        \item if $(\lambda f. \lambda x. t, p) \in \VALS{(A \onlyto B) \to A \onlyto B}$, then $(\tfix{f}{x}{t}, p) \in \TERMSP{A \onlyto B}$.
    \end{itemize}
\end{lemma}

\begin{proof}
    We prove the first statement; the second is identical with $A\onlyto B$ in place of $A\to B$.
    Suppose, for contradiction, that
    $(\tfix{f}{x}{t},p)\notin\TERMSP{A\to B}$.
    By Safety Property, part 2, there is an $n_0$ such that for every
    $n\ge n_0$,
    \[
        (\tfixn{n}{f}{x}{t},p)\notin\TERMSP{A\to B}.
    \]
    We now show, by induction on $n$, that every approximant
    $\tfixn{n}{f}{x}{t}$ is in $\TERMSP{A\to B}$ under resource $p$.
    For $n=0$, the approximant is $\lambda x.\texttt{div}$, which has type
    $A\to B$ because applying it to any argument diverges, and divergence is
    included in $\TERMSP{B}$.

    For the step case, assume
    $(\tfixn{n}{f}{x}{t},p)\in\TERMSP{A\to B}$.
    Since $\tfixn{n}{f}{x}{t}$ is a value, this means
    $(\tfixn{n}{f}{x}{t},p)\in\VALS{A\to B}$.
    From
    $(\lambda f.\lambda x.t,p)\in\VALS{(A\to B)\to A\to B}$,
    applying the outer arrow to this approximant gives
    \[
        (\lambda x.t[f\mapsto \tfixn{n}{f}{x}{t}],p)
        \in\TERMSP{A\to B}.
    \]
    This term is exactly $\tfixn{n+1}{f}{x}{t}$, so the induction is complete.
    Taking any $n\ge n_0$ contradicts the conclusion of S2.  Hence
    $(\tfix{f}{x}{t},p)\in\TERMSP{A\to B}$.

    For $A\onlyto B$, the base approximant $\lambda x.\texttt{div}$ satisfies
    the necessity-arrow type vacuously, because its body never belongs to
    $\TERMS{B}$.  The induction step uses the premise
    $(\lambda f.\lambda x.t,p)\in\VALS{(A\onlyto B)\to A\onlyto B}$ in exactly
    the same way.
\end{proof}

\begin{proof}
    (Proof of the $\text{T-Abs}$ and $\text{T-AbsN}$ cases in the soundness theorem)
    
    \begin{itemize}
    
        \item $\text{T-Abs}$. $\Gamma \vdash \Delta$ can be written as $\Gamma \vdash \tfix{f}{x}{t} : A \to B, \Delta'$. IH: $\Gamma, f : A \to B, x : A \vdash t : B, \Delta'$, with side condition $\Gamma\share(\Gamma,\Gamma)$.
        
        Suppose $\theta, \vec p \SATSL \Gamma$ and $\Delta'$ is not satisfied. We need to show $\theta, \sum \vec p \SATSP \tfix{f}{x}{t} : A \to B$. By definition, we need to show $(\tfix{f}{x}{t\theta}, \sum \vec p) \in \TERMSP{A \to B}$.

        The side condition $\Gamma\share(\Gamma,\Gamma)$ follows from the value-like, zero-potential shape of $\Gamma$ used in this rule; equivalently, it can be obtained from $VL(\Gamma)$.  Thus the formulas in $\Gamma$ can be duplicated and carried at zero resource in the closure proof.

        For all $(v_1, p_1) \in \VALS{A \to B}$ and all $(v_2, p_2) \in \VALS{A}$, let $\theta' = \theta[f \mapsto v_1][x \mapsto v_2]$. Since the recursive-function type is used without extra outer potential, $(v_1,0)\in\VALS{A\to B}$ also holds. By the preceding observation about $\Gamma$, we have $\theta',(\vec{0},0,p_2)\SATSL \Gamma,f:A\to B,x:A$. By IH, $\theta',p_2\SATSP t:B$, i.e. $(t\theta[f \mapsto v_1][x \mapsto v_2],p_2)\in\TERMSP{B}$, unless $\Delta'$ is already satisfied.

        Therefore $(\lambda x.t\theta[f \mapsto v_1], p_1) \in \VALS{A \to B}$.

        By definition of $\VALS{(A \to B) \to A \to B}$, we have $(\lambda f.\lambda x.t\theta, 0) \in \VALS{(A \to B) \to A \to B}$.

        By Fixpoint Closure Lemma, we have $(\tfix{f}{x}{t\theta}, 0) \in \TERMSP{A \to B}$.

        By SWL we have $(\tfix{f}{x}{t\theta}, \sum \vec p) \in \TERMSP{A \to B}$, which completes the proof.

        \item $\text{T-AbsN}$. $\Gamma \vdash \Delta$ can be written as $\Gamma \vdash \tfix{f}{x}{t} : A \onlyto B, \Delta'$. IH: $\Gamma, f : A \onlyto B, t : B \vdash x : A, \Delta'$, with side condition $\Gamma\share(\Gamma,\Gamma)$.

        Suppose $\theta,\vec p\SATSL\Gamma$ and $\Delta'$ is not satisfied. We need to show
        \[
            (\tfix{f}{x}{t\theta},\sum\vec p)\in\TERMSP{A\onlyto B}.
        \]
        As in the $\text{T-Abs}$ case, the sharing side condition $\Gamma\share(\Gamma,\Gamma)$ follows from the value-like, zero-potential form of $\Gamma$ and can be obtained from $VL(\Gamma)$.  Hence the captured formulas in $\Gamma$ can be used at zero resource in the closure proof.

        We first prove
        \[
            (\lambda f.\lambda x.t\theta,0)\in\VALS{(A\onlyto B)\to A\onlyto B}.
        \]
        Let $(v_1,p_1)\in\VALS{A\onlyto B}$.  It is enough to prove
        $(\lambda x.t\theta[f\mapsto v_1],p_1)\in\VALS{A\onlyto B}$.
        So take arbitrary $v_2,p_2$ and assume
        \[
            (t\theta[f\mapsto v_1][x\mapsto v_2],p_2)\in\TERMS{B}.
        \]
        Since the recursive-function type is used without extra outer potential, $(v_1,0)\in\VALS{A\onlyto B}$ also holds. With $\theta'=\theta[f\mapsto v_1][x\mapsto v_2]$, the left side of the induction hypothesis is satisfied by
        $\theta',(\vec{0},0,p_2)\SATSL \Gamma,f:A\onlyto B,t:B$.
        By IH, and excluding the already-handled $\Delta'$ alternative, we obtain
        $\theta',p_2\SATSP x:A$.
        Since $x\theta'=v_2$ is a value, this gives $(v_2,p_2)\in\VALS{A}$.
        Hence $(\lambda x.t\theta[f\mapsto v_1],p_1)\in\VALS{A\onlyto B}$.

        Therefore $(\lambda f.\lambda x.t\theta,0)\in\VALS{(A\onlyto B)\to A\onlyto B}$.
        By the Fixpoint Closure Lemma,
        $(\tfix{f}{x}{t\theta},0)\in\TERMSP{A\onlyto B}$.
        Finally, SWL gives
        $(\tfix{f}{x}{t\theta},\sum\vec p)\in\TERMSP{A\onlyto B}$, as required.

    \end{itemize}

\end{proof}

\subsection{The Safety Property S1}
\label{subsec:safety-property-s1}

\begin{definition}[Normal and stuck configurations]
    A configuration $(t,p)$ is \emph{normal} if there are no $t'$ and $p'$ such that $(t,p) \to (t',p')$.
    A normal configuration $(t,p)$ is \emph{stuck} if $t$ is not a value.
    Equivalently, stuck configurations are generated by the following redex forms, inside evaluation contexts:
    \begin{itemize}
        \item $v\ t$ where $v$ is not a fixpoint abstraction;
        \item $\tif{v}{t_2}{t_3}$ where $v$ is neither $\ttrue$ nor $\tfalse$;
        \item $\tmatchlist{v}{t_2}{hd}{tl}{t_3}$ where $v$ is neither $[]$ nor $v_1 :: v_2$;
        \item $\tmatchpair{v}{x}{y}{t}$ where $v$ is not a pair value;
        \item $\ttick{q}$ under resource $p$ where $p < q$.
    \end{itemize}
    Since reduction is deterministic, every closed configuration either diverges, reaches a value, or gets stuck.
\end{definition}

\begin{lemma}[Divergence transfer]\label{lem:s1-divergence-transfer}
    If $t_1 \lesssim t_2$, $t[x \mapsto t_2]$ is closed, and $(t[x \mapsto t_2],p) \to^\omega$, then
    $(t[x \mapsto t_1],p) \to^\omega$.
\end{lemma}

\begin{proof}
    We reason coinductively on the infinite reduction of $(t[x \mapsto t_2],p)$.
    If the infinite computation never evaluates an occurrence of $t_2$, then the same surrounding reduction steps are available from $(t[x \mapsto t_1],p)$, so the $t_1$-instance also diverges.

    Otherwise consider the first occurrence of $t_2$ that is forced, say under current resource $r$.
    If $(t_2,r)$ diverges, then $(t_1,r)$ must also diverge: if $(t_1,r)$ reached a normal configuration $(n,r')$, then $t_1\lesssim t_2$ would imply $(t_2,r)\to^*(n,r')$, contradicting divergence.
    Hence the whole $t_1$-instance diverges.

    If $(t_2,r)$ reaches a normal configuration $(n,r')$, then the infinite computation continues after this occurrence has become $n$.
    The $t_1$-instance either diverges while evaluating the corresponding occurrence, in which case we are done, or it reaches the same $(n,r')$ by $t_1\lesssim t_2$.
    In the latter case the number of already-forced occurrences has increased, and we apply the same coinductive argument to the infinite suffix.
\end{proof}

\begin{lemma}[Normal-form transfer]\label{lem:s1-normal-transfer}
    If $t_1 \lesssim t_2$, $t[x \mapsto t_2]$ is closed, and
    $(t[x \mapsto t_2],p) \to^* (n_2,p')$ with $(n_2,p')$ normal, then either
    $(t[x \mapsto t_1],p) \to^\omega$, or there is a term $n$ such that
    $n_2 = n[x \mapsto t_2]$ and
    $(t[x \mapsto t_1],p) \to^* (n[x \mapsto t_1],p')$.
\end{lemma}

\begin{proof}
    This is the resource-preserving version of the substitution-transfer lemma.
    The proof is by lexicographic induction on the length of
    $(t[x \mapsto t_2],p) \to^* (n_2,p')$ and the number of free occurrences of $x$ in $t$.
    Here, the first redex is \emph{blocked on $x$} when it is an occurrence of $x$ placed in the current evaluation context.

    If the sequence has length zero, take $n=t$.
    For a non-empty sequence, if the first redex is not blocked on $x$, then the one-step reduction can be uniquely pulled back through the substitution:
    $t \to t'$ and the reduct is $t'[x \mapsto t_2]$.
    The induction hypothesis applied to $t'$ gives the result, and prefixing the same one-step reduction to $t[x \mapsto t_1]$ preserves the resource.

    If the first blocked redex is an occurrence of $x$, then the original reduction first evaluates $t_2$ from some resource $r$ to a normal configuration $(m,r')$.
    If $t_1$ diverges from $r$, then so does the whole configuration.
    Otherwise, by $t_1 \lesssim t_2$ both $t_1$ and $t_2$ reach the same $(m,r')$.
    Replace this occurrence of $x$ by $m$ in the surrounding context and apply the induction hypothesis to the remaining suffix; the number of free occurrences of $x$ has strictly decreased.
    In every case the transferred computation has the same final resource $p'$ because each simulated step is the same small-step rule and the blocked occurrence reaches the same resource $r'$.
\end{proof}

\begin{lemma}[Value transfer]\label{lem:s1-value-transfer}
    If $t_1 \lesssim t_2$, $t[x \mapsto t_2]$ is closed, and
    $(t[x \mapsto t_2],p) \to^* (v_2,p')$ where $v_2$ is a value, then either
    $(t[x \mapsto t_1],p) \to^\omega$, or
    $(t[x \mapsto t_1],p) \to^* (v_1,p')$ for some value $v_1$ with the same outer value constructor as $v_2$.
    More precisely:
    \begin{itemize}
        \item if $v_2$ is $\tunit$, $\ttrue$, $\tfalse$, or $[]$, then $v_1=v_2$;
        \item if $v_2=(v_{21},v_{22})$, then $v_1=(v_{11},v_{12})$ and each $v_{1i}$ is obtained from the corresponding ancestor of $v_{2i}$ by the same transfer;
        \item if $v_2=v_{21}::v_{22}$, then $v_1=v_{11}::v_{12}$ and each component is transferred similarly;
        \item if $v_2=\tfix{f}{y}{s_2}$, then $v_1=\tfix{f}{y}{s_1}$ and there is a term $s$ such that $s_2=s[x \mapsto t_2]$ and $s_1=s[x \mapsto t_1]$.
    \end{itemize}
\end{lemma}

\begin{proof}
    Apply Lemma~\ref{lem:s1-normal-transfer} to the evaluation of $t[x \mapsto t_2]$.
    If the $t_1$-instance diverges, we are done.
    Otherwise it reaches $n[x \mapsto t_1]$ with $n[x \mapsto t_2]=v_2$ and the same resource $p'$.
    We then inspect the derivation that $n[x \mapsto t_2]$ is a value.
    If $n=x$, then $t_2$ is already the value $v_2$ under resource $p'$; by $t_1\lesssim t_2$, the term $t_1$ either diverges from $p'$ or reaches the same value with the same resource.
    If the outer constructor of $n$ is $\tunit$, $\ttrue$, $\tfalse$, or $[]$, the conclusion is immediate.
    If the outer constructor is a pair or list constructor, apply the same argument recursively to the value components; if any component diverges then the whole value construction diverges, and otherwise the components reach corresponding values with the same resource accounting.
    If the outer constructor is a fixpoint abstraction, the body is not evaluated, so the transferred value is the same abstraction with $t_1$ substituted in the body.
\end{proof}

\begin{lemma}[Stuck transfer]\label{lem:s1-stuck-transfer}
    If $t_1 \lesssim t_2$, $t[x \mapsto t_2]$ is closed, and $(t[x \mapsto t_2],p)$ gets stuck, then
    $(t[x \mapsto t_1],p)$ either diverges or gets stuck.
\end{lemma}

\begin{proof}
    Suppose $(t[x \mapsto t_2],p) \to^* (s_2,p')$ and $(s_2,p')$ is stuck.
    By Lemma~\ref{lem:s1-normal-transfer}, either $(t[x \mapsto t_1],p)$ diverges, or it reaches $(s_1,p')$ with $s_2=s[x \mapsto t_2]$ and $s_1=s[x \mapsto t_1]$.
    In the latter case, inspect the stuck redex inside $s_2$.
    If the stuck redex comes from an occurrence of $x$, then $t_2$ has reached a stuck normal configuration; by $t_1 \lesssim t_2$, $t_1$ either diverges or reaches the same stuck configuration.
    If the redex is supplied by the surrounding syntax, use Lemma~\ref{lem:s1-value-transfer} on the already-evaluated value subterms in that redex.
    Either one of those subcomputations diverges, in which case the whole $t_1$-instance diverges, or the same bad outer form is obtained: non-function application, non-boolean conditional guard, non-list match scrutinee, non-pair match scrutinee, or an insufficient-resource tick.
    The tick case uses the fact that Lemma~\ref{lem:s1-normal-transfer} preserves the final resource $p'$.
    Thus the $t_1$-instance either diverges or gets stuck.
\end{proof}

\paragraph{Strengthened statement.}
The theorem below is the strengthened form of S1 used for induction.  Clause
(i) implies the S1 clause from Section~3.3 by viewing a one-hole context
$C[-]$ as a term context with a fresh variable $x$.  Clause (ii) is an
auxiliary statement needed for the contravariant premise of necessity
arrows.  In this proof, $F$ is used as a metavariable for an annotated
finitely verifiable type, that is, for an instance of the grammar category
$A_F$ from Section~3.3; the notation is kept because it makes the strengthened
statement easier to read than repeatedly writing $A_F$.

\begin{theorem}
\label{thm:s1-enhanced}
    (Safety Property, part 1, enhanced) Let $t[x \mapsto t_2]$ be closed, and $t_1\lesssim t_2$. For all types $A$ and finitely verifiable types $F$, and for every resource $p$:
    \begin{itemize}
        \item (i) If $(t[x \mapsto t_2],p) \in \TERMSP{A}$, then $(t[x \mapsto t_1],p)\in \TERMSP{A}$.
        \item (ii) If $(t[x \mapsto t_2],p) \notin \TERMS{F}$, then $(t[x \mapsto t_1],p)\notin \TERMS{F}$.
    \end{itemize}
\end{theorem}

\begin{proof}
    We prove (i) for all safety-fragment types $A$ and (ii) for all finitely verifiable types $F$, simultaneously by induction on the type syntax.
    The two operational facts used throughout are Lemma~\ref{lem:s1-divergence-transfer}, which moves divergence from the $t_2$-instance to the $t_1$-instance, and Lemmas~\ref{lem:s1-value-transfer} and~\ref{lem:s1-stuck-transfer}, which transfer terminating value and stuck computations while preserving the final resource.

    \begin{itemize}
        \item $A=\TP{\Tunit}{q}$.
        For (i), assume $(t[x \mapsto t_2],p)\in\TERMSP{A}$.
        If the $t_2$-instance diverges, the $t_1$-instance diverges by Lemma~\ref{lem:s1-divergence-transfer}, hence it is in $\TERMSP{A}$.
        Otherwise it evaluates to $(\tunit,p')$ with $p'\ge q$.
        By Lemma~\ref{lem:s1-value-transfer}, the $t_1$-instance either diverges or evaluates to the same $(\tunit,p')$; in both cases it is in $\TERMSP{A}$.

        For (ii), assume $(t[x \mapsto t_2],p)\notin\TERMS{A}$.
        There are four possible reasons: it diverges, it gets stuck, it evaluates to a non-unit value, or it evaluates to $(\tunit,p')$ with $p'<q$.
        Divergence transfers by Lemma~\ref{lem:s1-divergence-transfer}; stuckness transfers by Lemma~\ref{lem:s1-stuck-transfer}; and the two value cases transfer by Lemma~\ref{lem:s1-value-transfer} while preserving both the outer value constructor and the final resource.
        Therefore $(t[x \mapsto t_1],p)\notin\TERMS{A}$.

        \item $A=\TP{\Tbool}{q}$.
        The proof is identical to the unit case, with $\ttrue$ and $\tfalse$ as the only successful values.

        \item $A=\TP{\Tok}{q}$.
        For (i), assume $(t[x \mapsto t_2],p)\in\TERMSP{A}$.
        If the $t_2$-instance diverges, the $t_1$-instance diverges by Lemma~\ref{lem:s1-divergence-transfer}.
        Otherwise the $t_2$-instance evaluates to some value $(v_2,p')$ with $p'\ge q$.
        By Lemma~\ref{lem:s1-value-transfer}, the $t_1$-instance either diverges or evaluates to a value with the same final resource $p'$.
        Since $\Tok$ accepts every value, both cases are in $\TERMSP{A}$.

        For (ii), assume $(t[x \mapsto t_2],p)\notin\TERMS{A}$.
        The possible failures are divergence, stuckness, or evaluation to a value with final resource $p'<q$.
        Divergence and stuckness transfer by Lemmas~\ref{lem:s1-divergence-transfer} and~\ref{lem:s1-stuck-transfer}; the terminating case transfers by Lemma~\ref{lem:s1-value-transfer}, preserving the final resource.
        Thus $(t[x \mapsto t_1],p)\notin\TERMS{A}$.

        \item $A=\TP{(A_1,A_2)}{q}$.
        For (i), assume $(t[x \mapsto t_2],p)\in\TERMSP{A}$.
        The divergent case is handled by Lemma~\ref{lem:s1-divergence-transfer}.
        Otherwise the $t_2$-instance evaluates to $((v_{21},v_{22}),p')$ and there are $p_1,p_2$ such that
        $(v_{21},p_1)\in\VALS{A_1}$, $(v_{22},p_2)\in\VALS{A_2}$, and $p'\ge q+p_1+p_2$.
        By Lemma~\ref{lem:s1-value-transfer}, the $t_1$-instance either diverges or evaluates to $((v_{11},v_{12}),p')$ where each $v_{1i}$ is the transferred counterpart of $v_{2i}$.
        Since values are terms, $(v_{21},p_1)\in\TERMSP{A_1}$ and $(v_{22},p_2)\in\TERMSP{A_2}$.
        Applying the induction hypothesis (i) to the two component contexts gives $(v_{11},p_1)\in\TERMSP{A_1}$ and $(v_{12},p_2)\in\TERMSP{A_2}$.
        Because $v_{11}$ and $v_{12}$ are values, this means $(v_{11},p_1)\in\VALS{A_1}$ and $(v_{12},p_2)\in\VALS{A_2}$.
        Together with $p'\ge q+p_1+p_2$, this proves the pair value is in $\VALS{A}$, so the whole $t_1$-instance is in $\TERMSP{A}$.

        For (ii), this case applies only when $A_1$ and $A_2$ are finitely verifiable.
        Assume $(t[x \mapsto t_2],p)\notin\TERMS{A}$.
        Divergence and stuckness transfer as above, and a non-pair value or a pair with $p'<q$ remains a failure by Lemma~\ref{lem:s1-value-transfer}.
        The remaining case is that the $t_2$-instance evaluates to $((v_{21},v_{22}),p')$ with $p'\ge q$, but no split $p_1,p_2$ satisfying $p'\ge q+p_1+p_2$ makes both components members of $\VALS{A_1}$ and $\VALS{A_2}$.
        Lemma~\ref{lem:s1-value-transfer} gives either divergence, or a transferred pair $((v_{11},v_{12}),p')$.
        For any candidate split $p_1,p_2$ with $p'\ge q+p_1+p_2$, at least one of $(v_{21},p_1)\notin\VALS{A_1}$ or $(v_{22},p_2)\notin\VALS{A_2}$ holds.
        Since the components are values and $A_1,A_2$ are finitely verifiable, this is equivalently non-membership in $\TERMS{A_1}$ or $\TERMS{A_2}$.
        By the induction hypothesis (ii), the corresponding transferred component is also not in $\TERMS{A_i}$, and hence not in $\VALS{A_i}$.
        Thus no split witnesses membership of the transferred pair in $\VALS{A}$, so $(t[x \mapsto t_1],p)\notin\TERMS{A}$.

        \item $A=\TP{\Tlist{A_0}}{q}$.
        For (i), assume $(t[x \mapsto t_2],p)\in\TERMSP{A}$.
        The divergent case is immediate.
        Otherwise the $t_2$-instance evaluates to a list value $[v_{21},\ldots,v_{2\ell}]$ with final resource $p'$, and there are resources $p_1,\ldots,p_\ell$ such that each $(v_{2i},p_i)\in\VALS{A_0}$ and $p'\ge q+\sum_i p_i$.
        Lemma~\ref{lem:s1-value-transfer} gives either divergence or the transferred list $[v_{11},\ldots,v_{1\ell}]$ with the same final resource.
        Applying the induction hypothesis (i) to each element context yields $(v_{1i},p_i)\in\VALS{A_0}$.
        Therefore the transferred list is in $\VALS{A}$, and the $t_1$-instance is in $\TERMSP{A}$.

        For (ii), this case applies only when $A_0$ is finitely verifiable.
        Assume $(t[x \mapsto t_2],p)\notin\TERMS{A}$.
        Divergence, stuckness, a non-list value, and a list value with $p'<q$ are handled as in the previous cases.
        In the remaining case the $t_2$-instance evaluates to a list $[v_{21},\ldots,v_{2\ell}]$ with final resource $p'\ge q$, but for every choice of $p_1,\ldots,p_\ell$ with $p'\ge q+\sum_i p_i$, at least one element $(v_{2i},p_i)$ is not in $\VALS{A_0}$.
        The transferred computation either diverges or evaluates to the corresponding list $[v_{11},\ldots,v_{1\ell}]$ with the same final resource.
        For any candidate resource assignment, choose an index $i$ witnessing failure for the original list.
        Since $v_{2i}$ is a value and $A_0$ is finitely verifiable, $(v_{2i},p_i)\notin\VALS{A_0}$ implies $(v_{2i},p_i)\notin\TERMS{A_0}$.
        The induction hypothesis (ii) gives $(v_{1i},p_i)\notin\TERMS{A_0}$, hence $(v_{1i},p_i)\notin\VALS{A_0}$.
        Thus the transferred list cannot be in $\VALS{A}$ for any assignment of element resources.

        \item $A=\TP{A_1\rightarrow A_2}{q}$.
        Only (i) is required, since function types are not finitely verifiable.
        Assume $(t[x \mapsto t_2],p)\in\TERMSP{A}$.
        If it diverges, use Lemma~\ref{lem:s1-divergence-transfer}.
        Otherwise it evaluates to $(\tfix{f}{y}{s_2},p')$ with $p'\ge q$ and
        \[
            \forall v,r.\; (v,r)\in\VALS{A_1}
            \Rightarrow
            (s_2[f\mapsto \tfix{f}{y}{s_2}][y\mapsto v],r)\in\TERMSP{A_2}.
        \]
        Lemma~\ref{lem:s1-value-transfer} gives either divergence or a value
        $\tfix{f}{y}{s_1}$ with the same final resource and a body template $s$ such that
        $s_2=s[x\mapsto t_2]$ and $s_1=s[x\mapsto t_1]$.
        Let $(v,r)\in\VALS{A_1}$ be arbitrary.
        The body of the original abstraction on this input is an instance of the context
        $s[f\mapsto \tfix{f}{y}{s}][y\mapsto v]$ with $t_2$ substituted for $x$.
        By the displayed premise it is in $\TERMSP{A_2}$, so by the induction hypothesis (i) the corresponding body with $t_1$ substituted for $x$ is also in $\TERMSP{A_2}$.
        Hence $(\tfix{f}{y}{s_1},p')\in\VALS{A}$, and the $t_1$-instance is in $\TERMSP{A}$.

        \item $A=\TP{A_1\onlyto F_0}{q}$, where $F_0$ is finitely verifiable by the safety-fragment grammar.
        Again only (i) is required.
        Assume $(t[x \mapsto t_2],p)\in\TERMSP{A}$.
        If it diverges, use Lemma~\ref{lem:s1-divergence-transfer}.
        Otherwise it evaluates to $(\tfix{f}{y}{s_2},p')$ with $p'\ge q$ and
        \[
            \forall v,r.\;
            (s_2[f\mapsto \tfix{f}{y}{s_2}][y\mapsto v],r)\in\TERMS{F_0}
            \Rightarrow
            (v,r)\in\VALS{A_1}.
        \]
        By Lemma~\ref{lem:s1-value-transfer}, the $t_1$-instance either diverges or evaluates to the transferred abstraction
        $(\tfix{f}{y}{s_1},p')$ with $s_2=s[x\mapsto t_2]$ and $s_1=s[x\mapsto t_1]$.
        To show this abstraction is in $\VALS{A}$, take arbitrary $v,r$ and assume
        \[
            (s_1[f\mapsto \tfix{f}{y}{s_1}][y\mapsto v],r)\in\TERMS{F_0}.
        \]
        By the contrapositive of the induction hypothesis (ii) for the finitely verifiable type $F_0$, the corresponding body with $t_2$ substituted for $x$ is also in $\TERMS{F_0}$.
        The semantic premise of the original necessity arrow then gives $(v,r)\in\VALS{A_1}$.
        Therefore $(\tfix{f}{y}{s_1},p')\in\VALS{A}$, and the $t_1$-instance is in $\TERMSP{A}$.
    \end{itemize}
\end{proof}

\subsection{Benchmarks}
\label{subsec:benchmarks-appendix}

This subsection lists the benchmark programs used by the implementation
experiments. The terms are written in the surface notation of the paper after
erasing implementation annotations such as \texttt{as}, \texttt{notas}, and
auxiliary terms used only to guide proof search. When the implementation uses
separate upper-bound and lower-bound versions that erase to the same term, we
list the term only once.

\newcommand{\bname}[1]{\underline{\texttt{#1}}}
\newcommand{\bkw}[1]{\texttt{#1}}
\newcommand{\benchdisplayspacing}{%
    \setlength{\abovedisplayskip}{0.25ex}%
    \setlength{\abovedisplayshortskip}{0.25ex}%
    \setlength{\belowdisplayskip}{0pt}%
    \setlength{\belowdisplayshortskip}{0pt}%
}
\renewcommand{\arraystretch}{0.92}

\paragraph{1-1 If.}
\par\nobreak\benchdisplayspacing
\begin{align*}
\begin{array}{@{}l@{}}
\bname{c\_if} := \tif{\ttrue}{\ttick{2}}{\ttick{3}}.
\end{array}
\end{align*}

\paragraph{1-2 Matchlist.}
\par\nobreak\benchdisplayspacing
\begin{align*}
\begin{array}{@{}l@{}}
\bname{c\_match} :=
    \bkw{match}\ [\tunit,\tunit]\ \{
    \bkw{nil} \hookrightarrow \ttick{1}
    \ |\ \bkw{cons}(h,xs) \hookrightarrow \ttick{4}\}.
\end{array}
\end{align*}

\paragraph{1-3 If-Fun.}
\par\nobreak\benchdisplayspacing
\begin{align*}
\begin{array}{@{}l@{}}
\bname{f2} := \tfix{f}{x}{\ttick{2}},\\
\bname{g4} := \tfix{g}{x}{\ttick{4}},\\
\bname{c\_choose} := (\tif{\ttrue}{\bname{f2}}{\bname{g4}})\ \tunit.
\end{array}
\end{align*}

\paragraph{1-4 Higher-Order.}
\par\nobreak\benchdisplayspacing
\begin{align*}
\begin{array}{@{}l@{}}
\bname{c\_higher} := \tfix{f}{x}{\tlet{u}{\ttick{2}}{x\ \tunit}}.
\end{array}
\end{align*}

\paragraph{1-5 Linear-Bool.}
\par\nobreak\benchdisplayspacing
\begin{align*}
\begin{array}{@{}l@{}}
\bname{lin\_bool} := \bkw{fix}\ f(x).\\
\quad \bkw{match}\ x\ \{\\
\quad\quad \bkw{nil} \hookrightarrow \ttick{5}
    \ |\ \bkw{cons}(x,xs) \hookrightarrow
    \tif{x}{\ttick{2}}{\ttick{3}};\ f\ xs\\
\quad \}.
\end{array}
\end{align*}

\paragraph{1-6 Early-Return.}
\par\nobreak\benchdisplayspacing
\begin{align*}
\begin{array}{@{}l@{}}
\bname{early\_return} := \bkw{fix}\ f(xs).\\
\quad \bkw{match}\ xs\ \{\bkw{nil} \hookrightarrow \ttick{1}
    \ |\ \bkw{cons}(h,t) \hookrightarrow
    \tif{h}{\ttick{1};\ f\ t}{\ttick{1}}\}.
\end{array}
\end{align*}

\paragraph{1-7 Infinite-Consumer.}
\par\nobreak\benchdisplayspacing
\begin{align*}
\begin{array}{@{}l@{}}
\bname{infinite\_consumer} := \tfix{f}{x}{\ttick{1};\ f\ x}.
\end{array}
\end{align*}

\paragraph{1-8 Squared.}
\par\nobreak\benchdisplayspacing
\begin{align*}
\begin{array}{@{}l@{}}
\bname{inner\_loop} := \bkw{fix}\ f(xs).\\
\quad \bkw{match}\ xs\ \{\bkw{nil} \hookrightarrow \tunit
    \ |\ \bkw{cons}(h,t) \hookrightarrow \ttick{1};\ f\ t\},\\[0.2ex]
\bname{squared} := \bkw{fix}\ f(rows).\\
\quad \bkw{match}\ rows\ \{\bkw{nil} \hookrightarrow \tunit
    \ |\ \bkw{cons}(row,rest) \hookrightarrow
    \bname{inner\_loop}\ row;\ f\ rest\}.
\end{array}
\end{align*}

\paragraph{1-9 Stuck.}
\par\nobreak\benchdisplayspacing
\begin{align*}
\begin{array}{@{}l@{}}
\bname{stuck\_term} := \tif{\tunit}{\tunit}{\tunit},\\[0.2ex]
\bname{stuck\_scan} := \bkw{fix}\ f(xs).\\
\quad \bkw{match}\ xs\ \{\bkw{nil} \hookrightarrow \tunit
    \ |\ \bkw{cons}(h,t) \hookrightarrow
    \tif{h}{\ttick{2};\ f\ t}{\ttick{3};\ \bname{stuck\_term}}\}.
\end{array}
\end{align*}

\paragraph{1-10 Divergence.}
\par\nobreak\benchdisplayspacing
\begin{align*}
\begin{array}{@{}l@{}}
\bname{diverge} := \tfix{d}{x}{d\ x},\\[0.2ex]
\bname{diverge\_scan} := \bkw{fix}\ f(xs).\\
\quad \bkw{match}\ xs\ \{\bkw{nil} \hookrightarrow \tunit
    \ |\ \bkw{cons}(h,t) \hookrightarrow
    \tif{h}{\ttick{2};\ f\ t}{\bname{diverge}\ \tunit}\}.
\end{array}
\end{align*}

\paragraph{2-1 Strcmp.}
\par\nobreak\benchdisplayspacing
\begin{align*}
\begin{array}{@{}l@{}}
\bname{eq} := \bkw{fix}\ f(p).\ \bkw{match}\ p\ \{(a,b) \hookrightarrow
    \ttick{1};\ \tif{a}{b}{\tif{b}{\tfalse}{\ttrue}}\},\\[0.2ex]
\bname{strcmp} := \bkw{fix}\ f(p).\\
\quad \bkw{match}\ p\ \{(s_1,s_2) \hookrightarrow\\
\quad\quad \bkw{match}\ s_1\ \{\\
\quad\quad\quad \bkw{nil} \hookrightarrow \ttrue
    \ |\ \bkw{cons}(h_1,t_1) \hookrightarrow\\
\quad\quad\quad\quad \bkw{match}\ s_2\ \{\\
\quad\quad\quad\quad\quad \bkw{nil} \hookrightarrow \tfalse
    \ |\ \bkw{cons}(h_2,t_2) \hookrightarrow\\
\quad\quad\quad\quad\quad\quad
    \tif{\bname{eq}\ (h_1,h_2)}{f\ (t_1,t_2)}{\tfalse}\\
\quad\quad\quad\quad \}\\
\quad\quad \}\\
\quad \}.
\end{array}
\end{align*}

\paragraph{2-2 Safe-Strcmp.}
\par\nobreak\benchdisplayspacing
\begin{align*}
\begin{array}{@{}l@{}}
\bname{eq} := \bkw{fix}\ f(p).\ \bkw{match}\ p\ \{(a,b) \hookrightarrow
    \ttick{1};\ \tif{a}{b}{\tif{b}{\tfalse}{\ttrue}}\},\\[0.2ex]
\bname{strcmp\_safe0} := \bkw{fix}\ f(p).\\
\quad \bkw{match}\ p\ \{(s_1,r_1) \hookrightarrow\\
\quad\quad \bkw{match}\ r_1\ \{(s_2,r_2) \hookrightarrow\\
\quad\quad\quad \bkw{match}\ r_2\ \{(l,flag) \hookrightarrow\\
\quad\quad\quad\quad \bkw{match}\ l\ \{\\
\quad\quad\quad\quad\quad \bkw{nil} \hookrightarrow\\
\quad\quad\quad\quad\quad\quad \bkw{match}\ s_1\ \{\\
\quad\quad\quad\quad\quad\quad\quad \bkw{nil} \hookrightarrow
    \bkw{match}\ s_2\ \{\bkw{nil} \hookrightarrow flag\\
\quad\quad\quad\quad\quad\quad\quad\quad
    |\ \bkw{cons}(s2h,s2t) \hookrightarrow \tfalse\}\\
\quad\quad\quad\quad\quad\quad\quad
    |\ \bkw{cons}(s1h,s1t) \hookrightarrow \tfalse\\
\quad\quad\quad\quad\quad\quad \}\\
\quad\quad\quad\quad\quad |\ \bkw{cons}(lh,lt) \hookrightarrow\\
\quad\quad\quad\quad\quad\quad \bkw{match}\ s_1\ \{\\
\quad\quad\quad\quad\quad\quad\quad \bkw{nil} \hookrightarrow
    \ttick{1};\ f\ (s_1,s_2,lt,\tfalse)\\
\quad\quad\quad\quad\quad\quad\quad |\ \bkw{cons}(s1h,s1t) \hookrightarrow\\
\quad\quad\quad\quad\quad\quad\quad\quad \bkw{match}\ s_2\ \{\\
\quad\quad\quad\quad\quad\quad\quad\quad\quad \bkw{nil} \hookrightarrow
    \ttick{1};\ f\ (s_1,s_2,lt,\tfalse)\\
\quad\quad\quad\quad\quad\quad\quad\quad\quad
    |\ \bkw{cons}(s2h,s2t) \hookrightarrow\\
\quad\quad\quad\quad\quad\quad\quad\quad\quad\quad
    \tif{\bname{eq}\ (s1h,s2h)}
        {f\ (s1t,s2t,lt,flag)}
        {f\ (s1t,s2t,lt,\tfalse)}\\
\quad\quad\quad\quad\quad\quad\quad\quad \}\\
\quad\quad\quad\quad\quad\quad \}\\
\quad\quad\quad\quad \}\\
\quad\quad\quad \}\\
\quad\quad \}\\
\quad \},\\[0.2ex]
\bname{strcmp\_safe} := \bkw{fix}\ f(p).\\
\quad \bkw{match}\ p\ \{(s_1,r_1) \hookrightarrow\\
\quad\quad \bkw{match}\ r_1\ \{(s_2,l) \hookrightarrow\\
\quad\quad\quad \bname{strcmp\_safe0}\ (s_1,s_2,l,\ttrue)\\
\quad\quad \}\\
\quad \}.
\end{array}
\end{align*}

\paragraph{2-3 Zip.}
\par\nobreak\benchdisplayspacing
\begin{align*}
\begin{array}{@{}l@{}}
\bname{zip} := \bkw{fix}\ f(p).\\
\quad \bkw{match}\ p\ \{(s_1,r_1) \hookrightarrow\\
\quad\quad \bkw{match}\ r_1\ \{(s_2,default) \hookrightarrow\\
\quad\quad\quad \bkw{match}\ s_1\ \{\\
\quad\quad\quad\quad \bkw{nil} \hookrightarrow\\
\quad\quad\quad\quad\quad \bkw{match}\ s_2\ \{\\
\quad\quad\quad\quad\quad\quad \bkw{nil} \hookrightarrow []
    \ |\ \bkw{cons}(h_2,t_2) \hookrightarrow\\
\quad\quad\quad\quad\quad\quad\quad
    (default,h_2) :: f\ ([],t_2,default)\\
\quad\quad\quad\quad\quad \}\\
\quad\quad\quad\quad |\ \bkw{cons}(h_1,t_1) \hookrightarrow\\
\quad\quad\quad\quad\quad \bkw{match}\ s_2\ \{\\
\quad\quad\quad\quad\quad\quad \bkw{nil} \hookrightarrow
    (h_1,default) :: f\ (t_1,[],default)\\
\quad\quad\quad\quad\quad\quad |\ \bkw{cons}(h_2,t_2) \hookrightarrow
    (h_1,h_2) :: f\ (t_1,t_2,default)\\
\quad\quad\quad\quad\quad \}\\
\quad\quad\quad \}\\
\quad\quad \}\\
\quad \}.
\end{array}
\end{align*}

\paragraph{2-4 Add.}
\par\nobreak\benchdisplayspacing
\begin{align*}
\begin{array}{@{}l@{}}
\bname{add} := \bkw{fix}\ f(p).\\
\quad \bkw{match}\ p\ \{(xs,ys) \hookrightarrow\\
\quad\quad \bkw{match}\ xs\ \{\\
\quad\quad\quad \bkw{nil} \hookrightarrow ys
    \ |\ \bkw{cons}(h,t) \hookrightarrow h :: f\ (t,ys)\\
\quad\quad \}\\
\quad \}.
\end{array}
\end{align*}

\paragraph{2-5 Compare.}
\par\nobreak\benchdisplayspacing
\begin{align*}
\begin{array}{@{}l@{}}
\bname{compare} := \bkw{fix}\ f(p).\\
\quad \bkw{match}\ p\ \{(a,b) \hookrightarrow\\
\quad\quad \bkw{match}\ a\ \{\\
\quad\quad\quad \bkw{nil} \hookrightarrow
    \bkw{match}\ b\ \{\bkw{nil} \hookrightarrow \ttrue
        \ |\ \bkw{cons}(bh,bt) \hookrightarrow \tfalse\}\\
\quad\quad\quad |\ \bkw{cons}(ah,at) \hookrightarrow
    \bkw{match}\ b\ \{\bkw{nil} \hookrightarrow \ttrue
        \ |\ \bkw{cons}(bh,bt) \hookrightarrow f\ (at,bt)\}\\
\quad\quad \}\\
\quad \}.
\end{array}
\end{align*}

\paragraph{2-6 Multi-Type.}
\par\nobreak\benchdisplayspacing
\begin{align*}
\begin{array}{@{}l@{}}
\bname{multi\_type} := \bkw{fix}\ f(p).\\
\quad \bkw{match}\ p\ \{(a,b) \hookrightarrow\\
\quad\quad \bkw{match}\ a\ \{\\
\quad\quad\quad \bkw{nil} \hookrightarrow \tunit\\
\quad\quad\quad |\ \bkw{cons}(ah,at) \hookrightarrow
    \bkw{match}\ b\ \{\bkw{nil} \hookrightarrow \ttrue
        \ |\ \bkw{cons}(bh,bt) \hookrightarrow f\ (at,bt)\}\\
\quad\quad \}\\
\quad \}.
\end{array}
\end{align*}

\paragraph{2-7 Max.}
\par\nobreak\benchdisplayspacing
\begin{align*}
\begin{array}{@{}l@{}}
\bname{ge0} := \bkw{fix}\ f(p).\\
\quad \bkw{match}\ p\ \{(a,b) \hookrightarrow\\
\quad\quad \bkw{match}\ a\ \{\\
\quad\quad\quad \bkw{nil} \hookrightarrow
    \bkw{match}\ b\ \{\bkw{nil} \hookrightarrow \ttrue
        \ |\ \bkw{cons}(bh,bt) \hookrightarrow \tfalse\}\\
\quad\quad\quad |\ \bkw{cons}(ah,at) \hookrightarrow
    \bkw{match}\ b\ \{\bkw{nil} \hookrightarrow \ttrue
        \ |\ \bkw{cons}(bh,bt) \hookrightarrow f\ (at,bt)\}\\
\quad\quad \}\\
\quad \},\\[0.2ex]
\bname{ge1} := \tfix{f}{p}{\ttick{1};\ \bname{ge0}\ p},\\[0.2ex]
\bname{max2} := \bkw{fix}\ f(p).\\
\quad \bkw{match}\ p\ \{(a,b) \hookrightarrow
    \tif{\bname{ge1}\ (a,b)}{a}{b}\},\\[0.2ex]
\bname{max\_list} := \bkw{fix}\ f(xs).\\
\quad \bkw{match}\ xs\ \{\\
\quad\quad \bkw{nil} \hookrightarrow f\ xs\\
\quad\quad |\ \bkw{cons}(h,t) \hookrightarrow
    \bkw{match}\ t\ \{\bkw{nil} \hookrightarrow h\\
\quad\quad\quad\quad
    |\ \bkw{cons}(th,rest) \hookrightarrow \bname{max2}\ (h,f\ t)\}\\
\quad \}.
\end{array}
\end{align*}

\paragraph{2-8 Max-Potential.}
\par\nobreak\benchdisplayspacing
\begin{align*}
\begin{array}{@{}l@{}}
\bname{compare} := \bkw{fix}\ f(p).\\
\quad \bkw{match}\ p\ \{(a,b) \hookrightarrow\\
\quad\quad \bkw{match}\ a\ \{\\
\quad\quad\quad \bkw{nil} \hookrightarrow
    \bkw{match}\ b\ \{\bkw{nil} \hookrightarrow \ttrue
        \ |\ \bkw{cons}(bh,bt) \hookrightarrow \tfalse\}\\
\quad\quad\quad |\ \bkw{cons}(ah,at) \hookrightarrow
    \bkw{match}\ b\ \{\bkw{nil} \hookrightarrow \ttrue
        \ |\ \bkw{cons}(bh,bt) \hookrightarrow f\ (at,bt)\}\\
\quad\quad \}\\
\quad \},\\[0.2ex]
\bname{strip} := \bkw{fix}\ f(xs).\\
\quad \bkw{match}\ xs\ \{\bkw{nil} \hookrightarrow []
    \ |\ \bkw{cons}(h,t) \hookrightarrow \tunit :: f\ t\},\\[0.2ex]
\bname{max\_potential} := \bkw{fix}\ f(xs).\\
\quad \bkw{match}\ xs\ \{\\
\quad\quad \bkw{nil} \hookrightarrow f\ xs\\
\quad\quad |\ \bkw{cons}(h,t) \hookrightarrow
    \bkw{match}\ t\ \{\bkw{nil} \hookrightarrow \bname{strip}\ h\\
\quad\quad\quad\quad |\ \bkw{cons}(th,rest) \hookrightarrow\\
\quad\quad\quad\quad\quad
    \tif{\bname{compare}\ (\bname{strip}\ h,f\ t)}
        {\bname{strip}\ h}
        {f\ t}\}\\
\quad \}.
\end{array}
\end{align*}

\paragraph{2-9 Stack.}
\par\nobreak\benchdisplayspacing
\begin{align*}
\begin{array}{@{}l@{}}
\bname{popn} := \bkw{fix}\ f(p).\\
\quad \bkw{match}\ p\ \{(stack,n) \hookrightarrow\\
\quad\quad \bkw{match}\ n\ \{\\
\quad\quad\quad \bkw{nil} \hookrightarrow stack\\
\quad\quad\quad |\ \bkw{cons}(nh,nt) \hookrightarrow
    \bkw{match}\ stack\ \{\bkw{nil} \hookrightarrow f\ (stack,n)\\
\quad\quad\quad\quad
    |\ \bkw{cons}(sh,st) \hookrightarrow \ttick{1};\ f\ (st,nt)\}\\
\quad\quad \}\\
\quad \},\\[0.2ex]
\bname{consume\_op} := \tfix{f}{x}{\tunit},\\[0.2ex]
\bname{stack\_op0} := \bkw{fix}\ f(p).\\
\quad \bkw{match}\ p\ \{(ops,stack) \hookrightarrow\\
\quad\quad \bkw{match}\ ops\ \{\\
\quad\quad\quad \bkw{nil} \hookrightarrow \tunit\\
\quad\quad\quad |\ \bkw{cons}(n,rest) \hookrightarrow\\
\quad\quad\quad\quad \bkw{match}\ n\ \{\\
\quad\quad\quad\quad\quad \bkw{nil} \hookrightarrow
    \ttick{1};\ f\ (rest,\tunit :: stack)\\
\quad\quad\quad\quad\quad |\ \bkw{cons}(nh,nt) \hookrightarrow\\
\quad\quad\quad\quad\quad\quad
    \tlet{consumed}{\bname{consume\_op}\ n}
        {f\ (rest,\bname{popn}\ (stack,n))}\\
\quad\quad\quad\quad \}\\
\quad\quad \}\\
\quad \},\\[0.2ex]
\bname{stack\_op} := \tfix{f}{ops}{\bname{stack\_op0}\ (ops,[])}.
\end{array}
\end{align*}

\clearpage
\subsection{Two-Sided Linear Type System}
\label{subsec:linear-study}

The main system combines two-sided typing with affine, numeric potential. Our technical report develops the parallel exploration summarized in \sec{linear}: what happens if the same two-sided viewpoint is applied to a \emph{linear} type system, where values themselves are non-replicable resources? The goal is not to replace the main AARA-based system, but to use the linear setting as a compact comparison point. Exact resource accounting makes closures more delicate, because closures may capture linear resources. Expressing the necessity direction for such closures leads naturally to \emph{dual} typing formulas. The comparison helps explain why the main system of Sections~\ref{sec:language} and~\ref{sec:semantics} uses affine contexts and ordered numeric potential.

\subsubsection{Language and Type System}
\label{sec:linear-language}

The language is a small call-by-value calculus with unit, Booleans, functions, and pairs. There is no separate resource token; in a linear language, values themselves are the resources. Figure~\ref{fig:linear-terms} gives the grammar.

\begin{figure}[!htb]
\centering
\begin{tcolorbox}[linearnontyping,title={Terms, Values, and Types}]
\begin{align*}
t ::= {}& x
      \mid \tunit \mid \ltletunit{t}{t}
      \mid \ttrue \mid \tfalse \mid \tif{t}{t}{t} \\
    & \mid \ltlambda{x}{t} \mid t\ t
      \mid (t,t) \mid \ltletpair{t}{x}{y}{t} \\
v ::= {}& \tunit \mid \ttrue \mid \tfalse \mid \ltlambda{x}{t} \mid (v,v) \\
T, A, B, C ::= {}& \Tunit \mid \Tbool \mid A \to B \mid A \onlyto B \mid (A,B)
\end{align*}
\end{tcolorbox}
\setlength{\abovecaptionskip}{4pt}
\setlength{\belowcaptionskip}{-4pt}
\caption{Terms, values, and types of the linear system.}
\label{fig:linear-terms}
\end{figure}

\paragraph{Evaluation.}
Evaluation is a small-step relation on configurations $(t,\theta)$, where $\theta$ is an ordinary finite map from variables to values. The map is linear: each binding has a distinct name, and variable lookup consumes that binding. Thus a closed term $t$ evaluates successfully to $v$ iff $(t,\emptyset)\to^*(v,\emptyset)$. A normal configuration is stuck if the term is not a value or if unused bindings remain. Our technical report gives the full reduction rules.

\paragraph{Formula conventions.}
Primitive formulas are generated by $F ::= t:T \mid \overline{t:T}$, and contexts $\Gamma$ and $\Delta$ are comma-separated lists of primitive formulas, considered modulo exchange. The metavariable $\Phi$ ranges over a single formula. As in \sec{typesystem}, semicolon is right-side syntactic sugar for conjunction: $t_1:A_1; t_2:A_2$ abbreviates $(t_1,t_2):(A_1,A_2)$. Semicolon binds more tightly than comma, and nested semicolons are read as nested pair formulas, so a single formula $\Phi$ may encode a right-side conjunction.

In the abstraction rules $\text{T-Abs}$ and $\text{T-AbsN}$, the captured context written as $\Gamma$ is implicitly required to be \emph{capturable}: every defined total interpretation of it, as defined in our technical report, must sum to an ordinary finite map $\theta$, rather than to a genuinely signed multimap. This is a side condition on the resources stored inside closures. General contexts elsewhere may still contain dual formulas, and dual formulas may also appear in the premises of abstraction rules as proof bookkeeping.

\paragraph{Typing rules.}
The system uses two-sided sequents $\Gamma \vdash \Delta$. The left side is multiplicative conjunction, the right side is multiplicative disjunction, and primitive formulas are ordinary typing formulas or their duals. Our technical report contains the full rule set; Figure~\ref{fig:linear-rules} shows the rules most relevant to the design discussion.

\begin{figure}[!t]
\centering
\begin{tcolorbox}[lineartyping,title={Typing Rules (Representative)}]
\begin{mathpar}
\text{(T-Id)} \inferrule{ }{t:A \vdash t : A}
\and
\text{(T-Ann)} \inferrule{\Gamma \vdash \Delta}{\Gamma, \overline{t:T}, t:T \vdash \Delta}
\and
\text{(T-Dual)} \inferrule{\overline{\Gamma} \vdash \overline{\Delta}}{\Gamma \vdash \Delta}
\end{mathpar}

\par\nointerlineskip\vspace{3pt}
\noindent\rule{\linewidth}{0.4pt}
\par\nointerlineskip\vspace{3pt}

\begin{mathpar}
\text{(T-Abs)}
\inferrule{\Gamma, x:A \vdash t : B,\Delta}
          {\Gamma \vdash \ltlambda{x}{t} : A \to B,\Delta}
\and
\text{(T-AbsN)}
\inferrule{\overline{\Gamma}, t:B \vdash x:A, \overline \Delta}
          {\Gamma \vdash \ltlambda{x}{t} : A \onlyto B,\Delta}
\end{mathpar}

\par\nointerlineskip\vspace{3pt}
\noindent\rule{\linewidth}{0.4pt}
\par\nointerlineskip\vspace{3pt}

\begin{mathpar}
\text{(T-Dis)} \inferrule{\Gamma\vdash t : B,\Delta \quad A \parallel B}
                         {\Gamma, t : A \vdash \Delta}
\and
\text{(T-BoolL)} \inferrule{\Gamma \vdash \Delta \quad b \in \{\ttrue, \tfalse\}}
                          {\Gamma, b : \Tbool \vdash \Delta}
\and
\text{(T-IfL)}
\inferrule{\Gamma_2,t_2:A \vdash \Phi \quad
           \Gamma_2,t_3:A \vdash \Phi \quad
           \Gamma_1,t_1:\Tbool,\Phi \vdash \Delta}
          {\Gamma_1,\Gamma_2,\tif{t_1}{t_2}{t_3}:A \vdash \Delta}
\and
\text{(T-AppL)}
\inferrule{\Gamma_1 \vdash t_1 : A \onlyto B; \Phi \quad
           \Gamma_2, t_2 : A,\Phi \vdash \Delta}
          {\Gamma_1, \Gamma_2, t_1\ t_2 : B \vdash \Delta}
\and
\text{(T-PairL)}
\inferrule{\Gamma, t_1:A, t_2:B \vdash \Delta}
          {\Gamma, (t_1, t_2) : (A, B) \vdash \Delta}
\and
\text{(T-LetPairL)}
\inferrule{\Gamma_2, t_2 : C \vdash (x,y) : (A, B) ; \Phi \quad
           \Gamma_1, t_1 : (A, B),\Phi \vdash \Delta}
          {\Gamma_1, \Gamma_2, \ltletpair{t_1}{x}{y}{t_2} : C\vdash \Delta}
\end{mathpar}
\end{tcolorbox}
\setlength{\abovecaptionskip}{4pt}
\setlength{\belowcaptionskip}{-4pt}
\caption{Representative rules for the linear type system.}
\label{fig:linear-rules}
\end{figure}

\paragraph{Dual formulas.}
The formula $\overline{t:T}$ is the dual of $t:T$, and the dual of a dual is the original formula. For a context, $\overline{\Gamma}$ dualizes every formula in $\Gamma$. Dualization negates resource weights, representing ``$-1$ copies'' of a formula rather than logical negation. The rule $\text{T-Dual}$ applies it to both contexts without exchanging sides: a derivation of $\Gamma\vdash\Delta$ gives $\overline{\Gamma}\vdash\overline{\Delta}$. For example, ordinary identity yields $\overline{x:A}\vdash\overline{x:A}$ by $\text{T-Dual}$, so no separate identity rule for dual formulas is needed.

Dual formulas are needed most visibly in $\text{T-AbsN}$. For an ordinary function, $\text{T-Abs}$ expresses the usual resource equation: the captured environment plus the argument resources produce the result resources. For a necessity arrow, the logical direction is reversed: a result of type $B$ should imply that the argument has type $A$. If we used the naive equation ``environment plus result equals argument'', captured resources would be counted on the wrong side. The dual context instead represents the equation as ``negative environment plus result equals argument'', equivalently ``environment plus argument equals result''.

For example, let
\[
\underline{\texttt{f}} = \ltlambda{x}{\ltlambda{y}{(x,y)}}.
\]
The following derivation shows that
\[
\vdash \underline{\texttt{f}} : \Tunit \to (\Tbool \onlyto (\Tunit,\Tbool)).
\]
\begin{mathpar}\small
\inferrule{
  \inferrule{
    \inferrule{
      \inferrule{
        y:\Tbool \vdash y:\Tbool
      }{\overline{x:\Tunit},x:\Tunit,y:\Tbool \vdash y:\Tbool}(\text{T-Ann})
    }{\overline{x:\Tunit},(x,y):(\Tunit,\Tbool) \vdash y:\Tbool}(\text{T-PairL})
  }{x:\Tunit \vdash \ltlambda{y}{(x,y)}:\Tbool \onlyto (\Tunit,\Tbool)}(\text{T-AbsN})
}{
  \vdash \underline{\texttt{f}}:\Tunit \to (\Tbool \onlyto (\Tunit,\Tbool))
}(\text{T-Abs})
\end{mathpar}

This example uses $\text{T-AbsN}$ with a non-empty captured context: the inner closure keeps the resource $x$. This is precisely where dual formulas matter. In the affine system from Sections~\ref{sec:language} and~\ref{sec:semantics}, functions are duplicable, so the corresponding abstraction rule can require an empty resource context. In the linear system, function values cannot be freely copied; a closure may carry resources that will be used when it is called once. The capturability side condition says that these captured resources must be ordinary runtime resources, while the dual formulas in the premise provide the bookkeeping needed for the necessity-arrow direction.

\subsubsection{Semantics and Soundness}
\label{sec:linear-semantics}

Figure~\ref{fig:linear-semantics} defines the type and term semantics used for the linear system. Types are interpreted over the same ordinary finite maps used by evaluation. We write $\theta_1 \uplus \theta_2$ for partial union of ordinary maps; all displayed uses of $\uplus$ carry the implicit requirement that the domains are disjoint.

\begin{figure}[!t]
\centering
\begin{tcolorbox}[lineartyping,title=Linear Semantics,fontupper=\small]
\begin{align*}
    \VALS{\Tunit} = & \ & \{(\emptyset, \tunit)\} \\
    \VALS{\Tbool} = & \ & \{(\emptyset, \ttrue),(\emptyset, \tfalse)\} \\
    \VALS{A \to B} = & \ & \{(\theta_1, \ltlambda{x}{t}): \forall v, \theta_2.\ (\theta_2, v)\in\VALS{A} \Rightarrow (\theta_1 \uplus \theta_2, \ltlambda{x}{t}\ v) \in \TERMS{B}\} \\
    \VALS{A \onlyto B} = & \ & \{(\theta_1, \ltlambda{x}{t}): \forall v, \theta.\ (\theta, \ltlambda{x}{t}\ v) \in \TERMS{B} \Rightarrow{} \\
    & & \qquad\qquad\qquad\qquad\qquad\exists \theta_2.\ \theta = \theta_1 \uplus \theta_2 \wedge (\theta_2, v)\in\VALS{A}\} \\
    \VALS{(A, B)} = & \ & \{(\theta_1 \uplus \theta_2, (v_1, v_2)): (\theta_1,v_1) \in \VALS{A} \wedge (\theta_2,v_2) \in \VALS{B}\}
\end{align*}

\par\nointerlineskip
\rule{\linewidth}{0.4pt}
\par\nointerlineskip

\begin{align*}
    \TERMS{A} = & \ & \{(\theta, t): \forall \theta_m.\ \exists \theta', v.\ (t,\theta \uplus \theta_m) \to^* (v,\theta' \uplus \theta_m) \wedge (\theta' , v) \in \VALS{A}\}\\
    \TERMSP{A} = & \ & \{(\theta, t): \forall \theta_m,\theta',v.\ (t,\theta \uplus \theta_m) \to^* (v,\theta' \uplus \theta_m) \Rightarrow (\theta' , v) \in \VALS{A}\}
\end{align*}
\end{tcolorbox}
\setlength{\abovecaptionskip}{4pt}
\setlength{\belowcaptionskip}{-4pt}
\caption{Semantics of the linear type system.}
\label{fig:linear-semantics}
\end{figure}

The definition of $\TERMS{A}$ is total-correctness: under any compatible surrounding map $\theta_m$, the term must evaluate to a value of type $A$ while leaving the surrounding map untouched. By contrast, $\TERMSP{A}$ is partial-correctness: if the term terminates under a compatible surrounding map, every terminating result must have type $A$. The clause for $A\onlyto B$ reads the arrow backward. Whenever an application can produce a result of type $B$, the successful resource map must contain the resources captured by the closure, and the remaining resources must type the argument at $A$.

\paragraph{Semantics of formulas and judgments.}
Typing judgments use signed multimaps rather than ordinary runtime maps. A signed multimap $\mu$ maps variables to values with integer weights; weight $0$ is identified with absence, and each variable may be associated with at most one value. For an ordinary map $\theta$, let $\langle\theta\rangle$ contain the same bindings, each with weight $1$, and let $-\mu$ negate every weight in $\mu$.
\begin{align*}
    \mu \SATS t : A \quad &\text{iff}\quad \exists \theta.\ \mu=\langle\theta\rangle \wedge (\theta,t)\in\TERMS{A}, \\
    \mu \SATS \overline{t : A} \quad &\text{iff}\quad \exists \theta.\ \mu=-\langle\theta\rangle \wedge (\theta,t)\in\TERMS{A}, \\
    \mu \SATSP t : A \quad &\text{iff}\quad \exists \theta.\ \mu=\langle\theta\rangle \wedge (\theta,t)\in\TERMSP{A}, \\
    \mu \SATSP \overline{t : A} \quad &\text{iff}\quad \exists \theta.\ \mu=-\langle\theta\rangle \wedge (\theta,t)\in\TERMSP{A}.
\end{align*}

$\vec{\mu} \SATSL \Gamma$ iff:
\begin{itemize}
    \item $\Gamma$ contains $n$ formulas, written as $F_1,\dots,F_n$;
    \item $\vec{\mu}=(\mu_1,\dots,\mu_n)$ is a vector of signed multimaps with the same length and order as $\Gamma$;
    \item $\mu_i \SATS F_i$ for every $i$.
\end{itemize}

For a vector interpreting a context $\Gamma$, we write $\bigoplus_\Gamma\vec{\mu}$ for the context-labelled partial sum. Its underlying signed-multimap sum is defined when no variable is associated with two different values among the components of $\vec{\mu}$. For each variable-value binding, the total positive multiplicity is at most $1$ and the total negative multiplicity is at most $1$. Thus a positive binding and a matching negative binding may cancel, but two positive copies, two negative copies, or any larger multiplicity are undefined even if later components would cancel the excess.

The subscript $\Gamma$ records which formula each component interprets. This label information matters only for cancellation: if two components interpret syntactically dual primitive formulas $F$ and $\overline F$, then their signed multimaps must be exact opposites. This prevents a positive occurrence of a formula from cancelling a negative occurrence justified by a different witness. When the labelled sum is defined, it drops zero-weight bindings. The empty sum is $\emptyset$. We omit the subscript when the context is clear.
\[
\mu \SATSR \Delta \quad\text{iff}\quad \exists F_j \in \Delta,\ \mu \SATSP F_j .
\]

\[
\Gamma\vDash\Delta \quad\text{iff for every } \vec{\mu}\text{ and }\mu',
\text{ if } \vec{\mu}\SATSL\Gamma \text{ and } \mu'=\bigoplus_\Gamma\vec{\mu}
\text{ is defined, then } \mu'\SATSR\Delta .
\]

For a concrete example, let $\theta_x=\{x\mapsto\tunit\}$ and $\mu_x=\langle\theta_x\rangle$. Evaluating $x$ consumes its binding and returns $\tunit$, leaving any compatible surrounding map untouched. Hence
\[
    \mu_x\SATS x:\Tunit,
    \qquad -\mu_x\SATS\overline{x:\Tunit}.
\]
For $\Gamma=x:\Tunit,\overline{x:\Tunit}$, the vector $(\mu_x,-\mu_x)$ satisfies the left context and has labelled sum $\emptyset$: the two occurrences of the binding cancel. Every defined sum for this context is empty, since dual formulas require opposite witnesses. The empty map satisfies $\tunit:\Tunit$ on the right, giving $\Gamma\vDash\tunit:\Tunit$. This is the cancellation used by $\text{T-Ann}$.

\begin{proposition}[Soundness of type disjointness]
    If $A \parallel B$, then there are no value $v$ and ordinary maps $\theta_A,\theta_B$ such that $(\theta_A,v)\in\VALS{A}$ and $(\theta_B,v)\in\VALS{B}$.
\end{proposition}

\begin{proposition}[Soundness of typing judgments]
    If $\Gamma \vdash \Delta$ is derivable, then $\Gamma \vDash \Delta$.
\end{proposition}

Our technical report records the full operational semantics, disjointness relation, and typing rules. Because this study is mainly exploratory, the soundness statements above are propositions that explain the intended semantic connection rather than a complete formal development. The key points are the revised $\onlyto$ semantics, which supports $\text{T-AppL}$, and the context-labelled sum, which gives the exact cancellation needed by the structural rules.

Because linear resources cannot be copied or discarded, it is tempting to conjecture that a sequent $s:A \vdash t:B$ expresses equality of resources and therefore should imply the reverse sequent $t:B \vdash s:A$. This is not the case. The formulas describe exact resources, but a sequent is still a directional entailment between formulas, and the proof rules are directional. For example, $x:\Tunit \vdash \ltlambda{y}{(x,y)}:\Tunit \to (\Tunit,\Tunit)$ is derivable by $\text{T-Abs}$ and $\text{T-Pair}$, but the reverse sequent is not derivable in the current system: there is no left rule that decomposes a lambda value of arrow type into the captured variable $x$.

\subsubsection{Discussion}
\label{sec:linear-discussion}

The linear system keeps the broad two-sided shape of Sections~\ref{sec:language} and~\ref{sec:semantics}: right rules explain how a term can be constructed with a type, left rules explain what must be true if a term has a type, and the semantic soundness proposition connects derivability with a semantic interpretation of formulas and sequents. Necessity arrows play the same conceptual role as before: they let a proof extract necessary information about inputs from information about outputs.

The main difference is the resource model. In the AARA-based system, resources are numeric potentials ordered by availability, so judgments naturally express upper and lower bounds using an affine discipline. In the linear system, every value is a resource and the accounting is exact. Evaluation uses ordinary maps because a running program owns concrete values, while sequent semantics uses signed multimaps because proofs must also represent cancellation between a formula and its dual.

This exact accounting explains why dual formulas become important here. They are not an arbitrary extension of the earlier system; they are the mechanism that lets a two-sided proof talk about closures that capture linear resources while still preserving the intended resource equation. The price is extra semantic bookkeeping, especially around captured contexts and cancellation.

Thus the linear system is best understood as a parallel exploration. It shows what a two-sided type system looks like when all resources are one-shot and exact, and it clarifies why the main AARA system uses a different discipline: reusable program values and ordered numeric resources rather than exact resource balance.

\clearpage
\subsubsection{Full Operational and Typing Rules}
\label{subsec:linear-definition}

This subsection collects the full operational semantics, disjointness relation, and typing rules for the two-sided linear type system introduced in our technical report.

\begin{tcolorbox}[linearnontyping,title=Evaluation]
$$
\begin{array}{c}
\text{(E-Var)} \inferrule{\ }{(x, \theta[x \mapsto v]) \to (v, \theta)} \\

\text{(E-LetUnit1)}
\inferrule{(t_1,\theta) \to (t_1',\theta')}
          {(\ltletunit{t_1}{t_2}, \theta) \to (\ltletunit{t_1'}{t_2}, \theta')}
\\
\text{(E-LetUnit2)}
\inferrule{\ }{(\ltletunit{\tunit}{t_2}, \theta) \to (t_2, \theta)} \\

\text{(E-If1)}
\inferrule{(t_1,\theta) \to (t_1',\theta')}
          {(\tif{t_1}{t_2}{t_3},\theta) \to (\tif{t_1'}{t_2}{t_3},\theta)}
\\
\text{(E-IfTrue)}
\inferrule{\ }{(\tif{\ttrue}{t_2}{t_3},\theta) \to (t_2,\theta)}
\\
\text{(E-IfFalse)}
\inferrule{\ }{(\tif{\tfalse}{t_2}{t_3},\theta) \to (t_3,\theta)} \\

\text{(E-App1)}
\inferrule{(t_1,\theta) \to (t_1',\theta')}
          {(t_1\ t_2,\theta) \to (t_1'\ t_2, \theta')}
\\
\text{(E-App2)}
\inferrule{(t_2,\theta) \to (t_2',\theta')}
          {(v_1\ t_2,\theta) \to (v_1\ t_2', \theta')} \\
\text{(E-AppAbs)}
\inferrule{\ }{(\ltlambda{x}{t}\ v_2,\theta) \to (t, \theta[x \mapsto v_2])} \\

\text{(E-Pair1)}
\inferrule{(t_1,\theta) \to (t_1',\theta')}
          {((t_1,t_2),\theta) \to ((t_1',t_2), \theta')}
\\
\text{(E-Pair2)}
\inferrule{(t_2,\theta) \to (t_2',\theta')}
          {((v_1,t_2),\theta) \to ((v_1,t_2'), \theta')} \\

\text{(E-LetPair1)}
\dfrac{(t_1,\theta)\to(t_1',\theta')}
      {\begin{aligned}
          (\ltletpair{t_1}{x}{y}{t_2},\theta) &\to{}\\
          &(\ltletpair{t_1'}{x}{y}{t_2},\theta')
       \end{aligned}}
\\
\text{(E-LetPair2)}
\dfrac{\ }{
       \begin{aligned}
          (\ltletpair{(v_1,v_2)}{x}{y}{t_2},\theta) &\to{}\\
          &(t_2,\theta[x\mapsto v_1][y\mapsto v_2])
       \end{aligned}}
\end{array}
$$
\end{tcolorbox}

\begin{tcolorbox}[linearnontyping,title=Disjointness Rules]
$$
\begin{array}{c}
\text{(D-Unit)}
\inferrule{T \in \{\Tbool, A \to B, A \onlyto B, (A, B)\}}
          {\Tunit \parallel T}
\\
\text{(D-Bool)}
\inferrule{T \in \{\Tunit, A \to B, A \onlyto B, (A, B)\}}
          {\Tbool \parallel T}
\\
\text{(D-Arrow)}
\inferrule{T_1 \in \{A \to B, A \onlyto B\} \\
           T_2 \in \{\Tunit, \Tbool, (A, B)\}}
          {T_1 \parallel T_2}
\\
\text{(D-Pair)}
\inferrule{T \in \{A \to B, A \onlyto B, \Tunit, \Tbool\}}
          {(A, B) \parallel T}
\end{array}
$$
\end{tcolorbox}

\begin{tcolorbox}[lineartyping,title=Structural Rules]
$$
\begin{array}{c}
\text{(T-Id)} \inferrule{ }{t:A \vdash t : A}
\quad
\text{(T-Ann)} \inferrule{\Gamma \vdash \Delta}{\Gamma, \overline{t:T}, t:T \vdash \Delta}
\\
\text{(T-Dual)} \inferrule{\overline{\Gamma} \vdash \overline{\Delta}}{\Gamma \vdash \Delta}
\end{array}
$$
\end{tcolorbox}

\begin{tcolorbox}[lineartyping,title=Right Rules]
As in our technical report, the context $\Gamma$ in the abstraction rules must be capturable.

$$
\begin{array}{c}
\text{(T-Unit)} \inferrule{ }{\emptyset \vdash \tunit : \Tunit}
\quad
\text{(T-Bool)} \inferrule{b \in \{\ttrue,\tfalse\}}{\emptyset \vdash b : \Tbool} \\

\text{(T-LetUnit)}
\inferrule{\Gamma_1 \vdash t_1 : \Tunit,\Delta_1 \\
           \Gamma_2 \vdash t_2 : A,\Delta_2}
          {\Gamma_1, \Gamma_2 \vdash \ltletunit{t_1}{t_2} : A, \Delta_1,\Delta_2}
\\
\text{(T-If)}
\inferrule{\Gamma_1 \vdash t_1:\Tbool,\Delta_1 \\
           \Gamma_2 \vdash t_2:A,\Delta_2 \\
           \Gamma_2 \vdash t_3:A,\Delta_2}
          {\Gamma_1,\Gamma_2 \vdash \tif{t_1}{t_2}{t_3}:A,\Delta_1,\Delta_2}
\\
\text{(T-Abs)}
\inferrule{\Gamma, x:A \vdash t : B,\Delta}
          {\Gamma \vdash \ltlambda{x}{t} : A \to B,\Delta}
\quad
\text{(T-AbsN)}
\inferrule{\overline{\Gamma}, t:B \vdash x:A, \overline \Delta}
          {\Gamma \vdash \ltlambda{x}{t} : A \onlyto B,\Delta}
\\
\text{(T-App)}
\inferrule{\Gamma_1 \vdash t_1 : A \to B,\Delta_1 \\
           \Gamma_2 \vdash t_2 : A,\Delta_2}
          {\Gamma_1, \Gamma_2 \vdash t_1\ t_2 : B,\Delta_1,\Delta_2}
\\
\text{(T-Pair)}
\inferrule{\Gamma_1 \vdash t_1 : A,\Delta_1 \\
           \Gamma_2 \vdash t_2 : B,\Delta_2}
          {\Gamma_1, \Gamma_2 \vdash (t_1, t_2) : (A, B),\Delta_1,\Delta_2}
\\
\text{(T-LetPair)}
\inferrule{\Gamma_1 \vdash t_1 : (A, B),\Delta_1 \\
           \Gamma_2, x:A, y:B \vdash t_2 : C,\Delta_2}
          {\Gamma_1, \Gamma_2 \vdash \ltletpair{t_1}{x}{y}{t_2} : C,\Delta_1,\Delta_2}
\end{array}
$$
\end{tcolorbox}

\begin{tcolorbox}[lineartyping,title=Left Rules]
$$
\begin{array}{c}
\text{(T-Dis)}
\inferrule{\Gamma\vdash t : B,\Delta \quad A \parallel B}
          {\Gamma, t : A \vdash \Delta}
\quad
\text{(T-BoolL)}
\inferrule{\Gamma \vdash \Delta \quad b \in \{\ttrue, \tfalse\}}
          {\Gamma, b : \Tbool \vdash \Delta}
\\
\text{(T-UnitL)}
\inferrule{\Gamma \vdash \Delta}
          {\Gamma, \tunit : \Tunit \vdash \Delta}
\quad
\text{(T-LetUnitL)}
\inferrule{\Gamma, t_1 : \Tunit, t_2 : A \vdash\Delta}
          {\Gamma, \ltletunit{t_1}{t_2} : A \vdash\Delta}
\\
\text{(T-IfL)}
\inferrule{\Gamma_2,t_2:A \vdash \Phi \\
           \Gamma_2,t_3:A \vdash \Phi \\
           \Gamma_1,t_1:\Tbool,\Phi \vdash \Delta}
          {\Gamma_1,\Gamma_2,\tif{t_1}{t_2}{t_3}:A \vdash \Delta}
\\
\text{(T-AppL)}
\inferrule{\Gamma_1 \vdash t_1 : A \onlyto B; \Phi \\
           \Gamma_2, t_2 : A,\Phi \vdash \Delta}
          {\Gamma_1, \Gamma_2, t_1\ t_2 : B \vdash \Delta}
\\
\text{(T-PairL)}
\inferrule{\Gamma, t_1:A, t_2:B \vdash \Delta}
          {\Gamma, (t_1, t_2) : (A, B) \vdash \Delta}
\\
\text{(T-LetPairL)}
\inferrule{\Gamma_2, t_2 : C \vdash (x,y) : (A, B) ; \Phi \\
           \Gamma_1, t_1 : (A, B),\Phi \vdash \Delta}
          {\Gamma_1, \Gamma_2, \ltletpair{t_1}{x}{y}{t_2} : C\vdash \Delta}
\end{array}
$$
\end{tcolorbox}

\end{document}